\documentclass[amsmath,amssymb,superscriptaddress,aps,prx,reprint, 10pt]{revtex4-2}

\usepackage{layouts}

\usepackage[english]{babel}
\usepackage[utf8]{inputenc}
\usepackage[T1]{fontenc}                        
\usepackage{verbatim}
\usepackage{soul}

\usepackage{graphicx}
\usepackage{xcolor}
\usepackage{sidecap} 
\sidecaptionvpos{figure}{t}                        
\usepackage[section,below]{placeins}            
\usepackage{mwe}
\usepackage{layouts}

\usepackage{tikz}
\usetikzlibrary{shapes.geometric, arrows}
\usetikzlibrary{decorations.markings}
\usetikzlibrary{decorations.pathmorphing}
\usetikzlibrary{decorations.pathreplacing}
\usetikzlibrary{shapes.symbols}

\usepackage{bm}
\usepackage{mathtools}                          
\usepackage{stmaryrd}
\usepackage{physics}
\allowdisplaybreaks                             

\usepackage{booktabs}
\usepackage{enumerate}
\usepackage{hyperref}
\hypersetup{hidelinks,
            colorlinks=true,
            allcolors=[RGB]{0,84,159}}

\usepackage[capitalise]{cleveref}               
\crefname{fig_a}{Fig.}{Fig.}                    
\crefname{fig_b}{Fig.}{Fig.}                    
\crefname{fig_c}{Fig.}{Fig.}                    
\crefname{fig_d}{Fig.}{Fig.}                    
\Crefname{fig_a}{Figure}{Figure}                
\Crefname{fig_b}{Figure}{Figure}
\Crefname{fig_c}{Figure}{Figure}
\Crefname{fig_d}{Figure}{Figure}
\creflabelformat{fig_a}{#2{#1(a)}#3}
\creflabelformat{fig_b}{#2{#1(b)}#3}
\creflabelformat{fig_c}{#2{#1(c)}#3}
\creflabelformat{fig_d}{#2{#1(d)}#3}

\newcommand*\subtxt[1]{_{\textnormal{#1}}}
\DeclareRobustCommand\_{\ifmmode\expandafter\subtxt\else\textunderscore\fi}

\newcommand*\supertxt[1]{^{\textnormal{#1}}}
\DeclareRobustCommand\^{\ifmmode\expandafter\supertxt\else\textasciicircum\fi}

\newcommand{\vast}{\bBigg@{4}}
\newcommand{\Vast}{\bBigg@{5}}

\newcommand{\bS}{{\bf S}}
\newcommand{\br}{{\bf r}}

\newcommand{\bk}{{\bf k}}

\newcommand{\bq}{{\bf q}}

\DeclareMathOperator{\arctanh}{arctanh}

\DeclarePairedDelimiter\floor{\lfloor}{\rfloor}

\newcommand{\affiliationRWTH}{
Institut f\"ur Theorie der Statistischen Physik, RWTH Aachen University and JARA-Fundamentals of Future Information Technology, 52056 Aachen, Germany
}
\newcommand{\affiliationMPSD}{
Max Planck Institute for the Structure and Dynamics of Matter,
Center for Free-Electron Laser Science (CFEL),
Luruper Chaussee 149, 22761 Hamburg, Germany
}

\newcommand{\affiliationMPIPKS}{
Max Planck Institute for the Physics of Complex Systems, 
N\"othnitzer Strasse 38, Dresden, 01187, Germany
}

\newcommand{\affiliationUPV}{
Nano-Bio Spectroscopy Group,
Departamento de F\'isica de Materiales,
Universidad del Pa\'is Vasco,
20018 San Sebastian, Spain
}

\newcommand{\affiliationICCQ}{
Initiative for Computational Catalysis (ICC),
Flatiron Institute, Simons Foundation,
New York City, NY 10010, USA
}

\begin{document}

\title{Cavity Control of Strongly Correlated Electrons Beyond Resonant Coupling}

\author{Lukas Grunwald}
\thanks{These authors contributed equally}
\affiliation{\affiliationRWTH}
\affiliation{\affiliationMPSD}

\author{Xinle Cheng}
\thanks{These authors contributed equally}
\affiliation{\affiliationMPSD}

\author{Emil Vi\~nas Bostr\"om}
\email{emil.bostrom@mpsd.mpg.de}
\affiliation{\affiliationMPSD}

\author{Michael Ruggenthaler}
\affiliation{\affiliationMPSD}

\author{Marios H. Michael}
\affiliation{\affiliationMPIPKS}
\affiliation{\affiliationMPSD}

\author{Dante M.~Kennes}
\email{dante.kennes@rwth-aachen.de}
\affiliation{\affiliationRWTH}
\affiliation{\affiliationMPSD}

\author{Angel Rubio}
\email{angel.rubio@mpsd.mpg.de}
\affiliation{\affiliationMPSD}
\affiliation{\affiliationICCQ}
\affiliation{\affiliationUPV}

\date{\today}

\begin{abstract}
    Interfacing materials with electromagnetic cavities offers a route to modify equilibrium properties through structured vacuum fluctuations. The coupling between light and correlated electrons lacks a characteristic energy scale, making vacuum induced ground state modifications of such systems inherently off-resonant and sensitive to the full photon mode structure.
    We develop a consistent cavity-QED formalism for dispersive electromagnetic environments in the Coulomb gauge, capturing both dynamical dressing via the vector potential and static screening by the dielectric. With this formalism, we perform a non-perturbative study of the cavity-induced modification of magnetism in the Hubbard model close to half-filling, including all cavity modes and with parameters determined from first principles.
    At half-filling, we show that the modification of the magnetic exchange interaction $J$ is controlled by a generalized Purcell factor, proportional to the frequency integrated photonic density of states relative to free space. This result identifies polaritonic surface cavities as promising platforms to modify correlated systems.
    For the surface cavity a competition between static screening and dynamical dressing via the vector potential leads to a net enhancement of $J$, directly observable in two-magnon Raman spectroscopy. The inclusion of screening is essential to obtain even qualitatively correct results.
    At weak doping, a variational exact diagonalization of the cavity-coupled $t$-$J$ polaron reveals that a low-frequency surface cavity can reverse the nodal-antinodal dichotomy of the bare polaron dispersion. This effect lies beyond mean field theory and is observable via ARPES measurements.
    Our framework establishes a concrete design principle linking cavity geometry to material response in the off-resonant regime, which will guide future experimental and theoretical explorations.
\end{abstract}

\maketitle

\section{Introduction}

\begin{figure}[t]
    \centering
    \includegraphics[width = \linewidth]{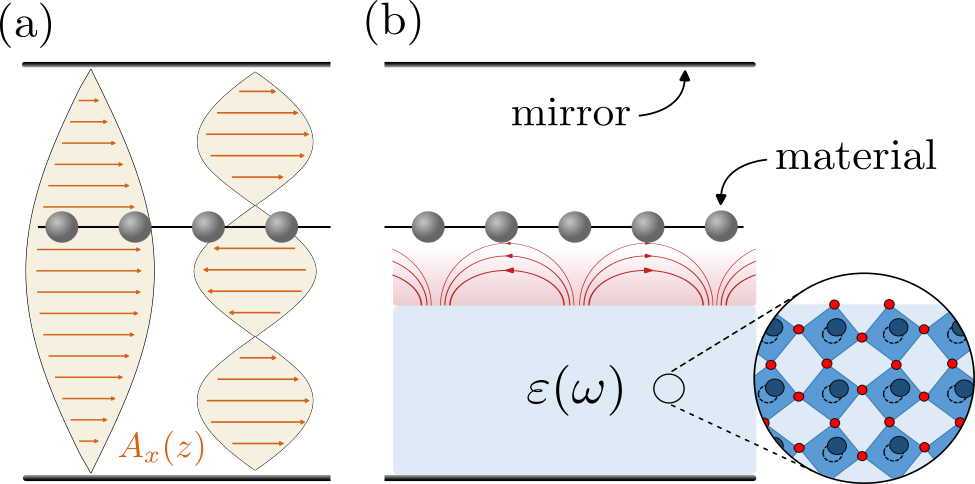}
    \caption{\textbf{Off-resonant ground state control of strongly correlated materials}. (a) Fabry-P\'erot resonator, illustrated as field lines of the modes $A_x(z)$, with a centrally embedded material.  The equally spaced modes redistribute spectral weight periodically in frequency. This modulation averages out upon frequency integration and off-resonant modifications stay negligible.
    (b) Surface polariton cavity interacting with a material above a dispersive dielectric substrate (blue, $\varepsilon(\omega)$). Substrate excitations (inset) hybridize with the electromagnetic field producing exponentially localized polaritonic surface modes at the vacuum-dielectric interface (red). The material couples to both longitudinal and transverse fields, creating an interplay of static screening and dynamical dressing that drives sizable ground-state modifications. At half-filling we find a percent-level enhancement of magnetic exchange, while at slight doping, we show a nodal-antinodal transition in the polaron dispersion.}
    \label{fig:graphical_abstract}
\end{figure}


Embedding quantum materials inside structured electromagnetic environments offers a route to modify equilibrium properties by coupling to the quantized photon field \cite{luCavityEngineering2025,schlawinCavityQuantum2022,kennesNewEraQuantum2022,garcia-vidalManipulatingMatter2021}. Unlike laser driven approaches, such modifications are non-transient, making them attractive for static on-chip material functionalization \cite{kippCavityElectrodynamics2025}. Recent experiments support this prospect, demonstrating dark cavity modifications of charge density wave \cite{jarcCavitymediatedThermal2023}, (fractional) quantum Hall \cite{paravicini-baglianiMagnetotransportControlled2019,appuglieseBreakdownTopological2022,enknerTunableVacuumfield2025}, and superconducting orders \cite{kerenCavityalteredSuperconductivity2026,Xu2026}. The nature of the light-matter coupling depends strongly on which degrees of freedom the cavity addresses. Energetically narrow excitations, such as optical phonons or excitons, couple resonantly at a well-defined frequency in the spirit of conventional polaritonic physics~\cite{Gao2018,Dirnberger2022,Dirnberger2023,Kim2025}. By contrast, correlated electrons, as paradigmatically appearing in the Mott insulating phase, lack a characteristic energy scale and hence probe the photonic environment across a broad spectral range. The coupling is therefore believed to be inherently off-resonant \cite{schaferRelevanceQuadratic2020,rokajFreeElectron2022}, a regime recently conceptualized as endyonic physics \footnote{{\it endyo} (``to put on, to be covered, to clothe/dress''), and is used to denote matter degrees of freedom dressed by vacuum fluctuations. It was coined during the Flagship School on ``Ab Initio Quantum Electrodynamics for Quantum Materials Engineering'' held at the Flatiron Institute in New York City, September 29–October 3, 2025.}, suggesting that a faithful theoretical description must account for the full photon mode spectrum.

Accurately describing such vacuum dressed states poses a major theoretical challenge. Most existing approaches reduce the electromagnetic environment to a single cavity mode with a phenomenologically assigned coupling constant, though microscopic derivations exist for specific cavity geometries \cite{svendsenTheoryQuantum2023}. While often providing qualitative insights, such single-mode descriptions obscure the role of the full photon mode structure and offer limited guidance on which cavity geometries can produce measurable modifications of material ground states. A multi-mode treatment retaining the full cavity spectrum is therefore essential for connecting cavity design to material response. Such a treatment of quantum electrodynamics (QED) in cavities has so far been lacking for correlated electron systems.

Here, we develop a theory of cavity QED and build a non-perturbative framework to study cavity modifications of the magnetic exchange $J$ and spin-polaron dispersion, in the strongly correlated Hubbard model close to half-filling, within a Born Oppenheimer approximation that neglects light-matter interactions of nuclear degrees of freedom \cite{luCavityEngineering2025,flickAtomsMolecules2017}.  These results are relevant for cuprate parent compounds such as La$_2$CuO$_4$~\cite{Lee2006}. Through a non-perturbative resummation over all photon modes, we derive the cavity-dressed $J$ to all orders in the light-matter coupling, which is determined from first principles. We consider Fabry-P\'erot cavities, consisting of two planar metallic mirrors, as well as surface polariton cavities, where optically active substrate excitations hybridize with the electromagnetic field to produce spatially localized modes at the vacuum-substrate interface (see \cref{fig:graphical_abstract}). For the latter, we develop a Coulomb gauge Hopfield quantization scheme that consistently accounts for both dynamical vector potential dressing, and the static screening of the Coulomb interaction by the dielectric environment. The screening contribution is absent in existing frameworks, and its omission can lead to qualitatively incorrect results: for the system considered here, neglecting it reverses the sign of the modification in $J$.

Within this framework, we find that off-resonant vacuum modifications of the magnetic exchange are governed by a generalized Purcell factor: the relevant figure of merit is the frequency-integrated photonic density of states (PDOS), rather than the PDOS at a single transition frequency as in the standard Purcell effect \cite{purcellSpontaneousEmission1964}. This principle immediately discriminates between different cavity architectures: standard Fabry-P\'erot resonators, which redistribute spectral weight periodically across a broad frequency range, show near-complete cancellation upon integration and result in negligible modifications within the approximation considered here. In contrast, polaritonic surface cavities concentrate spectral weight into a narrow peak near the surface mode frequency, which survives integration. This is specific to the off-resonant regime. Resonant pathways, such as the optical phonons or excitons mentioned above, sample the PDOS locally and can benefit from conventional Fabry-Pérot resonators. For a surface cavity consisting of a gold substrate, the competition between dynamical dressing and static screening yields a net enhancement of $J$ at the percent level for nanometer scale separations. This is directly measurable via two-magnon Raman spectroscopy. We further show that the strongly peaked surface PDOS admits a rigorous single-mode approximation, providing {\it ab initio} coupling constants for effective single-mode models.

Extending the analysis to the underdoped regime of the strongly correlated Hubbard model, we solve for the dispersion of the magnetic polaron in the downfolded $t$-$J$ model. While the cavity modification of exchange $J$ is suppressed by an incomplete cancellation between dynamical dressing and static screening, the polaron dispersion is primarily sensitive to the former. Here, the cavity qualitatively changes the polaron dispersion, reversing the nodal-antinodal dichotomy in the strongly-coupled regime. Our findings thus establish off-resonant cavity engineering as an effective and unique way to control strongly correlated electronic systems.

  
The paper is organized as follows: In \cref{sec:hopfield}, we develop a Coulomb gauge Hopfield quantization scheme and derive the resulting light-matter Hamiltonian. \cref{sec:low_energy} presents the low-energy effective theory, obtained via a cavity Schrieffer-Wolff transformation, and the non-perturbative calculation of the magnetic exchange modification resulting from all cavity modes. In \cref{sec:magnetic_exchange}, we evaluate the general expressions for a Fabry-Pérot and a plasmonic surface cavity, demonstrating the failure of the former and the viability of the latter to generate substantial magnetic modifications. We show that the modification of the magnetic exchange is observable in the two-magnon Raman spectrum, and in the dynamical spin susceptibility. In \cref{sec:tj}, we extend the analysis to the single-hole $t$-$J$ problem and show that a surface cavity can qualitatively reshape the polaron dispersion. We conclude in \cref{sec:conclusion} with a discussion of implications for cavity material engineering and experimental signatures.

\section{Cavity Quantum Electrodynamics in the Dynamical Weyl Gauge}
\label{sec:hopfield}

A complete theory of cavity QED requires two ingredients: the quantization of electromagnetic modes in a given photonic environment \cite{ruggenthalerUnderstandingPolaritonic2023}, and their coupling to external charged matter. We begin with the first. If the scalar potential $\phi$ is assumed to be zero, the vector potential can be decomposed as \cite{luCavityEngineering2025}
\begin{align}
    \vb{A}(\vb{r}) = \sum_\lambda
    \sqrt{\frac{\hbar}{2\epsilon_0 \omega_\lambda}}
    \left(
        \vb{f}^*_\lambda(\vb{r}) a_\lambda +
        \vb{f}_{\lambda}(\vb{r}) a_{\lambda}^\dagger
    \right),
    \label{eq:gauge_vector_decomp}
\end{align}
where $\lambda$ is a mode index, and $a_\lambda$ ($a_\lambda^\dagger$) is a bosonic ladder operator that annihilates (creates) a photon in the corresponding mode. They satisfy the canonical commutation relations $\comm*{a_\lambda}{a_{\lambda'}^\dagger} = \delta_{\lambda \lambda'}$. The mode function $\vb{f}_\lambda(\vb{r})$ and its associated frequency $\omega_\lambda$ encode all geometric and material properties of the photonic environment, and are obtained as solutions of the frequency-space wave equation in the quantization volume $\Omega$ %
\footnote{In the literature usually denoted as Helmholtz eigenvalue equation \cite{jacksonClassicalElectrodynamics2009}.}
\begin{equation}
    \frac{\omega_\lambda^2}{c^2}\varepsilon(\vb{r}, \omega_\lambda)\vb{f}_\lambda(\vb{r}) = \nabla\times\nabla\times\vb{f}_\lambda(\vb{r}), 
    \quad \vb{r} \in \Omega.
    \label{eq_eig_f}
\end{equation}
This equation is to be supplemented by appropriate boundary conditions on the surface of $\Omega$, and can be shown to full-fill the generalized Coulomb gauge constraint $\nabla \cdot [\varepsilon\,\vb{f}_\lambda] = 0$ (see \cref{sec:appdx_mode_functions}). For non-dispersive media, the relative dielectric function $\varepsilon(\vb{r}, \omega)$ is frequency independent, describing an inert, non-absorbing medium. The mode functions are orthonormal under the inner product $\int_\Omega \dd{\vb{r}} \varepsilon\,\vb{f}_\lambda^* \cdot\vb{f}_{\lambda'} = \delta_{\lambda\lambda'}$, and the free-field Hamiltonian reduces to $H_\mathrm{ph} = \sum_\lambda \hbar\omega_\lambda\, a_\lambda^\dagger a_\lambda$. In vacuum, with periodic boundary conditions, the modes are transverse plane waves $\vb{f}_\lambda \sim |\Omega|^{-1/2} \vb{e}^{\pm}_{\vb{k}}e^{-i \vb{r} \cdot \vb{k}}$ with $\vb{e}^{\pm}_{\vb{k}}$ perpendicular to momentum the $\vb{k}$ and with dispersion $\omega = c|\vb{k}|$. A cavity reshapes this spectrum while preserving the structure of \cref{eq:gauge_vector_decomp,eq_eig_f}.

In a dielectric substrate, the situation is more involved but well established. The frequency dependence of $\varepsilon(\vb{r}, \omega)$ reflects the presence of optically active excitations in the substrate, plasmons or optical phonons, that hybridize with the electromagnetic field.  A consistent quantization must then explicitly include the corresponding substrate degrees of freedom as dynamical variables. Modeling them via harmonic polarization fields yields the Hopfield quantization scheme~\cite{glauberQuantumOptics1991,lenacQuantumOptics2003,gubbinRealspaceHopfield2016}, sometimes referred to as quantization in Weyl gauge~\cite{bechler2006path, ferreira2020quantization} since the modes are described entirely through the vector potential $\vb{A}$. The Hopfield scheme preserves the structure of \cref{eq:gauge_vector_decomp,eq_eig_f}, but introduces a modified normalization condition $\int_\Omega \mathrm{d}\vb{r}\,(\varepsilon+\frac{\omega}{2} \partial_\omega\varepsilon)\,|\vb{f}_\lambda|^2 = 1$ that accounts for the energy stored in the material polarization. Because the substrate excitations carry longitudinal dipole fields
\footnote{This is not unique for dispersive media. Longitudinal fields can also appeasr for non-dispersive, but spatially varying dielectrics.}, the mode functions $\vb{f}_\lambda$ now acquire both transverse and longitudinal components.
 
Existing Hopfield treatments have been formulated for a free-standing dielectric, without considering the coupling to additional charged matter. This is the second ingredient required for a complete cavity-QED description. When external electrons are introduced, the Coulomb interaction between those electrons and the dielectric must be treated consistently. In vacuum, the dynamical electromagnetic field involves only transverse degrees of freedom, and the longitudinal part of the vector potential gives rise to the instantaneous Coulomb interaction [see~\cref{sec:weyl_gauge_vacuum}]. In contrast, in a polaritonic cavity the longitudinal component hybridizes with substrate excitations, and through this acquires a finite frequency and becomes dynamical. The static Coulomb interaction between charged particles close to such a dielectric therefore deviates from its vacuum form. Here, we develop a full cavity-QED framework that
accounts for this situation, with details given in~\cref{sec:appdx_mirror_charge}.

We consider a system consisting of three parts: the free electromagnetic field, an optically active excitation of the dielectric medium, and the external electronic system that we aim to modify. The optical excitation -- either a plasmon or an optical phonon -- is described by the displacement $\vb{Q}_i$ and the physical momentum $\vb{P}_i$, labeled by an index $i$. To pass to a continuum description, we coarse-grain over a mesoscopically large region $\Omega_{\vb{r}}$ around position $\vb{r}$, denoting the average of $\vb{Q}_i$ and $\vb{P}_i$ in $\Omega_{\vb{r}}$ by $\bar{\vb{Q}}(\vb{r})$ and $\bar{\vb{P}}(\vb{r})$. We then define a dipolar vector field $\vb{X}(\vb{r}) = \rho^{1/2}\bar{\vb{Q}}(\vb{r})$ and its conjugate momentum field $\vb{\Pi}(\vb{r})=\rho^{-1/2}n\bar{\vb{P}}(\vb{r})$, where $\rho$ and $n$ are the mass density and number density of the quasi particles. $\vb{X}$ and $\vb{\Pi}$ satisfy the canonical commutation relation $[X^i(\vb{r}), \Pi^j(\vb{r}')] = i \hbar \delta_{ij}\delta(\vb{r}-\vb{r}')$. To characterize the strength of the optical response for the medium, we introduce the plasma frequency $\omega_p = \sqrt{Z^2e^2/\epsilon_0\rho}$, where $Ze$ is the effective charge of the particle. With these definitions, the electric polarization equals $\omega\_p\vb{X}$. In the Coulomb gauge, the Hamiltonian of this system reads
\begin{equation}
    H = H_\mathrm{e} + V + H_\mathrm{ph} +\int_\Omega\mathrm{d}\vb{r}\left[\rho_\mathrm{e}\phi-\vb{J}_\mathrm{e}(\vb{A}_\mathrm{T})\cdot\vb{A}_\mathrm{T}\right] ,    \label{eq_ham_cavity_qed}
\end{equation}
where $H_\mathrm{e}$ describes the non-interacting electron system, and $\vb{J}_\mathrm{e}(\vb{A}_\mathrm{T})$ and $\rho_\mathrm{e}$ are the (effective) electronic current %
\footnote{For Dirac electron, $\vb{J}\_e$ is the physical current $-ev_F\bar{\psi}\bm{\gamma}\psi$, and is independent of $\vb{A}$. However in general the light-matter coupling is nonlinear, $\vb{J}\_e$ is a function of $\vb{A}$ and differs from the physical current. Any local light-matter coupling can be written in such a form, and we adopt this form instead of $H_\mathrm{l-m}$ for notational concreteness.}
%
and charge density operators, respectively. Throughout, the subscripts $\mathrm{T}$ and $\mathrm{L}$ denote the transverse ($\vb*{\nabla} \cdot \vb{A}\_T = 0$) and longitudinal ($\vb*{\nabla} \cross \vb{A}\_L = 0$) components of a vector field. The bare Coulomb repulsion $V$ between the electrons is supplemented by the scalar potential $\phi$, which mediates the electrostatic coupling between the electronic charge density and the longitudinal component of the dielectric displacement $\vb{X}$. We note that this is the contribution missed in previous discussions. Explicitly,
\begin{align}
  \phi({\vb{r}}) &= -\frac{1}{4\pi \sqrt{\epsilon_0}} \int_\Omega\mathrm{d}\vb{r}' \frac{\nabla'\cdot\left[\omega\_p (\vb{r}')\vb{X}(\vb{r}')\right]}{|\vb{r}-\vb{r}'|} \\
  H_\mathrm{ph} &= \frac{1}{2} \int_\Omega\mathrm{d}\vb{r}\,\left[\epsilon_0\vb{E}_\mathrm{T}^2+\mu_0^{-1}(\nabla\times\vb{A}_\mathrm{T})^2 \right] \nonumber \\
  &+ \frac{1}{2} \int_\Omega\mathrm{d}\vb{r}\, \left[(\vb{\Pi} - \sqrt{\epsilon_0}\omega\_p \vb{A}_\mathrm{T})^2+\omega^2_{\rm TO} \vb{X}^2 + (\omega\_p \vb{X})_\mathrm{L}^2 \right], \nonumber
\end{align}
with $\omega_\mathrm{TO}$ denoting the transverse optical (TO) phonon frequency and $\omega\_{p}$ the plasma frequency, both are generally spatially dependent. The last term in $H_{\rm ph}$ comes from the Coulomb self-interaction of the phonon field, and can be written in the same form as the electronic Coulomb interaction $V$. Together with $\phi$ and $V$, these terms constitute the full Coulomb interaction of the total charge density of the system, which comprises both electronic and dielectric polarization contributions.
The transverse fields fulfill $[E^i_\mathrm{T}(\vb{r}),A^j_\mathrm{T}(\vb{r}')] = i\hbar\epsilon_0^{-1} \delta_{ij}\delta\_{T}(\vb{r}-\vb{r}')$, with $\delta\_T$ the transversal Dirac $\delta$-function \cite{zinn-justinQuantumFieldTheory2002}.

The polaritonic Hamiltonian $H_\mathrm{ph}$, which is used to
dress the electronic system, is quadratic in the photon and phonon
fields and can be diagonalized by introducing normal mode
operators $a_\lambda$. These operators are linear combinations of the transverse electromagnetic fields and the canonical variables of the polarization field, with coefficients determined by demanding that the equation of motion $[a_\lambda, H_\mathrm{ph}] = -i \omega_\lambda a_\lambda$ and the bosonic commutation relations $\comm*{a_\lambda}{a_{\lambda'}^\dagger} = \delta_{\lambda \lambda'}$ are satisfied. One can introduce a longitudinal vector potential $\vb{A}_\mathrm{L}$ to capture the polarization field, and $\vb{A} = \vb{A}_\mathrm{T}+\vb{A}_\mathrm{L}$ then becomes the vector potential in \cref{eq:gauge_vector_decomp}. The modes functions $\vb{f}_\lambda$ for $\vb{A}$ satisfy \cref{eq_eig_f} and the corresponding Hopfield normalization condition (see \cref{sec:appdx_mirror_charge}). The polaritonic Hamiltonian can be written as $H_\mathrm{ph} = \sum_\lambda \omega_\lambda a_\lambda^\dagger a_\lambda$. In this way, the current framework reproduces the Hopfield quantization for a free-standing medium, but in general introduces an additional light-matter coupling via the scalar potential $\phi(\vb{r})$, that captures the longitudinal excitations of the substrate. 

To express the light-matter coupling entirely in terms of the vector
potential, i.e. obtain a theory in Coulomb gauge with $\phi = 0$, which is computationally convenient, we remove the scalar potential $\phi$ from
\cref{eq_ham_cavity_qed} by a unitary transformation 
\begin{equation}
    U = \exp\left[\frac{i}{4\pi\hbar}\int_\Omega\mathrm{d}\vb{r}\rho_\text{e}(\vb{r})\int_\Omega\mathrm{d}\vb{r}'\frac{\nabla'\cdot\vb{A}\_L(\vb{r}')}{|\vb{r}-\vb{r}'|}\right].
    \label{eq:unitary_u}
\end{equation}
The transformed Hamiltonian $H' = U^\dagger H U$ takes the form
\begin{equation}
     H' =  H_\text{e} + W + H_\mathrm{ph} -\int_\Omega \dd{r}\vb{J}_\text{e}(\vb{A})\cdot\vb{A},
    \label{eq_trans_ham}
\end{equation}
where $\vb{A}(\vb{r})$ retains the mode decomposition of \cref{eq:gauge_vector_decomp} in terms of the mode functions used to diagonalize $H\_{ph}$. The transformation absorbs $\phi$ into a screened Coulomb interaction with interaction kernel $W(\vb{r},\vb{r}')$. It satisfies the Poisson equation with the static dielectric function $\varepsilon(\vb{r},0)$ (see \cref{appdx:electrostatics})
\begin{equation}
    \nabla\cdot\left[\varepsilon(\vb{r},0)\nabla W(\vb{r},\vb{r}')\right]  = -\delta(\vb{r}-\vb{r}')/\epsilon_0,
\end{equation}
and hence precisely describes the interaction of two point charges embedded in the static dielectric environment described by $\varepsilon(\vb{r},0)$, viz. can be obtained from classical electrostatics.  Crucially, this screening effect has been overlooked in previous cavity-QED treatments of condensed matter systems, and as we demonstrate below, its inclusion is essential to obtain even qualitatively correct predictions.

To summarize, we started from the standard Coulomb gauge QED Hamiltonian, with substrate excitations treated explicitly, and diagonalized the photon-phonon sector to obtain polaritonic normal modes. We then removed the additional scalar potential, accounting for the Coulomb interaction between matter and substrate by a unitary transformation. The result, \cref{eq_trans_ham}, is a cavity-QED Hamiltonian that reproduces the Hopfield quantization for a free-standing dielectric, while consistently incorporating the coupling to external charged matter. The result is cast in a form directly amenable to standard condensed matter many-body techniques, where all dynamical light-matter coupling proceeds through the vector potential $\vb{A}$, while the dielectric environment enters through the mode structure and the screened Coulomb interaction $W$. While formally all our results are in the Coulomb gauge, the final Hamiltonian is reminiscent of the expression in the Weyl gauge (apart from the screened Coulomb interaction), and we therefore use the term {\it dynamical Weyl gauge} to refer to this form of the light-matter coupling. 

However, unlike the standard Weyl gauge in vacuum, where longitudinal degrees of freedom are non-dynamical and removed by constraints, the longitudinal components here originate from optical excitations of the medium and remain physical. The explicit starting point of the quantization is the Coulomb gauge [see \cref{sec:weyl_gauge_vacuum}].
In the following, we apply this framework to perform a non-perturbative calculation of the cavity-dressed magnetic exchange interaction, incorporating all cavity modes with coupling constants determined from first principles.



\section{Off-resonant cavity modifications}
\label{sec:low_energy}

\subsection{Electronic theory}
We focus on strongly correlated matter systems, and consider the paradigmatic single band Hubbard model at half-filling and on the square lattice, positioned parallel to the substrate(s). This model is approximately realized by cuprate parent compounds like La$_2$CuO$_4$~\cite{Lee2006}. It can be obtained by downfolding the continuum minimal-coupling Hamiltonian in the dynamical Weyl gauge \cref{eq_trans_ham} to a single-band lattice description and introducing light–matter coupling via Peierls substitution within the long-wavelength approximation \cite{luttingerEffectMagnetic1951,dmytrukGaugeFixingStrongly2021}, as is standard in the literature \cite{eckhardtSurfacemediatedUltrastrong2024,sentefQuantumClassical2020,eckhardtQuantumFloquet2022,passettiCavityLightMatter2023}. This procedure  guarantees charge conservation to all orders ~\cite{distefanoResolutionGauge2019,stefanoReplyGauge2024,garzianoGaugeInvariance2020,stokesGaugeNoninvariance2024,savastaGaugePrinciple2021,liElectromagneticCoupling2020}. 
The matter Hamiltonian is given by
\begin{align}
    H\_{m} = \sum_{\expval{ij}\sigma}
       \left(t_{ij} e^{-i \hat\theta_{ij}} c^\dagger_{i\sigma} c_{j\sigma}+h.c.\right)  + U \sum_i n_{i \uparrow} n_{i \downarrow},
    \label{eq:cavity_hubbard}
\end{align}
with hopping matrix elements $t_{ij}$, screened Hubbard interaction $U = U_0 + \Delta U$ and the Peierls phase in dipole approximation $\hat\theta_{ij} = (e/\hbar) (\vb{R}_i -\vb{R}_j) \cdot \hat{\vb{A}} = \sum_\lambda ( g_\lambda a_\lambda^\dagger + g_\lambda^* a_\lambda)$. Here and in the following, $U_0$ denotes the unscreened Hubbard interaction, $\Delta U$ the static screening contribution, and $U$ the resulting screened interaction. We have introduced the {\it ab initio} light-matter coupling constants
\begin{align}
    g_\lambda(\vb{R}_i) = \sqrt{\frac{e^2 a_{ij}^2}{2 \hbar \epsilon_0 \omega_\lambda}}
    \left[ \vb{e}_{ij} \cdot \vb{f}_\lambda(\vb{R}_i) \right] \equiv g_\lambda
    \label{eq:resum_coupling_constant}
\end{align}
with bond length and direction $ (\vb{R}_i - \vb{R}_j) = a_{ij} \vb{e}_{ij}$ and volume scaling $g_\lambda \sim |\Omega|^{-1/2}$. \Cref{eq:cavity_hubbard} maps the strongly correlated continuum theory to a simpler lattice model, while retaining the full cavity mode structure.

In strongly correlated compounds the magnetic exchange $J$ controls the low energy physics, and determines the magnetic order as well as the bandwidth of the magnon dispersion, making it a key characteristic for such systems. This later feature makes it directly experimentally accessible via two-magnon Raman spectroscopy. To lowest order in $t/U \ll 1$, it can be evaluated as the energy difference between the singlet and triplet states of a two-site (dimer) model $J = E_T - E_S$, both of which are dressed by light-matter interactions.

\subsection{Variational calculation}
\label{sec:main_variational_J}
Before embarking on a full calculation of how a cavity modifies $J$ for a half-filled Hubbard model, we introduce a variational approach that is numerically more tractable, and provides us with physical insight. In the strong coupling limit $t \ll U$, the magnetic exchange can be captured with a two-site model, where the two sites are labeled $i$ and $j$ and located at $\vb{R}_i$ and $\vb{R}_j$. Denoting the vacuum state of the cavity by $\ket{0}$, a variational ansatz for the spin triplet state is $\ket{T} = 2^{-1/2} \left(\ket{i\uparrow, j\downarrow} + \ket{i\downarrow, j\uparrow}\right)\ket{0}$,
which has an energy $E_T = \frac{1}{2}\sum_\lambda\omega_\lambda$ equal to the cavity zero-point motion. Taking the cavity to be in its vacuum state is motivated by the fact that the triplet states span one-dimensional Hilbert spaces~\cite{grunwaldCavitySpectroscopy2024}, leading to an effective decoupling of light and matter. For the spin singlet state, we consider the variational wavefunction
\begin{equation}
\begin{aligned}
    \ket{S} &= \frac{1}{\sqrt{2}} \big( \ket{i\uparrow, j\downarrow} - \ket{i\downarrow, j\uparrow} \big) \ket{0}\\
    &+ \frac{\beta}{\sqrt{2}} \big( \ket{i\uparrow, i\downarrow}\ket{G} + \ket{j\uparrow, j\downarrow}\ket{\bar{G}} \big),
\end{aligned}
\end{equation}
with Gaussian states $\ket{G} = e^{-i s \hat\theta_{ij}}\ket{0}$ and $\ket{\bar{G}} = e^{i s \hat\theta_{ij}}\ket{0}$ \cite{shi2018variational}. The real parameters $\beta$ and $s$ are found by minimizing the energy $E_S = \bra{S}(H_\mathrm{m} + H_\mathrm{ph})\ket{S}/\left<S|S\right>$. Initially consider the minimization with respect to $\beta$, to  obtain
\begin{equation}
    \underset{\beta}{\mathrm{min}}\, (E_S  - E_T) = \frac{1}{2}\left(U_\mathrm{eff}-\sqrt{U^2_\mathrm{eff}+16t_\mathrm{eff}^2}\right).
\end{equation}
This is of the same form as for a Hubbard dimer in vacuum, but with $t$ and $U$ replaced by the effective parameters $t_\mathrm{eff} = t\exp\left[-\frac{1}{4}(1-s)^2\sum_\lambda |g_\lambda|^2\right]$ and $U_\mathrm{eff} = U + \frac{1}{2}s^2\sum_\lambda \omega_\lambda |g_\lambda|^2$. When $t\ll U$, we find that $t_\mathrm{eff}\ll U_\mathrm{eff}$, and the magnetic exchange $J$ can be approximated as
\begin{equation}
    J  \approx \underset{s}{\mathrm{max}}\, \frac{4t^2 e^{-\frac{1}{2}(1-s)^2 \sum_\lambda |g_\lambda|^2}}{U + \frac{1}{2}s^2\sum_\lambda\omega_\lambda |g_\lambda|^2}.
    \label{eq:variational J}
\end{equation}

As discussed in Appendix~\ref{sec:screening}, the screening correction to $U$ is $\Delta U = -\frac{1}{2}\sum_\lambda \omega_\lambda |g_{\lambda,\mathrm{L}}|^2$, the $\rm{L}$ indicating the longitudinal component of the mode functions in the light-matter couplings \cref{eq:resum_coupling_constant}. Substituting this into Eq.~(\ref{eq:variational J}), and taking $s=1$, we obtain the lower bound
\begin{equation}
    J > \frac{4t^2}{U_0+\frac{1}{2}\sum_\lambda\omega_\lambda(|g_\lambda|^2 - |g_{\lambda,\mathrm{L}}|^2)}.
    \label{eq:variational_result}
\end{equation}
This is a lower bound within the family of variational states considered here. As shown in~\cref{sec:appdx_surface_cavity}, the longitudinal modes dominate for a surface cavity, and therefore $g_{\lambda,\mathrm{L}} = g_\lambda$ and $J > J_0 = 4t^2/U_0$. Meanwhile, if $\Delta U$ is disregarded in Eq.~(\ref{eq:variational J}), one finds $J < J_0$. This shows that the static screening is crucial to correctly determine the sign of the magnetic exchange correction.

\subsection{Effective spin Hamiltonian}
\label{ssec:effective_hamiltonian}

The exact calculation of $J$, even without the cavity, is generally challenging. Therefore, $J$ is usually calculated perturbatively in the parameter $t/U$, in the strong coupling limit $t \ll U$ where electrons effectively localize to the lattice sites. Here, we use a similar approach, and perform a cavity Schrieffer-Wolff transformation (see \cref{sec:appdx_sw_transform}) to perturbatively eliminate doubly occupied sites, and effectively map the correlated electronic problem onto a photon dressed Heisenberg model~\cite{weberCavityrenormalizedQuantum2023,grunwaldCavitySpectroscopy2024,sentefQuantumClassical2020,vinasbostromControllingMagnetic2023}
\begin{align}
    \label{eq:sw_hamiltonian}
    H\_{s} &= \sum_{\expval{ij}} J_{ij}({\bf a}^{\dagger}, {\bf a})
    \left(
        \vb{S}_i \cdot \vb{S}_j - \frac{1}{4}
    \right) + \mathcal{O}\Big(\frac{t}{U}\Big)^3, \\
    J_{ij}^{\vb{n}\vb{m}} &= \frac{\tilde{J}}{2} \sum_{\vb{k}} \phi_{ij}^{\vb{n}\vb{k}} \phi^{\vb{k}\vb{m}}_{ji} \left( \frac{U}{U + \vb*{\omega} \cdot (\vb{k} - \vb{n})} \!+ [\vb{n} \leftrightarrow \vb{m}] \right).
    \nonumber
\end{align}
We have introduced the Peierls phase matrix elements $\phi_{ij}^{\vb{n}\vb{k}} = \mel*{n_1, n_2, \dots}{e^{-i \theta_{ij}}}{k_1, k_2, \dots}$ in terms of the vectors of number occupations $\ket{\vb{n}} = \ket{n_1, n_2, \dots}$ and cavity resonance frequencies $\vb*{\omega} = \left(\omega_1, \omega_2, \dots \right)$. For vanishing light-matter coupling, $g_\lambda \to 0$, \cref{eq:sw_hamiltonian} reduces to the bare magnetic exchange $J_0 = 4 |t_{ij}|^2 / U_0$, while at finite coupling the screened interaction $U$ [see \cref{sec:appdx_mirror_charge}] enters through $\tilde{J} = 4|t_{ij}|^2 / U$.

We focus on the renormalization of the magnetic exchange coming from cavity vacuum fluctuations, setting $\vb{n} = \vb{m} = 0$ while summing over all possible occupations $\ket{\vb{k}}$ of the virtual state. This dark cavity limit is justified within the strong coupling expansion, where $\ket{\psi\_{GS}} = \ket{\psi\_{e}} \otimes \ket{\vb{n} = 0} + \order{t/U}$, so that the leading-order contribution stems from the zero-photon sector, while finite photon occupations are sub-leading in $t / U$ \cite{weberCavityrenormalizedQuantum2023,grunwaldCavitySpectroscopy2024, welakuhNonperturbativeMass2025}. This is further justified by noting that for a cavity of macroscopic in-plane extent $A$, the single-mode couplings $g_\lambda \to 0$ as $A \to \infty$, such that each mode is infinitesimally coupled.

\subsection{Cavity modifications to all orders}

The magnetic exchange interaction in the dark cavity limit can be written as a coupled summation over the mode occupations $\vb{k}$ (with $k_i \in \mathbb{N}_0$), given by
\begin{align}
    J = \tilde{J} e^{-\sum_\lambda |g_\lambda|^2}
    \sum_{\vb{k}}
    \left[ 
    \prod_\lambda \frac{|g_\lambda|^{2k_\lambda}}{k_\lambda!}
    \right]
    \frac{U}{U + \vb*{\omega} \cdot \vb{k} }.
    \label{eq:resum_magnetic_J_multinomial}
\end{align}
The expression inside the braces stems from the Peierls phase product $\phi_{ij}^{\vb{0}\vb{k}} \phi^{\vb{k}\vb{0}}_{ji}$, evaluated using the Baker-Campbell-Hausdorff lemma~\cite{Li2020}. The final fraction couples the summations over different $k_i$ and $k_j$, making a direct evaluation exponentially hard. It can however be decoupled via a Laplace transformation~%
\footnote{An algebraic factorization of the summations at finite $\tilde{U}$ is impossible. Formally, this follows from the observation that $f(x_1, x_2) = f_1(x_1) f_2(x_2)$ if and only if $\partial_{x_1} \partial_{x_2} \log(f(x_1, x_2)) = 0$. For our function, this relation is not fulfilled, implying that it cannot be factorized as an algebraic equation.}
\begin{align}
    \frac{U}{U + \vb*{\omega} \cdot \vb{k} } = \int_0^\infty \dd{x} e^{-x(1 + \vb*{\omega} \cdot \vb{k} / U)},
    \label{eq:resum_laplace_decoupling}
\end{align}
which upon exchanging summation and integration~%
\footnote{Here regularity is enforced by an explicit Gaussian cutoff, which we sent to zero at the end of our calculation.}
allows us to sum the Laurent series' over $k_i$ and write \cref{eq:resum_magnetic_J_multinomial} as a product of exponentials. As a result, the cavity modified exchange can be written as
%
\begin{align}
    J = \tilde{J} \int_0^\infty \dd{x} e^{-x}
    \exp[ \sum_\lambda |g_\lambda|^2 \left(e^{-x \omega_\lambda/U} - 1 \right) ].
    \hspace{-0.1cm}
    \label{eq:resum_laplace_intermediate}
\end{align}
This trades the exponentially hard coupled summation over cavity occupations [\cref{eq:resum_magnetic_J_multinomial}] for a single mode sum and a one-dimensional auxiliary integration. An alternative derivation based on the volume scaling of the light-matter interactions is given in \cref{ssec:appdx_resummation_scaling}.

It is convenient to rewrite the mode summation over an arbitrary function $F(\omega_\lambda)$ in terms of an integral over the photonic density of states (PDOS) as
\begin{align}
    \sum_\lambda |g_\lambda|^2 F(\omega_\lambda) = P_0 \int_0^\infty \dd{\omega} \omega^{-1} F(\omega) \rho_{x}(\omega).
    \label{eq:resum_mode_sum_pdos}
\end{align}
Here, we have introduced the universal prefactor $P_0 = (ea_{ij})^2 / (2 \hbar \epsilon_0)$, and defined the PDOS along a bond $\vb{e}_{ij}$ by
\begin{align}
    \rho_{ij}(\omega, \vb{R}_i) = \sum_\lambda \left|\vb{f}_\lambda(\vb{R}_i) \cdot \vb{e}_{ij} \right|^2 \delta(\omega - \omega_\lambda).
    \label{eq:photonic_density_of_states}
\end{align}
Depending only on the geometric details of the photonic environment, this expression can be evaluated explicitly for a given cavity setup. 

Empirically, one finds that materials and optical elements do not modify the electromagnetic environment far above their characteristic energy scales, as e.g. cavity mirrors in a Fabry-P\'erot setup become transparent above the plasma frequency of the constituent materials \cite{svendsenInitioCalculations2024}. Hence, the cavity PDOS reduces to that of free space in the ultraviolet (UV) limit, $\rho(\omega) \to \rho_0(\omega) = \rho_0 \omega^2$ as $\omega \to \infty$, with $\rho_0 = 1 / (3 \pi^2 c^3)$. To avoid divergences of \cref{eq:resum_laplace_intermediate} in this limit, it is necessary to introduce a method of regularization.

To that end, we note that it is the coupling of electrons to the free electromagnetic field that generates the Coulomb interaction and a finite electronic mass \cite{ruggenthalerUnderstandingPolaritonic2023,luCavityEngineering2025}. Consequently, this coupling is implicitly included in the electronic parameters of the Hubbard model. In other words, if we start from a theory with finite $t_{ij}$ and $U$, we have already included the coupling to the free electromagnetic field. To avoid double counting, the vector potential and the associated photonic density of states must be treated as difference modes with respect to free space $\rho(\omega) \to \rho(\omega) - \rho_0(\omega)$, leading to a UV regular theory in which the high-energy behavior is absorbed into the bare electronic mass~\cite{luCavityEngineering2025,ruggenthalerUnderstandingPolaritonic2023} (see \cref{ssec:appdx_resummation_regularization} for details). This regularization is consistent with state-of-the-art approaches to cavity material engineering~\cite{eckhardtSurfacemediatedUltrastrong2024,bostromEquilibriumNonlinear2024,eckhardtCavitronicsLowDimensional2024,svendsenTheoryQuantum2023,fanCavityControl2026,chengAnyonicChern2025} and, for surface cavities, naturally restricts the PDOS to the substrate's surface modes (see \cref{ssec:appdx_bulk_modes}).

Summarizing the discussion above, the magnetic exchange interaction is determined via the integral
\begin{align}
    J_{ij} = J_0 \frac{U_0}{U_0 + \Delta U} \int_0^\infty \dd{x} e^{-x} e^{M_{ij}(x)},
    \label{eq:resum_J}
\end{align}
given in terms of the cavity modification function $M_{ij}(x)$ along a bond in the $\vb{e}_{ij}$ direction. This function is defined by
\begin{align}
    M_{ij}(x) = P_0 \int_0^\infty 
    \hspace{-0.05cm}
    \dd{\omega} \frac{\Delta \rho_{ij}(\omega)}{\omega}
    \left( e^{- \omega x/U}  - 1 \right) g_\eta(\omega),
    \label{eq:resum_cavity_mod}
\end{align}
with the relative PDOS $\Delta \rho_{ij}(\omega) = \rho_{ij}(\omega) - \rho_{0, ij}(\omega)$ and the Gaussian regularization $g_\eta(\omega) = \exp(-\eta^2 \omega^2)$, included for completeness. We note that the cavity only enters the function $M_{ij}(x)$ through its PDOS, and so its effect on $J_{ij}$ is entirely contained in $\Delta \rho_{ij}(\omega)$.

Before proceeding with the evaluation of $J$ for specific cavity structures, we develop some physical intuition for the derived formalism. We consider a perturbative expansion of \cref{eq:resum_cavity_mod} in the small parameter $\theta = \hbar \omega_\star / U$, where $\omega_\star$ is the characteristic cavity frequency. For typical THz cavities, $\theta \sim 10^{-3}$, justifying a truncation at leading order. The resulting expression for the magnetic exchange can be evaluated analytically as (see \cref{ssec:appdx_resum_single_mode}), and gives
\begin{gather}
    \label{eq:resum_J_taylor_J}
    J = \frac{\tilde{J}}{1 + \theta \tilde{g}^2 + \order{\theta}^2} \\
    \tilde{g}^2 = P_0 \omega_\star^{-1} \int_{0}^{\infty} \dd{\omega}
    \Big[ \rho(\omega) - \rho_0(\omega) \Big] g_\eta(\omega).
    \label{eq:resum_taylor_g_eff}
\end{gather}
Here the dimensionless effective coupling $\tilde{g}$ dictates the vacuum fluctuation induced modification. Notably, \cref{eq:resum_taylor_g_eff} acts as a generalized Purcell coefficient. Unlike resonantly coupled models that probe the PDOS at a single frequency, $\tilde{g}$ depends on the \textit{integrated} PDOS relative to free space, reflecting the off-resonant nature of the coupling. Significant modifications of $J$ therefore require cavity architectures producing strong spectral weight redistributions, which survive frequency integration. Note the resemblance of \cref{eq:resum_taylor_g_eff} with our variational result \cref{eq:variational_result}.


\section{Magnetic Exchange Interaction}
\label{sec:magnetic_exchange}

\subsection{Fabry-Pérot Cavity}
\begin{figure}[t]
    \centering
    \includegraphics{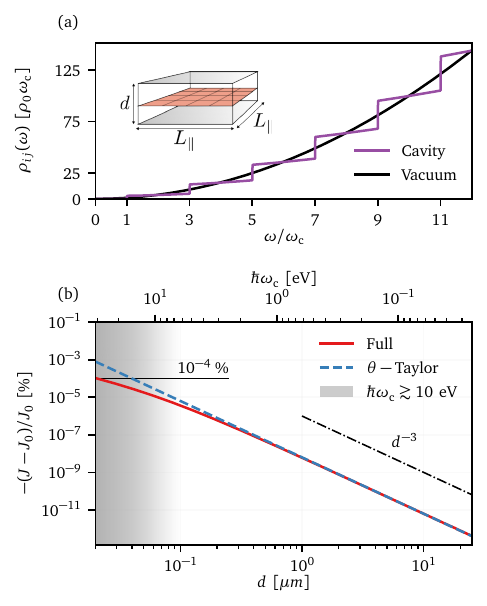}
    \caption{\textbf{Planar cavity dressed magnetic exchange} for a Fabry-P\'erot (FP) cavity with perfect mirrors separated by distance $d$ (inset, top panel) that interacts with a material placed at the midpoint $z = d/2$.
    (a) Photonic density of states [\cref{eq:photonic_density_of_states}] at the cavity center for a FP resonator (purple) and free space (black) in the thermodynamic limit $L_\parallel \to \infty$. The oscillatory cavity PDOS nearly averages to zero over each resonance interval.
    (b) Relative modification of magnetic exchange [\cref{eq:resum_J}] as a function of cavity mirror distance $d$ and fundamental cavity resonance energy $\hbar \omega\_c$ (top axis) for $a_{ij} = 6\;$\AA~ and $U = 5\;$eV. The full calculation (red) follows a $d^{-3}$ scaling, showing excellent agreement with the perturbative limit (blue, \cref{eq:resum_taylor_g_eff}). The gray region ($d \lesssim 0.1~\mu\rm{m}$) marks the breakdown of the perfect-conductor approximation, where $\omega\_c$ approaches the plasma frequency of typical metals. Predicted modifications remain below current experimental resolution.}
    \label{fig:main_magnetic_modification_fp}
    \label[fig_a]{fig:main_magnetic_modification_fp_a}
    \label[fig_b]{fig:main_magnetic_modification_fp_b}
\end{figure}
To develop a concrete understanding of the vacuum-modified magnetic exchange, we first examine the paradigmatic case of a Fabry-P\'erot (FP) cavity. It consists of two planar mirrors separated by a distance $d$ with the material centrally embedded [inset of \cref{fig:main_magnetic_modification_fp_a}]. Since the mirrors are far from the material, screening contributions are negligible and the mirrors can be modeled as perfect conductors entering as boundary conditions in \cref{eq_eig_f}. The resulting mode functions are derived analytically in \cref{ssec:appdx_pdos_fp}, and yield the PDOS shown in \cref{fig:main_magnetic_modification_fp_a} in the thermodynamic limit $L_\parallel \to \infty$. Photons propagate freely in the mirror plane but form standing waves in the perpendicular direction. This produces discrete resonances at integer multiples of the fundamental cavity frequency $\omega\_c = c\pi/d$, each with locally quadratic dispersion $\omega(n, \vb{k}_\parallel)^2 = {(n\omega_c)^2 + c^2 \vb{k}_\parallel^2}$. The resulting PDOS, illustrated in \cref{fig:main_magnetic_modification_fp_a}, is gapped, with spectral weight periodically redistributed to higher energies.

\Cref{fig:main_magnetic_modification_fp_b} shows the relative modification of the magnetic exchange [\cref{eq:resum_J}] as a function of the cavity spacing $d$ and the fundamental resonance energy $\hbar \omega_c$. The exchange is systematically suppressed as $d$ decreases, with no resonance features, yet the magnitude of the modification remains small across the entire physically relevant parameter range. This can be understood intuitively from the spectral weight redistribution described by \cref{eq:resum_taylor_g_eff}. While the FP cavity significantly alters the photon density of states (PDOS) at specific frequencies, it is remarkably inefficient at shifting the \textit{integrated} spectral weight relative to free space --- the average PDOS over a resonance interval $[(2n + 1)\omega_c, (2n + 3)\omega_c]$ nearly coincides with the free-space value, leading to a small effective coupling $\tilde{g}$ and scaling $|\Delta J| \sim d^{-3}$.

These results establish that direct electronic coupling to a standard FP cavity is insufficient to meaningfully modify magnetic order in the ground state. This is a consequence of the off-resonant nature of the interaction, which probes the cavity PDOS only through a broad frequency average of scale $\sim U$. Coupling to resonant degrees of freedom that sample $\rho(\omega)$ over a narrow frequency window, such as IR-active optical phonons, could circumvent this limitation.

\subsection{Surface Cavity}

\begin{figure}[t]
    \centering
    \includegraphics[width = \linewidth]{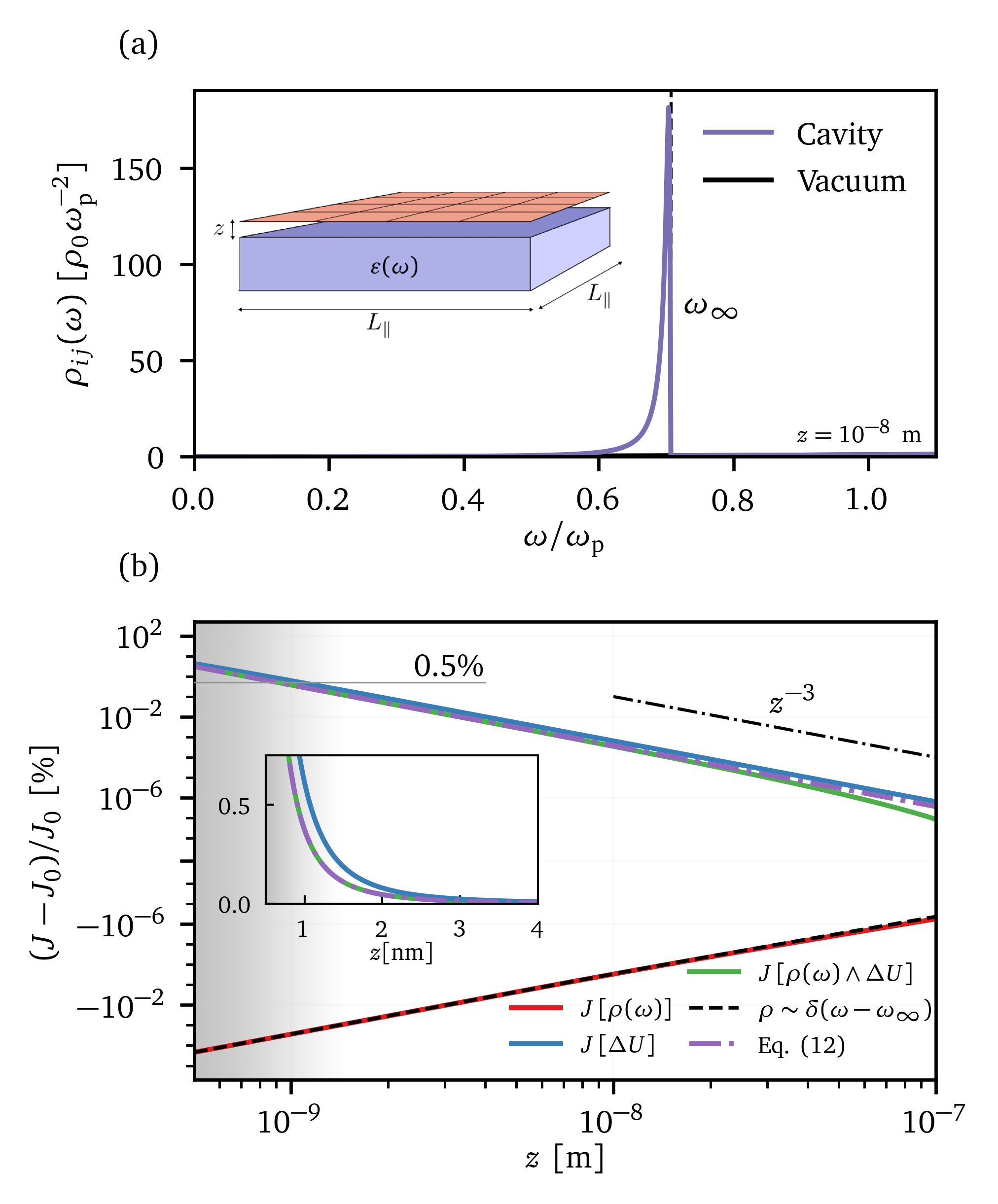}
    \caption{\textbf{Surface cavity dressed magnetic exchange} for a material at a distance $z$ above a gold substrate (top panel inset), modeled via a Lorentzian dielectric function $\varepsilon(\omega)$ with plasma frequency $\hbar \omega\_p = 9.45$~eV [\cref{eq:quant_surface_dielectric}]. (a) Photonic density of states [\cref{eq:photonic_density_of_states}] of the gold substrate (purple) and in free space (black; nearly zero on this scale) in the thermodynamic limit $L_\parallel \to \infty$, evaluated at $z = 10~$nm above the substrate.
    The exponentially localized surface mode produces a strongly peaked enhancement at the limit frequency $\omega_\infty$ [\cref{eq:quant_surface_limit_freq}]. (b) Contribution-resolved relative modification of the magnetic exchange [\cref{eq:resum_J}] as a function of substrate distance $z$ in a sym-logarithmic scale for $a_{ij} = 6\;$\AA~ and $U = 5\;$eV. Result from PDOS in the top panel indicated by the color-matched arrow. The dynamical dressing via the vector potential (red) suppresses $J$, while the static screening (blue) enhances it. The single-mode approximation (black dashed) accurately captures the dynamical contribution. Both mechanisms scale as $z^{-3}$, leading to strong cancellations and net effect (green) with a percent-level enhancement at nanometer distances, well captured by the variational approach [\cref{eq:variational J}]. The gray region ($z < 1~\rm{nm}$) marks the breakdown of the macroscopic dielectric description.}
    \label{fig:main_magnetic_modification_surf}
    \label[fig_a]{fig:main_magnetic_modification_surf_a}
    \label[fig_b]{fig:main_magnetic_modification_surf_b}
\end{figure}

To achieve large modifications in the off-resonant regime, a promising direction is offered by polaritonic surface cavities whose consistent Coulomb gauge quantization we developed in \cref{sec:hopfield}. The setup [\cref{fig:main_magnetic_modification_surf_a}, inset] consists of a material placed at a distance $z$ above a substrate, effectively described by a dielectric function $\varepsilon(\vb{r}, \omega)$. Optically active excitations of the substrate hybridize with the electromagnetic field at the interface, forming polaritonic modes that act as our cavity. The resulting mode functions and PDOS are derived in \cref{sec:appdx_surface_cavity}. In addition to bulk modes propagating throughout all of space, such a cavity hosts surface modes that are exponentially localized at the vacuum-dielectric interface. These feature a nearly flat dispersion $\omega(\vb{k}) \approx \omega_\infty = \omega\_p/\sqrt{2}$, producing a large concentration of spectral weight at $\omega_\infty$. The PDOS is shown in \cref{fig:main_magnetic_modification_surf_a} for a gold substrate, modeled via a Lorentzian dielectric function with plasma frequency $\hbar\omega\_p = 9.45$~eV. The surface modes are the dominant contribution to the PDOS close to the interface ($z \lesssim 2\pi c / \omega_\infty$) by many orders of magnitude (\cref{sec:appdx_surface_cavity}).

The result of coupling the plasmonic cavity modes to a strongly correlated material is shown in \cref{fig:main_magnetic_modification_surf_b} as a function of material-substrate distance $z$. The gauge-consistent treatment of the surface cavity reveals two competing contributions to the magnetic exchange: First, a dynamical dressing of the electron hopping through the vector potential $\vb{A}$, which suppresses $J$, and is remarkably well-captured by a single-mode approximation $\Delta \rho_{ij}(\omega) \approx \tilde{g}^2 \omega_\infty \delta(\omega - \omega_\infty)$ reflecting the strongly peaked nature of the surface PDOS (see \cref{ssec:appdx_single_mode} for details). Second, a static screening of the Hubbard $U$ intuitively understood via substrate mirror charges, which enhances the magnetic exchange. Both mechanisms show a scaling $\sim z^{-3}$, and their competition determines the net modification. We generally find an increase of the total magnetic exchange, which for the gold substrate considered here reaches the few percent level at nanometer distances. THis is well within the reach of Raman spectroscopy.

The balance between dynamical dressing and static screening can be tuned via the dielectric properties of the substrate. While for small plasma frequencies $\hbar \omega\_p \lesssim U$ the dynamical dressing contribution only varies weakly, it is strongly quenched for large plasma frequencies $\hbar \omega\_p \gg U$ such that the total effect in this limit is dominated by static screening (see \cref{ssec:scaling_surface}). This distinct dependence of both contributions on substrate parameters provide a sensitive framework for engineering magnetic order in strongly correlated systems.

\subsection{Spectroscopic signatures}

\begin{figure}
    \centering
    \begin{tikzpicture}
    \node at ( 0.0, 2.2) {\includegraphics[width=0.50\linewidth]{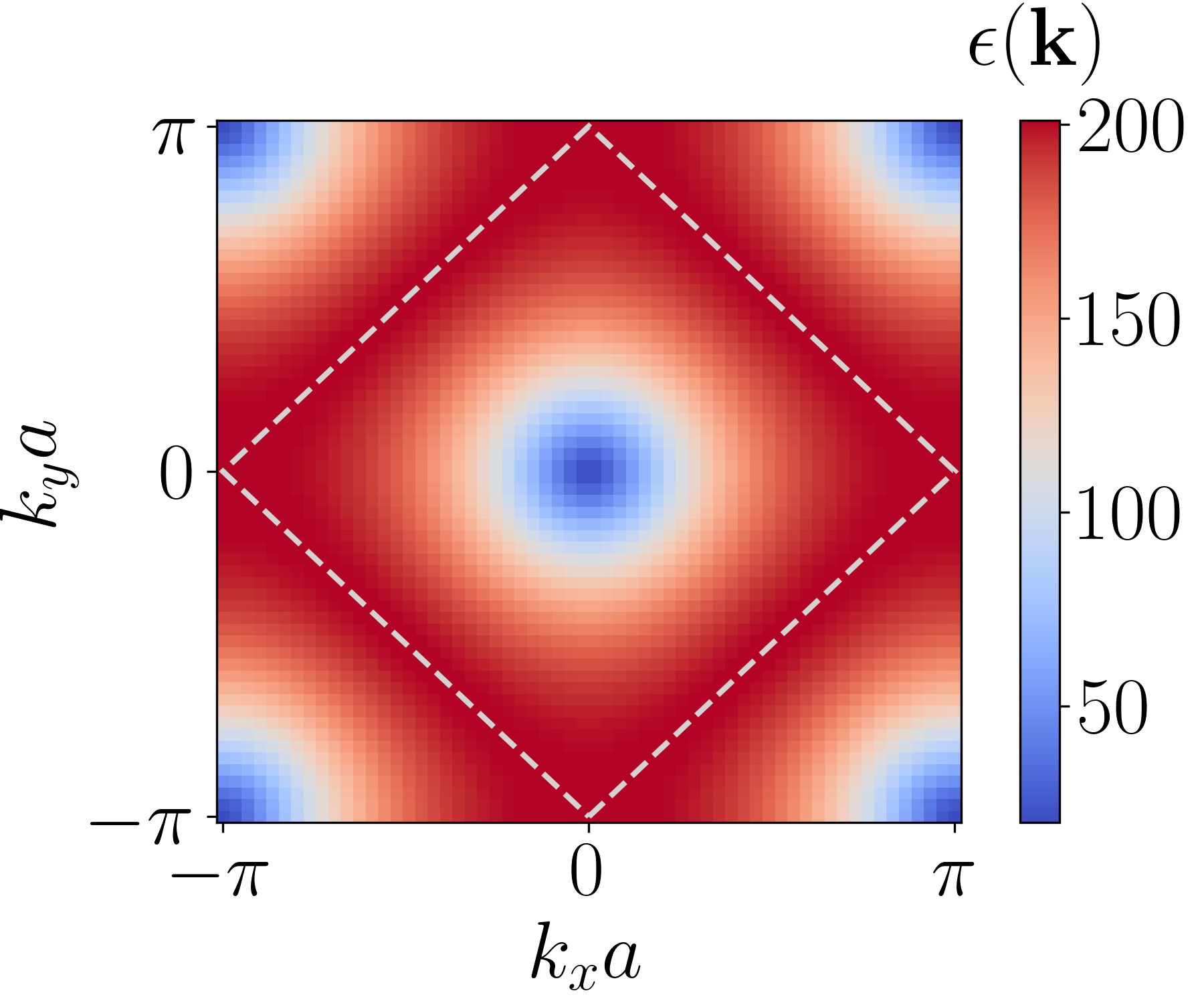}};
    \node at (-0.0,-1.6) {\includegraphics[width=0.51\linewidth]{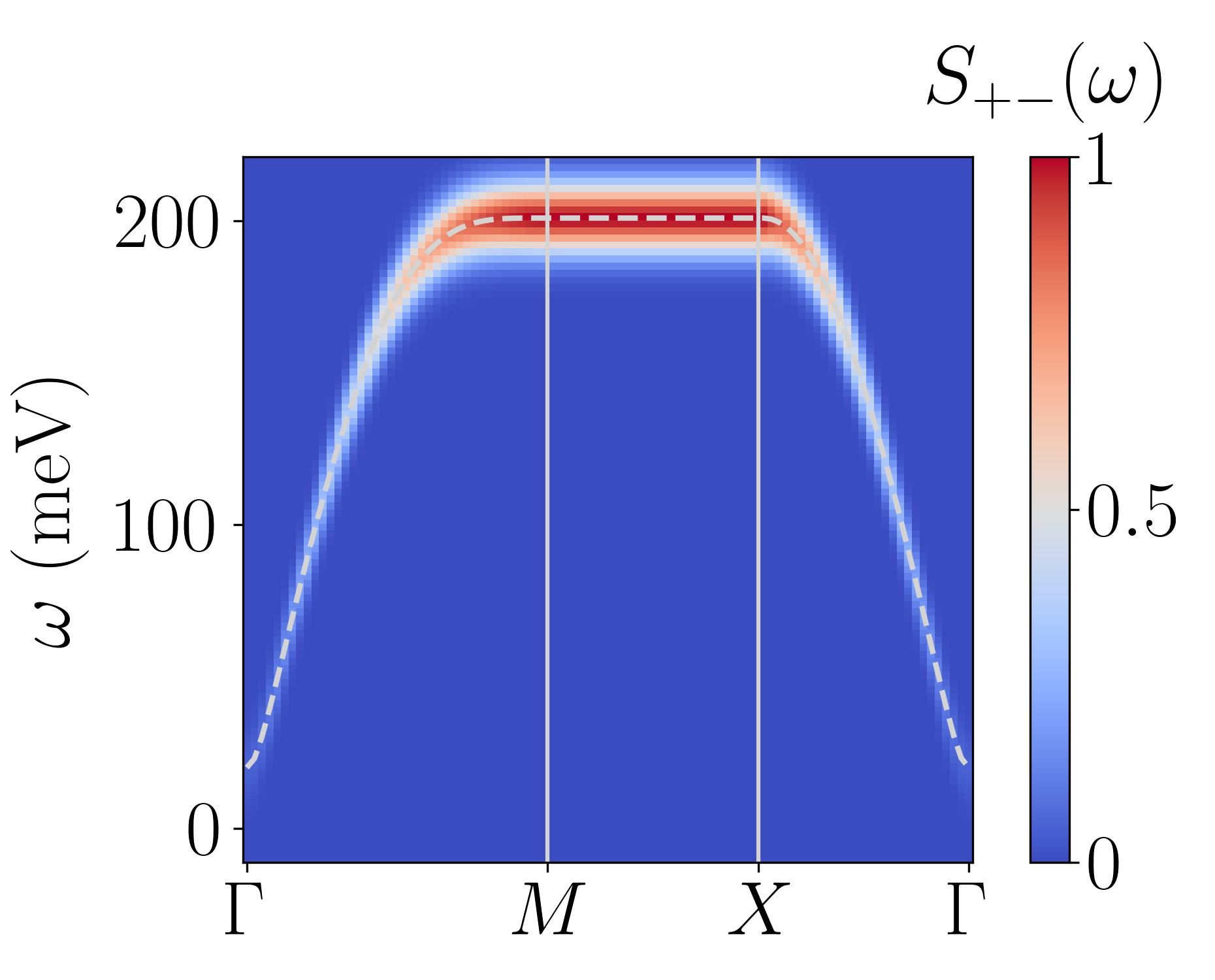}};
    \node at ( 4.4, 0.0) {\includegraphics[width=0.50\linewidth]{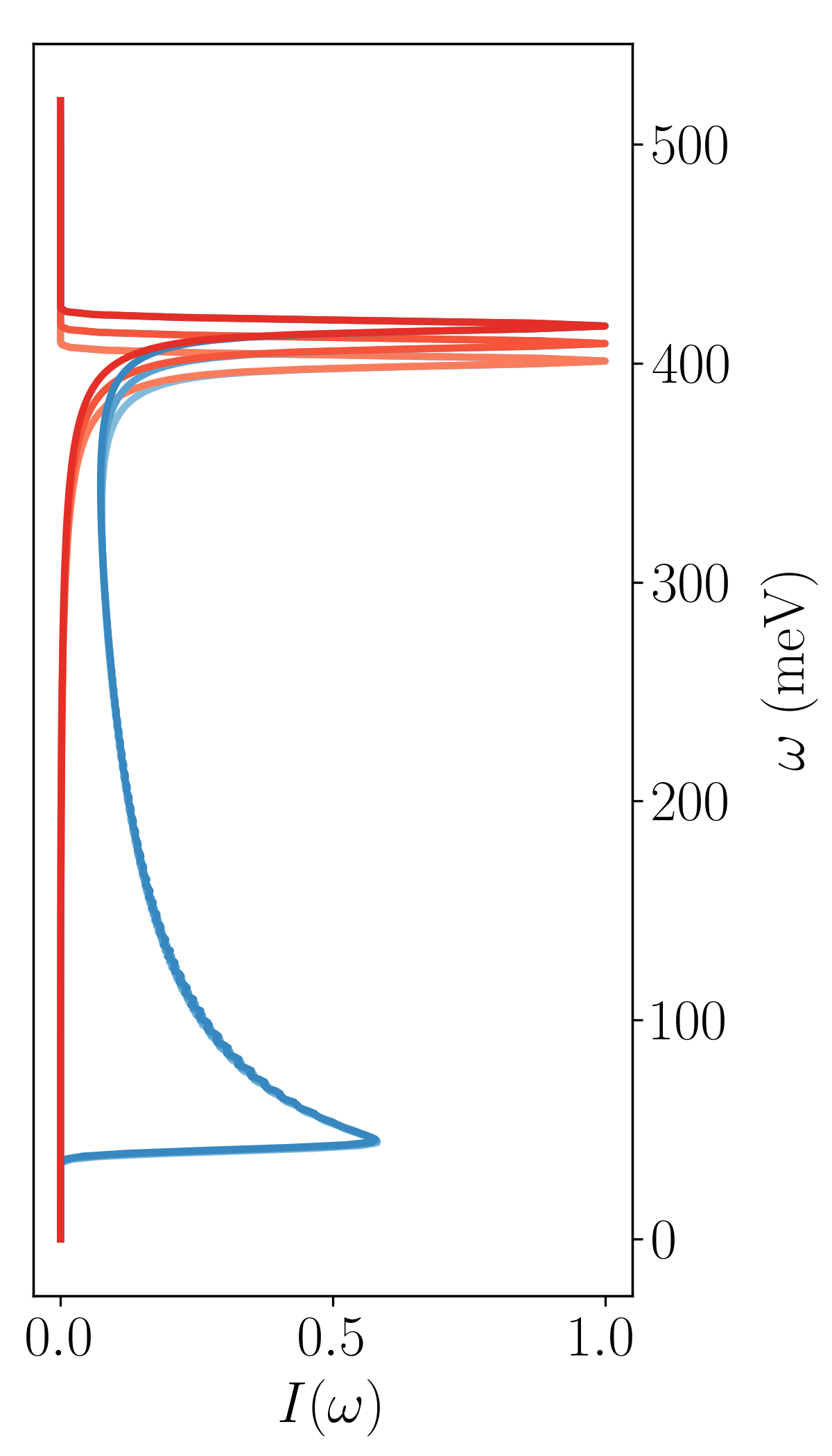}};
    \draw[->,>=stealth,thick] (4.8,1.7) -- (4.8,2.3);
    \node at ( 4.3, 2.4) {$4\Delta J$};

    \node at (-2.0, 3.1) {(a)};
    \node at (-2.0,-0.1) {(b)};
    \node at ( 5.0, 3.1) {(c)};
    \end{tikzpicture}
    \caption{{\bf Magnon signatures of cavity renormalization.} (a) Magnon dispersion for a square lattice antiferromagnet. The light gray dashed line indicates the magnetic Brillouin zone. (b) Transverse dynamical spin structure factor $S_{+-}(\omega)$ along the high symmetry lines of the Brillouin zone. The high symmetry points are $\Gamma = (0,0)$, $M = (\pi, 0)$ and $X = (\pi/2,\pi/2)$. (c) Two-magnon Raman spectrum $I(\omega)$ of a square lattice antiferromagnet. The blue lines show the parallel polarization component $I_{xx} = I_{yy}$, and the red lines show the cross polarization signal $I_{xy}$. The lightest lines are computed using a bare exchange value of $J_0 = 100$ meV, comparable to that of the parent cuprate La$_2$CuO$_4$. The modification from the cavity is taken to be $\Delta J = 2$~\% or $\Delta J = 4$~\%, corresponding to the darker lines and increasing in the direction of the arrow. The spectra are computed using an intrinsic Raman linewidth of $2$ meV.}
    \label{fig:raman}
\end{figure}
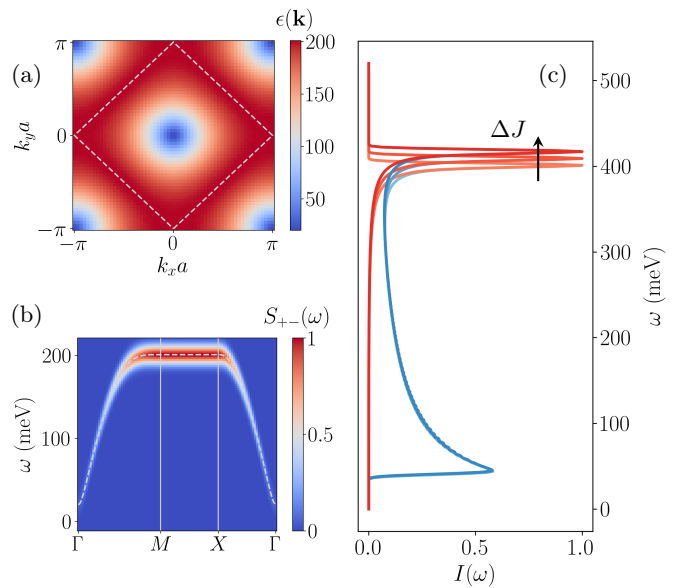

The cavity-induced modification of the magnetic exchange can be directly accessed via Raman or resonant inelastic X-ray scattering (RIXS) measurements, probing the system's magnon excitations. In antiferromagnets with a combined $\mathcal{TI}$ symmetry, as is the case for a N\'eel state on the square lattice, the magnon branches are degenerate by symmetry. For such systems, single magnon optical excitations are highly suppressed, and the dominant contribution to the Raman signal comes from magnon pair creation processes (see Appendix~\ref{app:raman}). Complementary, single magnon dispersion can be probed by RIXS measurements of the transverse component $S_{+-}$ of the dynamical spin structure factor. 

As seen in Fig.~\ref{fig:raman}a, the magnon dispersion for the Heisenberg model described by Eq.~\ref{eq:sw_hamiltonian} is flat along the $M$ -- $X$ -- $M$ direction, leading to a large density of states at $\omega = 2J$ and a corresponding peak in $S_{+-}$ (see Fig.~\ref{fig:raman}b). Similarly, the large magnon DOS at the zone edges leads to a large Raman peak at $\omega = 4J$, originating from creation of two magnons with energy $2J$. While this feature is present both in the parallel and cross polarization configuration (see Fig.~\ref{fig:raman}c), it is more pronounced in the latter. The cavity modification of $J$ is thus directly accessible to spectroscopic measurements. Due to the very high frequency resolution of Raman measurements, typically of the order of $5$ cm$^{-1}$ (equivalently $0.5$ meV), it is possible to resolve changes down to $0.1\%$ assuming a bare exchange of $J_0 \approx 100$ meV. In Fig.~\ref{fig:raman}c, we show the calculated Raman spectra for different values of $\Delta J = J - J_0$, showing a shift of the main peak by a value $4\Delta J$.

\section{t-J Polaron in a Surface Cavity}
\label{sec:tj}

We extend the discussion to the weakly doped Hubbard model and investigate cavity induced modifications of the polaron dispersion. Away from half-filling, light-matter coupling enters both the hopping and spin-exchange channels. The cavity-coupled low-energy theory can be obtained via a cavity Schrieffer-Wolff transformation [\cref{sec:appdx_sw_transform}], and takes the form of a $t$-$J$ model
\begin{align}
\label{eq:tJ_model}
    H\_{m,eff} &= \sum_{\expval{ij}\sigma}
       t_{ij} e^{-i \theta_{ij}} \tilde{c}^\dagger_{i\sigma} \tilde{c}_{j\sigma}  \\
       & \hspace{1cm}+
    \sum_{\expval{ij}} 
    J_{ij}(a^{\dagger}, a)
    \left(
    \vb{S}_i \cdot \vb{S}_j - \frac{1}{4}n_i n_j 
    \right),
    \nonumber
\end{align}
with the projected creation operator $\tilde{c}^\dagger_{i\sigma} = c^\dagger_{i\sigma}(1-n_{i\bar{\sigma}})$. The magnetic exchange $J_{ij}(a^\dagger, a)$ is approximately given by the same expression as in the half-filled limit [\cref{eq:sw_hamiltonian}], and as shown in section.~\ref{sec:magnetic_exchange}, it is typically within $1\%$ of it's vacuum value. We can therefore take  $J_{ij}(a^\dagger,a) \approx J$ in the following calculations.

Focusing on the single-polaron limit of the nearest-neighbor $t$-$J$ model, corresponding to the the weakly underdoped regime, we solve the single-hole problem using a boson-augumented exact diagonalization procedure. This is an extension of the approach introduced in~\cite{bonvca2007numerical}. Starting from a parent state $\ket{\psi_0}$ with zero photons and a single hole at the origin, a reduced basis is generated by repeated applications of the kinetic energy term. This generates the so-called string states 
\begin{equation}
     \bigg[\sum_{\expval{ij}\sigma}
        e^{-i s\theta_{ij}} \tilde{c}^\dagger_{i\sigma} \tilde{c}_{j\sigma}\bigg]^l\ket{\psi_0} = \mathrm{span}\Big\{\ket{h,f}\ket{\bm{r_h}}_\mathrm{ph}\Big\},
    \label{eq:tJ_states}
\end{equation}
where $l\leq N_h$ is the number of hole steps, $h$ denotes the final hole position and $f$ is the string of spin flips relative to the N\'eel background. The photons are described by displaced photonic Gaussian states $\ket{\bm{r}_h}_\mathrm{ph}$, defined as
\begin{equation}
    \ket{\vb{r}_h}_{\mathrm{ph}}
    = \exp\left[is\frac{e}{\hbar}\vb{r}_h\cdot\hat{\vb{A}}\right]\ket{0}_{\mathrm{ph}}
\end{equation}
within the long-wavelength limit. 

Similarly to the approach in Sec.~\ref{sec:main_variational_J}, we introduce a variational parameter $s\in[0,1]$ that interpolates between the weak ($s \approx 0$) and strong ($s \approx 1$) coupling limit. In practice, we use a variational basis that contains both the maximally displaced states with $s=1$, and the photon vacuum ($s=0$). \Cref{eq:tJ_states} defines a computational basis in which the $t-J$ Hamiltonian [\cref{eq:tJ_model}] is expanded and subsequently diagonalized. The calculation is performed in Bloch sectors at fixed momentum $\vb{k}$, directly yielding the polaron dispersion. Since the translated string states are not orthonormal, the overlap matrix is retained explicitly. The detailed construction and explicit form of all matrix elements is given in \cref{app:tj_truncated_ed}.

\begin{figure}[t!]
    \centering
    \includegraphics[width=\linewidth]{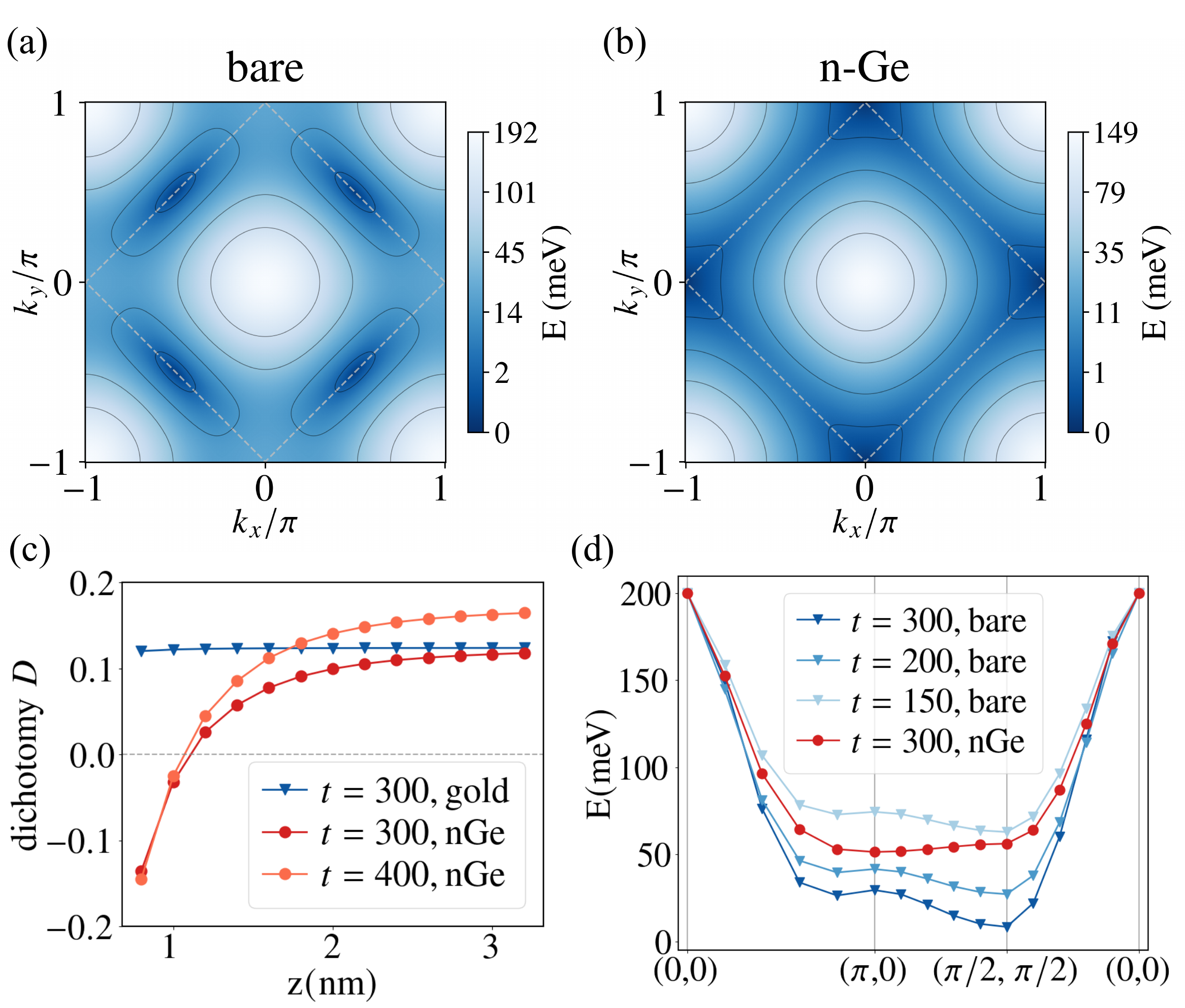}
    \caption{\textbf{Cavity control of the single-hole $t$-$J$ polaron.} (a) Two-dimensional dispersion of the $t$-$J$ polaron without a cavity, with $t=300\,\mathrm{meV}$ and $J=100\,\mathrm{meV}$. The gray dashed lines mark the boundary of the magnetic Brillouin zone. (b) Two-dimensional dispersion of the $t$-$J$ polaron in an electron-doped Ge surface cavity, again with $t=300\,\mathrm{meV}$ and $J=100\,\mathrm{meV}$. The band minimum is shifted to the antinodal point. (c) Dichotomy $D$ as a function of distance $z$ to the surface for gold and n-Ge substrates. (d) Polaron dispersion along the $(0,0)$--$(\pi,0)$--$(\pi/2,\pi/2)$--$(0,0)$ path, comparing different bare hoppings without a cavity to the electron-doped Ge cavity. In all the calculations we use $N_h=6$.}
    \label{fig:tj}
    \label[fig_a]{fig:tj_a}
    \label[fig_b]{fig:tj_b}
    \label[fig_c]{fig:tj_c}
    \label[fig_d]{fig:tj_d}
\end{figure}

The resulting single-polaron dispersion is shown in \cref{fig:tj} for $J = 100$ meV and, unless otherwise stated, $t = 300$ meV. In the absence of a cavity, the polaron minimum lies at the nodal point $(\pi/2,\pi/2)$, about $20$ meV below the antinodal point $(\pi,0)$ [\cref{fig:tj_a}]. This nodal--antinodal dichotomy is a central characteristic of the $t$-$J$ polaron and is associated with a Fermi-contour pocket around $(\pi/2,\pi/2)$ and enhanced quasiparticle weight at the Fermi energy \cite{zhou2004dichotomy,marshall1996unconventional}. Coupling to a cavity modifies this strongly. Placing the material $1$ nm above an electron-doped Ge ($n$-Ge) substrate with $\hbar\omega_\infty=100$ meV \cite{frigerio2016tunability}, acting as our surface cavity, the dispersion along the magnetic Brillouin-zone boundary is strongly flattened and the band minimum shifts to the antinodal point [\cref{fig:tj_b}]. Thus, the surface cavity changes the momentum structure of the lowest quasiparticle state instead of merely rescaling the polaron bandwidth.

To quantify this effect, we define the dichotomy
\begin{align}
    D=\frac{E_{\mathrm{AN}}-E_{\mathrm{N}}}{E_{\Gamma}-E_{\mathrm{AN}}},
    \label{eq:tj_dichotomy}
\end{align}
where $E_{\mathrm{N}}$, $E_{\mathrm{AN}}$, and $E_{\Gamma}$ are the polaron energies at $(\pi/2,\pi/2)$, $(\pi,0)$, and $(0,0)$, respectively. As shown in \cref{fig:tj_c}, decreasing the distance to the $n$-Ge substrate drives $D$ through zero, realizing a cavity-induced reversal of the nodal-antinodal dichotomy. 

A qualitative modification of the polaron dichotomy requires strong coupling $\tilde{g}^2 \sim 1$. Since the effective light-matter coupling scales as $\tilde{g}^2 \sim \omega_\infty^{-1}$ [cf. \cref{eq:resum_taylor_g_eff}], this favors substrates with low plasma frequency. For the previously considered gold substrate we find $\tilde{g}^2\simeq0.01$ at distance a $z = 1$ nm, leading no qualitative modifications of the polaron dispersion. The plasma frequency dependence of the dichotomy is therefore distinct from our results for the magnetic exchange, where we found that modification is controlled by $\tilde{g}^2\omega_\infty/U$ [\cref{eq:resum_taylor_g_eff}], eliminating the leading order dependence on $\omega_\infty$.

Finally, \cref{fig:tj_d} compares the cavity-induced change in dispersion with a direct reduction of the bare hopping $t$ in the absence of the cavity. Reducing $t$ decreases $E_{\Gamma}-E_{\mathrm{N}}$ but leaves the dichotomy essentially unchanged. The surface cavity therefore provides an independent control mechanism for doped strongly correlated systems, capable of reshaping the internal momentum structure of the magnetic polaron rather than simply renormalizing a microscopic hopping amplitude, as one would expect from a meanfield treatment. The reshaping of momentum structure evidenced by Fig.~\ref{fig:tj} can be directly probed in ARPES experiments.

\section{Conclusion}
\label{sec:conclusion}

We developed a non-perturbative, multi-mode theory for the vacuum modification of magnetic exchange interactions, for strongly correlated electron systems embedded in electromagnetic cavities. By non-perturbatively resumming all cavity modes we showed that off-resonant vacuum modifications are described by a generalized Purcell factor. The relevant figure of merit is the frequency-integrated photonic density of states relative to free space, rather than the spectral density at any single resonance frequency. This establishes a clear conceptual distinction from resonant coupling scenarios, where the physics is determined by the PDOS at a specific frequency, and provides a complementary physical framework for understanding material control driven by vacuum fluctuations.

This principle reveals a fundamental distinction between different cavity architectures. Standard Fabry-Pérot resonators produce negligible off-resonant effects due to cancellations of spectral weight redistribution upon frequency integration, while polaritonic surface cavities achieve significant modifications through their strongly peaked spectral density. Crucially, a consistent Coulomb gauge treatment of such cavities reveals that the dynamical dressing of electron hopping by the vector potential is accompanied by a static screening of the Coulomb interaction via substrate mirror charges. Both contributions scale identically with the substrate-material distance and partially cancel, making the inclusion of screening essential for obtaining even qualitatively correct results. For a gold substrate, their competition yields a net enhancement of $J$ at the few percent level for nanometer-scale distances, which is optically accessible via two-magnon Raman spectroscopy. We further showed that for surface cavities, the strongly peaked PDOS admits a faithful single-mode approximation, providing a rigorous foundation for effective single-mode models.

Beyond the half-filled limit, the $t$-$J$ calculation shows that surface cavities can reshape the dispersion of a doped magnetic polaron. In particular, an electron-doped Ge substrate can reverse the nodal--antinodal dichotomy of the bare $t$-$J$ model, whereas a direct reduction of the hopping amplitude leaves this dichotomy essentially unchanged. This demonstrates that cavity vacuum fluctuations provide a control knob for doped correlated systems that is distinct from conventional parameter renormalization.

Our results establish a concrete design principle linking cavity geometry to material response in the off-resonant regime, providing a quantitative foundation for endyonic material control. This framework has already been applied to obtain quantitative predictions for specific correlated materials using a single-mode approximation grounded in the full multi-mode theory \cite{fanCavityControl2026}. Natural extensions include more complex matter models, additional material platforms, finite-density doped systems beyond the single-polaron limit, and cavity architectures that further enhance the integrated spectral weight redistribution.

Beyond the directly tractable geometries considered here, the PDOS of arbitrary cavity structures can be obtained, e.g. via macroscopic QED and inserted directly into \cref{eq:resum_J}, enabling systematic screening of candidate platforms \cite{scheelMacroscopicQuantum2008}. Extending the matter description to multi-band models would further connect to materials where modest changes in exchange and polaron dispersion shift phase boundaries. Finally, while our resummation treats the dark cavity limit, considering a finite quantization volume with a physical mode cutoff, e.g. at the plasma frequency \cite{svendsenInitioCalculations2024}, would allow \cref{eq:resum_laplace_intermediate} to be evaluated for a large but finite set of thermally populated modes, opening a route toward exploring finite-temperature cavity effects.

\section{Contributions}

E.V.B. conceived the original resummation idea and supervised the project together with A.R. and D.M.K. X.C. constructed the Hopfield-Coulomb quantization framework, and developed the variational method for exchange $J$ and the polaron dispersion. L.G. derived the low-energy model, developed the resummation framework and its regularization, and implemented the numerical calculations. M.R. and M.M. provided input on the regularization, resummation, and Hopfield quantization. M.R. additionally provided input from ab initio theories. All authors contributed to the interpretation of results. L.G., X.C., and E.V.B. wrote the manuscript with input from all authors.

\section{Data Availability}

The data that support the findings of this study are available from the corresponding author upon reasonable request.

\section{Acknowledgements}
We acknowledge stimulating and instructive discussions with Simone Latini, Mark Kamper Svendsen, Omar Mehio, Martin Claassen, Lukas Weber and Dominik Sidler.
We acknowledge support from the Cluster of Excellence “CUI: Advanced Imaging of Matter”–EXC 2056–project ID 390715994, SFB-925 “Light induced dynamics and control of correlated quantum systems”–project ID 170620586 of the Deutsche Forschungsgemeinschaft (DFG), the European Research Council (ERC-2024-SyG-UnMySt–101167294), and the Max Planck-New York City Center for Non-Equilibrium Quantum Phenomena. D.M.K. acknowledges support by the DFG via Germany's Excellence Strategy-Cluster of Excellence Matter and Light for Quantum Computing (ML4Q, Project No. EXC 2004/1, Grant No.390534769) and individual Grant No. 508440990. The Flatiron Institute is a division of the Simons Foundation.

\newpage 
\appendix

\section{Coulomb Gauge Hopfield Quantization}
\label{sec:appdx_mirror_charge}
In this section we provide details for the Hopfield quantization and show that the screened Coulomb interaction $W$ can be obtained from electrostatic calculation. For notational simplicity we take the reduced units $\hbar=\epsilon_0=c=1$ in \cref{sec:appdx_mirror_charge} and \cref{sec:weyl_gauge_vacuum}.

We start from the diagonalization of $H_\mathrm{ph}$. The polaritonic operator can be written as a linear combination
\begin{equation}
    a_\lambda = \int\mathrm{d}\vb{r}(\bm{\alpha}_{\mathrm{T},\lambda}\cdot\vb{E}_\mathrm{T} + \bm{\beta}_{\mathrm{T},\lambda}\cdot\vb{A}_\mathrm{T}+\bm{\gamma}_\lambda\cdot\vb{\Pi}+\bm{\eta}_\lambda\cdot\vb{X}).
    \label{eq:a_lambda}
\end{equation}
satisfies $[a_\lambda, H_\mathrm{ph}] = -i \omega_\lambda a_\lambda$, which leads to the following equations of motion for the spatially dependent coefficients
\begin{equation}
    \begin{aligned}
        &-i\omega_\lambda\bm{\alpha}_{\mathrm{T},\lambda} = -\bm{\beta}_{\mathrm{T},\lambda},\\
        &-i\omega_\lambda\bm{\beta}_{\mathrm{T},\lambda}  = \nabla\times\nabla\times\bm{\alpha}_{\mathrm{T},\lambda} + (\omega_p^2\bm{\alpha}_{\mathrm{T},\lambda})_\mathrm{T} - (\omega_p\bm{\eta}_\lambda)_\mathrm{T},\\
        &-i\omega_\lambda\bm{\gamma}_\lambda = \bm{\eta}_\lambda - \omega_p\bm{\alpha}_{\mathrm{T},\lambda},\\
        &-i\omega_\lambda\bm{\eta}_\lambda = -\omega_o^2\bm{\gamma}_\lambda - \omega_p(\omega_p\bm{\gamma}_\lambda)_\mathrm{L}.
    \label{eq_motion}
    \end{aligned}
\end{equation}
We then introduce a new variable
\begin{equation}
    \vb{f}_\lambda = \sqrt{2\omega_\lambda}\left[i\bm{\alpha}_\mathrm{T} - \omega_\lambda^{-1}(\omega_p\bm{\gamma}_\lambda)_\mathrm{L}\right].
\end{equation} 
In the following we will show that $\vb{f}(\vb{r})$ is the mode function of vector potential in Weyl gauge. All the bosonic variables can then be expressed purely with $\vb{f}_\lambda$:
\begin{equation}
    \begin{aligned}
        &\bm{\alpha}_{\mathrm{T},
        \lambda} = -\frac{i}{\sqrt{2\omega_\lambda}}\vb{f}_{\mathrm{T},\lambda},\\
        &\bm{\beta}_{\mathrm{T},\lambda} = \sqrt{\frac{\omega_\lambda}{2}} \vb{f}_{\mathrm{T},\lambda},\\
        &\bm{\gamma}_\lambda = \frac{\omega_p}{\omega_o^2-\omega_\lambda^2}\sqrt{\frac{\omega_\lambda}{2}}\vb{f}_\lambda,\\
        &\bm{\eta}_\lambda = \frac{-i\omega_\lambda\omega_p}{\omega_o^2-\omega_\lambda^2}\sqrt{\frac{\omega_\lambda}{2}}\vb{f}_\lambda-\frac{i\omega_p}{\sqrt{2\omega_\lambda}}\vb{f}_{\mathrm{T},\lambda}.
    \end{aligned}
\end{equation}
And the generalized eigen-equation for $\vb{f}_\lambda$ is
\begin{equation}
\omega_\lambda^2\varepsilon(\omega_\lambda)\vb{f}_\lambda = \nabla\times\nabla\times\vb{f}_\lambda. \label{eq_eig_f_app}
\end{equation}
Notice that this equation automatically implies the constraint $(\varepsilon\vb{f})_\mathrm{L} = 0$ for $\omega_\lambda\neq0$, which is often called the generalized Coulomb condition. To determine the normalization of $\vb{f}$, we use the condition $[a_\lambda, a_\lambda^\dagger] = 1$, which gives
\begin{equation}
    2\int\mathrm{d}\bm{r} \mathrm{Im}\left(\bm{\beta}_{\mathrm{T},\lambda}\cdot\bm{\alpha}_{\mathrm{T},\lambda}^* + \bm{\gamma}_\lambda\cdot\bm{\eta}_\lambda^*\right) = 1.
\end{equation}
Combined with the constraint $(\varepsilon\vb{f})_\mathrm{L} = 0$, we arrive at
\begin{equation}
    \int\mathrm{d}\bm{r}\left[1+\frac{\omega_p^2}{\omega_o^2-\omega_\lambda^2}+\frac{\omega_\lambda^2\omega_p^2}{(\omega_o^2-\omega_\lambda^2)^2}\right]|\vb{f}_\lambda(\vb{r})|^2 = 1.
\end{equation}
Or equivalently, $\int\mathrm{d}\bm{r}(\varepsilon+\frac{\omega}{2}\frac{\partial\varepsilon}{\partial\omega})|\vb{f}|^2 = 1$. To sum up, we have shown that the bosonic Hamiltonian can be diagonalized as $H_\mathrm{ph} = \sum_\lambda\omega_\lambda (a_\lambda^\dagger a_\lambda+\frac{1}{2})$. And using decompositions such as $\vb{X} = \sum_\lambda[\vb{X},a^\dagger_\lambda]a_\lambda-\sum_\lambda [\vb{X},a_\lambda]a^\dagger_\lambda$, all the bosonic variables can be expanded in $a_\lambda$ and $a^\dagger_\lambda$ as
\begin{equation}
\begin{aligned}
    &\vb{E}_\mathrm{T} = \sum_\lambda(i\bm{\beta}_{\mathrm{T},\lambda}^* a_\lambda+h.c.),\\
    &\vb{A}_\mathrm{T} = \sum_\lambda(-i\bm{\alpha}_{\mathrm{T},\lambda}^* a_\lambda+h.c.),\\
    &\vb{\Pi} = \sum_\lambda(-i\bm{\eta}^*_\lambda a_\lambda+h.c.),\\
    &\vb{X} = \sum_\lambda(i\bm{\gamma}_\lambda^* a_\lambda+h.c.).
\end{aligned}
\end{equation}
We define the longitudinal part of the vector potential 
\begin{equation}
    \mathbf{A}_\mathrm{L} = -\sum_\lambda \left[\omega_\lambda^{-1}(\omega_p\bm{\gamma^*}_\lambda)_\mathrm{L}a_\lambda + h.c.\right]
\end{equation}

Under the unitary $U$ defined in Eq.~(\ref{eq:unitary_u}), the transformation of $a_\lambda$ and $a_\lambda^\dagger$ then reads
\begin{equation}
    U^\dagger a_\lambda U = a_\lambda+iF_\lambda, \quad U^\dagger a_\lambda^\dagger U = a^\dagger_\lambda - iF_\lambda.
\end{equation}

To understand the transformation of $\psi_{\vb{r}}$ under $U$, we calculate the commutation relations of $\vb{A}_\mathrm{L}$ with $\vb{A}_\mathrm{T}$, and $\vb{A}_\mathrm{L}$ with itself,
\begin{equation}
\begin{aligned}
    &[\vb{A}_\mathrm{L}(\vb{r}), \vb{A}_\mathrm{T}(\vb{r}')] = \sum_\lambda \frac{i}{2\omega_\lambda}\mathrm{Im}\left[\vb{f}^*_{\mathrm{L},\lambda}(\vb{r})\otimes\vb{f}_{\mathrm{T},\lambda}(\vb{r}')\right],\\
    &[\vb{A}_\mathrm{L}(\vb{r}), \vb{A}_\mathrm{L}(\vb{r}')] = \sum_\lambda\frac{i}{2\omega_\lambda}\mathrm{Im}\left[\vb{f}^*_{\mathrm{L},\lambda}(\vb{r})\otimes\vb{f}_{\mathrm{L},\lambda}(\vb{r}')\right].
    \label{eq_commutator}
\end{aligned}
\end{equation}
Combining Eq.~(\ref{eq_eig_f_app}) and the fact that $\varepsilon(\omega)$ is a real function, we find that the matrix $\vb{S}(\vb{r},\vb{r}') = \sum_\lambda \omega_\lambda^{-1}\vb{f}_\lambda(\vb{r})\vb{f}_\lambda^\dagger(\vb{r}')$ is real, thus both commutators in Eq.~(\ref{eq_commutator}) vanish. Since field operators $\vb{A}_\mathrm{L}(\vb{r})$ at different $\vb{r}$ all commute, the transformation of a fermionic field $\psi_{\vb{r}}$ under $U$ is
\begin{equation}
    U^\dagger \psi_{\vb{r}}U = \exp\left[\frac{ie}{4\pi}\int\mathrm{d}\vb{r}'\frac{\nabla'\cdot\vb{A}_\mathrm{L}(\vb{r}')}{|\vb{r}-\vb{r}'|}\right]\psi_{\vb{r}}.
    \label{eq_trans_psi}
\end{equation}
Now we can investigate the transformation of covariant derivative $\psi^\dagger_{\vb{r}}(-i\nabla+e\vb{A}_\mathrm{T})\psi_{\vb{r}}$ under $U$, using Eq.~(\ref{eq_trans_psi}) and the fact that $\vb{A}_\mathrm{L}(\vb{r})$ commutes with $\vb{A}_\mathrm{T}$, we obtain
\begin{equation}
    U^\dagger \psi^\dagger_{\vb{r}}(-i\nabla+e\vb{A}_\mathrm{T})\psi_{\vb{r}} U = \psi^\dagger_{\vb{r}}(-i\nabla + e\vb{A})\psi_{\vb{r}},
\end{equation}
where $\vb{A} = \sum_{\lambda}\sqrt{\frac{1}{2\omega_\lambda}}(\vb{f}^*_\lambda a_\lambda+\vb{f}_\lambda a_\lambda^\dagger)$. 

Using the transformation of $a_\lambda$ and $\psi_{\vb{r}}$ under $U$, the transformed Hamiltonian $H_\mathrm{Weyl} = U^\dagger H U$ reads
\begin{equation}
     H_\mathrm{Weyl} =  H^0_\text{e} + W +H_\mathrm{ph} -\int\mathrm{d}\vb{r}\vb{J}_\text{e}(\vb{A})\cdot\vb{A}.
\end{equation}
where $\vb{J}_\mathrm{e}(\vb{A})$ is an effective current that can be separated out for notational concreteness, in particular $\vb{J}_\mathrm{e}(\vb{A}) = -ev_F\bar{\psi}\bm{\gamma}\psi$ for Dirac fermion, and $\vb{J}_\mathrm{e}(\vb{A}) = \frac{e}{2m}\left[i\hbar(\psi^\dagger\nabla\psi - \psi\nabla\psi^\dagger) - e\vb{A}\right]$ for Schr\"odinger fermion. $\vb{A}(\vb{r})$ retains the mode decomposition of \cref{eq:gauge_vector_decomp} in terms of the mode functions used to diagonalize $H\_{ph}$. The transformation absorbs $\phi$ into a screened Coulomb interaction
\begin{equation}
    \begin{aligned}
        W = V - \frac{1}{2\epsilon_0}\sum_\lambda F^\dagger_\lambda F_\lambda,
        \label{eq:v_screened}
    \end{aligned}
\end{equation}
with
\begin{equation}
     F_\lambda = \frac{1}{4\pi}\int_\Omega\mathrm{d}\vb{r}\rho_\text{e}(\vb{r})\int_\Omega\mathrm{d}\vb{r}'\frac{\nabla'\cdot \vb{f}^*_{\lambda,\mathrm{L}}(\vb{r}')}{|\vb{r}-\vb{r}'|}.
     \label{eq:F_lambda}
\end{equation}
As an example, we now calculate the screened Coulomb interaction $W$ for a surface polaritonic cavity. Under the deep sub-wavelength limit~\cite{chengAnyonicChern2025}, the surface modes have a fixed frequency $\omega_s = \sqrt{\omega^2_o+\omega_p^2/2}$, and the mode function reads [see \cref{sec:appdx_mode_functions} for a derivation]
\begin{equation}
    \vb{f}^*_{\vb{q}}(\vb{r}) = -\frac{1}{2}\frac{\omega_p}{\omega_s}\sqrt{\frac{q}{S}}e^{\pm qz+i\vb{q}\cdot\vb{r}_\parallel}(i\hat{\vb{q}} \pm \hat{\vb{z}}),
\end{equation}
where $\vb{q}$ is the in-plane momentum of the surface mode, $\vb{r}_\parallel = (x,y)$ is the in-plane coordinate, $+$ and $-$ signs are for the vector potential in $z<0$ and $z>0$ region. Substituting $\vb{f}^*_{\vb{q}}$ into Eq.~(\ref{eq:F_lambda}) gives
\begin{equation}
    F_{\vb{q}} = \frac{1}{2\sqrt{S}}\frac{\omega_p}{\omega_s}\int\mathrm{d}\vb{r}\rho_\mathrm{e}(\vb{r})\frac{e^{-qz+i\vb{q}\cdot\vb{r}}}{\sqrt{q}}.
\end{equation}
And the screening potential between two electrons above the surface is thus given by
\begin{equation}
\begin{aligned}
    \Delta V &= W-V=-\frac{1}{2}\sum_{\vb{q}} F_{\vb{q}}^\dagger F_{\vb{q}}\\
    & = \frac{-1}{16\pi}\frac{\omega_p^2}{\omega_s^2}\int\mathrm{d}\vb{r}\int\mathrm{d}\vb{r}'\frac{\rho_\mathrm{e}(\vb{r})\rho_\mathrm{e}(\vb{r}')}{\sqrt{(z+z')^2+|\vb{r}_\parallel - \vb{r}'_\parallel|^2}}.
\end{aligned} 
\end{equation}
Using the fact that the static dielectric constant of the media is $\varepsilon = 1+\frac{\omega_p^2}{\omega_o^2}$, $\Delta V$ is equal to the Coulomb interaction between a charge of $-e$ located at $(x,y,z)$ and a charge of $\frac{\varepsilon-1}{\varepsilon+1}e$ located at $(x',y',-z')$. This is the well-known mirror charge potential from electrostatics. For a plasmonic cavity, $\omega_p\gg\omega_o$; when the two electrons are both distance $z$ from the surface, and $\delta r$ apart in the horizontal direction, $\Delta V$ reads
\begin{equation}
    \Delta V(\delta r) = -\frac{\alpha}{\sqrt{4z^2+\delta r^2}}.
\end{equation}
The correction of Hubbard $U$ then reads
\begin{equation}
\begin{aligned}
    \Delta U &=\Delta V(0) - \Delta V(d)\\
    &=-\frac{\alpha}{2l} + \frac{\alpha}{\sqrt{4l^2 + d^2}}\approx -\frac{\alpha d^2}{16 l^3} = -\frac{1}{2}\tilde{g}^2\omega_s.
\end{aligned}
\end{equation}

\section{Obtain the Screened Interaction from Electrostatics}
\label{appdx:electrostatics}
The fact that the screened potential $W$ can be obtained from electrostatics is not specific to the case of surface polaritonic cavity. We first prove it for a general dielectric function $\varepsilon(\omega)$ that is spatially homogeneous. 

The screening potential $\Delta V$ can be written as
\begin{equation}
\begin{aligned}
    \Delta V = -\frac{1}{2}\int\mathrm{d}\vb{r}\int\mathrm{d}\vb{r}'\rho\_e(\vb{r})\rho\_e(\vb{r}')G(\vb{r},\vb{r'}),
\end{aligned}
\end{equation}
where
\begin{equation}
\begin{aligned}
    G(\vb{r},\vb{r}') &= \frac{1}{(4\pi)^2}\int\mathrm{d}\vb{r}_1\int\mathrm{d}\vb{r}_2\times\\
    &\nabla\frac{1}{|\vb{r}_1-\vb{r}|}\cdot\sum_\lambda\vb{f}_\lambda(\vb{r}_1) \vb{f}^\dagger_\lambda(\vb{r}_2)\cdot\nabla'\frac{1}{|\vb{r}_2-\vb{r}'|}
\end{aligned}
\end{equation}
For a spatially homogeneous dielectric function, the longitudinal modes can be labeled by its wave vector $\vb{q}$ and a branch index $\alpha$, the mode functions take the form $\vb{f}_\mathrm{L,\vb{q}\alpha} = \mathcal{N}_{\alpha}\hat{\vb{q}}e^{-i\vb{q}\cdot\vb{r}}$, where $\mathcal{N}_\alpha^{-1} = \sqrt{\Omega}\omega_\alpha\partial_\omega\varepsilon(\omega_\alpha)/2$ and the mode frequency $\omega_\alpha$ satisfies $\varepsilon(\omega_\alpha) = 0$. The completeness relation thus reads
\begin{equation}
    \sum_\lambda \vb{f}_{\lambda,\mathrm{L}}(\vb{r}_1)\vb{f}_{\lambda,\mathrm{L}}^\dagger(\vb{r}_2) = \frac{1}{2\pi}\sum_\alpha\frac{\nabla_1\nabla_2 |\vb{r}_1-\vb{r}_2|^{-1}}{\partial_\omega(\omega\varepsilon)|_{\omega=\omega_\alpha}}.
\end{equation}
To work out the summation over $\alpha$, we consider the contour integral $\mathcal{I} = \frac{1}{2\pi i}\oint_C\mathrm{d}\omega\omega^{-1}\varepsilon^{-1}(\omega)$, where $C$ is a large circle at infinity going counter-clockwise. We assume that the dielectric function is real in both frequency and time domain (which is only violated if time-reversal symmetry is broken), thus $\varepsilon(\omega) = \varepsilon(-\omega)$. The zeros of $\varepsilon$ are symmetric on the positive half and negative half of the frequency axis, and only the positive half are kept as $\omega_\alpha$. Using the residue theorem we obtain
\begin{equation}
    \mathcal{I} = 2\sum_\alpha\frac{1}{{\partial_\omega(\omega\varepsilon)|_{\omega_\alpha}}}+ \frac{1}{\varepsilon(0)},
    \label{eq:residue_identity}
\end{equation}
where the factor 2 comes from summation over positive and negative halves of the $\omega$ axis. On the other hand, the optical response of any material is suppressed at very high frequency, namely $\epsilon(\infty) = 1$, thus $\mathcal{I} = 1$, and combining with Eq.~\ref{eq:residue_identity} gives $2\sum_\alpha \left[\partial_\omega\varepsilon(\omega)\right]^{-1}_{\omega_\alpha} = 1-\varepsilon(0)^{-1}$. The interaction kernel $G$ for $\Delta V$ thus reads $G(\vb{r},\vb{r}') = \left[(4\pi\varepsilon(0))^{-1}-(4\pi)^{-1}\right]|\vb{r}-\vb{r}'|^{-1}$, and it follows naturally that the screened interaction is $(4\pi\varepsilon(0))^{-1}|\vb{r}-\vb{r}'|^{-1}$, namely Coulomb interaction with the static dielectric constant $\varepsilon(0)$.

Now we move on to the most general case where $\varepsilon(\vb{r},\omega)$ depends both on frequency and position. Defining an operator $T(\omega) = \omega^2\varepsilon(\vb{r},\omega) - \nabla\times\nabla\times$, we construct the contour integral $\mathcal{I}=\frac{1}{2\pi i}\oint_C \mathrm{d}\omega \omega T^{-1}(\omega)$, since $T(\omega)\rightarrow\omega^2 I$ as $\omega\rightarrow\infty$, $\mathcal{I}$ is just the identity operator $I$. On the other hand, the integral can be calculated as a sum of residues, and because $\varepsilon(\omega) = \varepsilon(-\omega)$, the eigen frequencies distribute symmetrically on the $\omega$ axis, and opposite eigen frequencies contribute the same residue for $\mathcal{I}$. We denote the positive eigen frequencies as $\omega_\lambda$. As $\omega\rightarrow\omega_\lambda$, $T^{-1}(\omega)$ has an asymptotic form $T^{-1}\sim Z^{-1}\frac{\ket{\lambda} \bra{\lambda}}{\omega-\omega_\lambda}$, where $\ket{\lambda} = \vb{f}_\lambda(\vb{r})$. Using the relation $\bra{\lambda}TT^{-1}\ket{\lambda} = \left<\lambda|\lambda\right>$, $Z$ is found to be $\bra{\lambda}T'(\omega_\lambda)\ket{\lambda}$. The residue theorem for integral $\mathcal{I}$ thus reads
\begin{equation}
    \mathcal{I} = \sum_\lambda \frac{2\omega_\lambda\ket{\lambda}\bra{\lambda}}{\bra{\lambda}T'(\omega_\lambda)\ket{\lambda}} + \mathrm{Res}\left[\omega T^{-1}(\omega)\right]|_{\omega=0}.
    \label{eq:operator_residue}
\end{equation}
Only second-order pole of $T^{-1}(\omega)$ at $\omega=0$ will contribute to the residue, for which $\varepsilon(\vb{r},\omega)$ can be taken to be $\varepsilon(\vb{r},0)$. Defining $\mathbb{L}$ as the subspace of longitudinal functions, $\mathbb{L}$ is the null space of $\nabla\times\nabla\times$, but generally not invariant under $\varepsilon(0)$. Using $\mathcal{P}_\mathrm{L}$ to denote the projector onto $\mathbb{L}$, the residue in Eq.~\ref{eq:operator_residue} at $\omega=0$ reads
\begin{equation}
    \mathrm{Res}\left[\omega T^{-1}(\omega)\right]|_{\omega=0} = \mathcal{P}_\mathrm{L}\left[\mathcal{P}_\mathrm{L}\varepsilon(0)\mathcal{P}_\mathrm{L}\right]_\mathbb{L}^{-1}\mathcal{P}_\mathrm{L},
\end{equation}
where $\varepsilon(0)$ is regarded as an operator, and $\mathcal{P}\varepsilon(0)\mathcal{P}$ is assumed to be invertible on $\mathbb{L}$. We will denote this residue by $\tilde{\varepsilon}(0)^{-1}$, as it is the pseudo-inverse of $\mathcal{P}_\mathrm{L}\varepsilon(0)\mathcal{P}_\mathrm{L}$. We further denote $(4\pi)^{-1}\nabla_{\bm{\xi}}|\vb{r}-\bm{\xi}|^{-1}$ as $\ket{{\bm{\xi}}}$, which lies within $\mathbb{L}$. The kernel $G(\vb{r},\vb{r}')$ reads
\begin{equation}
    G(\vb{r},\vb{r}') = \frac{1}{4\pi|\vb{r}-\vb{r}'|} - \bra{\vb{r}}\tilde{\varepsilon}(0)^{-1}\ket{\vb{r}'}.
\end{equation}
And the kernel $W(\vb{r},\vb{r}')$ for screened interaction is $\bra{\vb{r}}\tilde{\varepsilon}(0)^{-1}\ket{\vb{r}'}$. $W(\vb{r},\vb{r}')$ satisfies the following equation
\begin{equation}
\begin{aligned}
    \nabla\cdot\left[\varepsilon(\vb{r},0)\nabla W(\vb{r},\vb{r}')\right] &= \nabla\cdot\mathcal{P}_\mathrm{L}\varepsilon(0)\mathcal{P}_\mathrm{L}\tilde{\varepsilon}(0)^{-1}\ket{\vb{r}'}\\
    &= -\nabla\cdot\ket{\vb{r}'} = -\delta(\vb{r}-\vb{r}'),
\end{aligned}
\end{equation}
which is the Poisson equation in an dielectric media with $\varepsilon(\vb{r},0)$. Therefore, the screened interaction $W$ calculated from Eq.~\ref{eq:v_screened} is exactly the same as the electrostatic interaction in an dielectric media with $\varepsilon(\vb{r},0)$. 
 
\section{Screening of the Hubbard Interaction}\label{sec:screening}
Now we show how to determine $\Delta U$ for a two-site Hubbard model, with sites $i$ and $j$ located at $\vb{R}_i$ and $\vb{R}_j$. Assuming that the Wannier orbital at each site has a radius much smaller than $|\vb{R}_i - \vb{R}_j|$, we can approximate its charge distribution by a point charge. When $\rho_\mathrm{e}(\vb{r}) = -e\delta(\vb{r}-\vb{R})$, we denote $F_\lambda$ in Eq.~(\ref{eq:F_lambda}) as $F_
\lambda(\vb{R}$), which can be expanded around $\vb{R}_0$ as
\begin{equation}
\begin{aligned}
    F_\lambda (\vb{R}) &\approx F_\lambda(\vb{R}_0) + (\vb{R}-\vb{R}_0)\cdot \nabla_{\vb{R}}F_\lambda(\vb{R}_0)\\
    &= F_\lambda(\vb{R}_0) + e(\vb{R}-\vb{R}_0)\cdot \vb{f}^*_{\lambda,\mathrm{L}}(\vb{R}_0).
\end{aligned}
\end{equation}
Further using this in Eq.~(\ref{eq:v_screened}), we find the screening correction to $U$ to be $\Delta U = -\frac{1}{2}\sum_\lambda \omega_\lambda |g_{\lambda,\mathrm{L}}|^2$, where $g_{\lambda,\mathrm{L}}$ is the coupling constants to longitudinal part of the modes, defined as
\begin{equation}
g_{\lambda,\mathrm{L}} = \sqrt{\frac{e^2}{2 \hbar \epsilon_0 \omega_\lambda}}
    (\vb{R}_j-\vb{R}_i) \cdot \vb{f}_{\lambda,\mathrm{L}}\left(\frac{\vb{R}_j+\vb{R}_i}{2}\right).
\end{equation}

\section{Weyl Gauge in Vacuum}
\label{sec:weyl_gauge_vacuum}
In this section we discuss the canonical quantization of QED in vacuum, and compare with the dynamical Weyl gauge we introduce in this work. We start from the QED Lagrangian
\begin{equation}
    L = \int\mathrm{d}^4 x \,\frac{1}{2}(E^2-B^2)-\rho\phi+\vb{A}\cdot\vb{J}+\mathcal{L}^0_\mathrm{e}.
\end{equation}
By choosing Weyl gauge where $\phi=0$, the Lagrangian reads
\begin{equation}
\begin{aligned}
    L &= \int\mathrm{d}^4 x\,\frac{1}{2}\left(|\partial_t\vb{A}_\mathrm{T}|^2+|\partial_t\vb{A}_\mathrm{L}|^2-|\nabla\times\vb{A}_\mathrm{T}|^2\right)\\
    &+ \int\mathrm{d}^4 x\,\vb{A}_\mathrm{T}\cdot\vb{J}_\mathrm{T} + \vb{A}_\mathrm{L}\cdot\vb{J}_\mathrm{L} +\mathcal{L}^0_\mathrm{e}.
    \label{eq:weyl_lagrangian}
\end{aligned}
\end{equation}
To proceed, a first approach is to integrate out $\vb{A}_\mathrm{L}$. By using charge conservation equation $\nabla\cdot\vb{J}_L = -\partial_t\rho$, we can formally write $\vb{J}_{\mathrm{L}}(\vb{k},\omega) = \frac{\omega}{k}\hat{\vb{k}}\rho(\vb{k},\omega)$ in reciprocal space. Integrating out $\vb{A}_\mathrm{L}$ then gives rise to an interaction $V_\mathrm{coulomb} = \frac{1}{2V}\sum_{\vb{k}}\frac{1}{k^2}\rho_{\vb{k}}\rho_{\vb{-k}}$, recovering the Hamiltonian in Coulomb gauge. However, this approach is not rigorous, as the subtlety of initial condition is not properly treated in reciprocal formulation.

A more rigorous approach is to notice that Eq~(\ref{eq:weyl_lagrangian}) allows a direct canonical quantization, by introducing the conjugate coordinate $\vb{E} = -\partial_t\vb{A}$, which satisfies the commutation relation $[E^i(\vb{r}),A^j(\vb{r}')] = i\delta_{ij}\delta(\vb{r}-\vb{r}')$. The corresponding Hamiltonian is
\begin{equation}
    H_\mathrm{QED} =  H^0_\mathrm{e}+\int\mathrm{d}\vb{r}\,\frac{1}{2}|\vb{E}|^2 + \frac{1}{2}|\nabla\times\vb{A}_\mathrm{T}|^2 - \vb{A}\cdot\vb{J}.
    \label{eq:H_qed}
\end{equation}
We then define a unitary transformation $\tilde{U}$ as
\begin{equation}
    \tilde{U} = \exp\left[\frac{-i}{4\pi}\int\mathrm{d}\vb{r}\rho(\vb{r})\int\mathrm{d}\vb{r}'\frac{\nabla'\cdot\vb{A}_\mathrm{L}(\vb{r}')}{|\vb{r}-\vb{r}'|}\right],
\end{equation}
which is essentially the inverse of the unitary $U$ in Eq.~(\ref{eq:unitary_u}). $\vb{E}(\vb{r})$ transforms under $\tilde{U}$ as
\begin{equation}
    \tilde{U}^\dagger \vb{E}(\vb{r}) \tilde{U} = \vb{E}(\vb{r}) + \frac{1}{4\pi}\int\mathrm{d}\vb{r}\rho(\vb{r}')\frac{\vb{r}-\vb{r}'}{|\vb{r}-\vb{r}'|^3}.
\end{equation}
And $H_\mathrm{QED}$ transforms under $\tilde{U}$ as
\begin{equation}
\begin{aligned}
    &\tilde{U}^\dagger H_\mathrm{QED} \tilde{U} = H^0_\mathrm{e}+\int\mathrm{d}\vb{r}\,\frac{1}{2}|\vb{E}_\mathrm{T}|^2 + \frac{1}{2}|\nabla\times\vb{A}_\mathrm{T}|^2\\
    &+ \int\mathrm{d}\vb{r}\,\frac{1}{2}|\vb{E}_\mathrm{L}|^2 + \phi \rho - \vb{A}_\mathrm{T}\cdot\vb{J}_\mathrm{T} +V_\mathrm{coulomb},
\end{aligned} 
\end{equation}
where $\phi(\vb{r}) = \frac{1}{4\pi}\int\mathrm{d}\vb{r}'\frac{\vb{r}'-\vb{r}}{|\vb{r}'-\vb{r}|^3}\cdot\vb{E}_\mathrm{L}(\vb{r}')$ and $V_\mathrm{coulomb} = \frac{1}{8\pi}\iint\mathrm{d}\vb{r}\mathrm{d}\vb{r}'\frac{\rho(\vb{r})\rho(\vb{r}')}{|\vb{r}-\vb{r}'|}$.

Since $\vb{E}(\vb{r})$ commutes with $\tilde{H}_\mathrm{QED} = \tilde{U}^\dagger H_\mathrm{QED} \tilde{U}$, it is in fact a classical field. Minimizing $\tilde{H}_\mathrm{QED}$ with respect to $\vb{E}_\mathrm{L}$ gives $\vb{E}_\mathrm{L}(\vb{r}) = -\frac{1}{4\pi}\int\mathrm{d}\vb{r}\bar{\rho}(\vb{r}')\frac{\vb{r}-\vb{r}'}{|\vb{r}-\vb{r}'|^3}$, where $\bar{\rho}(\vb{r})$ is the expectation value of charge density. The energy $E_\mathrm{aux}$ contributed by the terms $ \int\mathrm{d}\vb{r}\left(\frac{1}{2}|\vb{E}_\mathrm{L}|^2 + \phi \rho\right)$ is $-\frac{1}{8\pi}\iint\mathrm{d}\vb{r}\mathrm{d}\vb{r}'\frac{\bar\rho(\vb{r})\bar\rho(\vb{r}')}{|\vb{r}-\vb{r}'|}$, which exactly cancels the Hartree part of Coulomb energy contributed by $V_\mathrm{coulomb}$. However, $E_\mathrm{aux}$ must vanish for a physical state in order to recover the result of Coulomb gauge. Thus we conclude that the ground state of $\tilde{H}_\mathrm{QED}$ is not a physical state. The physical states actually live within the null space of $\vb{E}_\mathrm{L}(\vb{r})$, and because $\vb{E}_\mathrm{L}$ is a good quantum number for $\tilde{H}_\mathrm{QED}$, this constraint is always satisfied as long as it is satisfied as an initial condition. For the system before the transformation $\tilde{U}$, the constraint reads
\begin{equation}
    \vb{E}_\mathrm{L}\ket{\mathrm{phsical}} = \frac{1}{4\pi}\int\mathrm{d}\vb{r}\rho(\vb{r}')\frac{\vb{r}-\vb{r}'}{|\vb{r}-\vb{r}'|^3}\ket{\mathrm{physical}},
\end{equation}
which is simply the Gauss' law $(\nabla\cdot\vb{E}-\rho)\ket{\mathrm{physcial}} =0$.

To sum up, the canonical quantization of QED in vacuum under the Weyl gauge must be supplemented by a constraint of Hilbert space, in order to fix the longitudinal, zero energy modes $\vb{E}_\mathrm{L}$. Meanwhile, in the dynamical Weyl gauge, the longitudinal modes acquire finite frequencies, and are physical degrees of freedom originating from the matter excitation inside the media. In fact, if one adds a term $\frac{1}{2}\int\mathrm{d}\vb{r}\omega_\mathrm{LO}^2|\vb{A}_\mathrm{L}|^2$ to the Hamiltonian~(\ref{eq:H_qed}), it will be able to describe the QED inside a media, coupled to a longitudinal optical phonon with frequency $\omega_\mathrm{LO}$. In this case, $\vb{A}_\mathrm{L}$ describes physical degrees of freedom, and no constraint on Hilbert space is needed.

\section{Cavity Schrieffer Wolff Transformation}
\label{sec:appdx_sw_transform}

We briefly outline the steps used to derive \cref{eq:sw_hamiltonian}. For an in depth derivation we refer to the Supplemental Material Section III of our previous work \cite{grunwaldCavitySpectroscopy2024}. The Schrieffer Wolff (SW) transformation is a unitary block diagonalization transformation that eliminates the coupling between energetically separated low- and high-energy sectors of the Hilbert space. It is implemented via an anti-unitary SW operator $S = -S^\dagger$ as \cite{schriefferRelationAnderson1966,bravyiSchriefferWolff2011,grunwaldCavitySpectroscopy2024}
%
\begin{align}
    \tilde{H} = e^{S} H e^{-S} = H + [S, H] + \frac{1}{2!} [S, [S, H]] + \dots.
\end{align}
Low- and high-energy Hilbert space sectors are represented by projectors $\mathcal{P}$ (low energy) and $\mathcal{Q}$ (high energy) with $\mathcal{P} + \mathcal{Q} = 1$ and $\mathcal{P} \mathcal{Q} = 0$. The new basis is defined by $\mathcal{P} \tilde{H} \mathcal{Q} = 0 = \mathcal{Q} \tilde{H} \mathcal{P}$, viz. the condition that there is no coupling between the sectors, defining an effective low-energy theory.

The SW operator $S$ can in general not be determined in closed form, and is instead computed as a perturbation expansion. For the Hubbard model in the main text it is determined as an expansion in $t/U \ll 1$. The coupling between low- and high-energy sectors, here spanned by states with doubly occupied sites, is eliminated order-by-order. By expanding the Hamiltonian \cref{eq:cavity_hubbard} in the photon number basis, we obtain the leading order SW operator \cite{grunwaldCavitySpectroscopy2024}
\begin{align}
    \label{eq:sw_operator}
    S^{\vb{n}\vb{n'}} = - \Big(&
        \mathcal{P} H_t^{\vb{n}\vb{n'}} \mathcal{Q} \frac{1}{U + \vb*{\omega} \cdot (\vb{n'} - \vb{n})}
        \\
        &+
        \mathcal{Q} H_t^{\vb{n}\vb{n'}} \mathcal{P} \frac{1}{-U + \vb*{\omega} \cdot (\vb{n'} - \vb{n})}
        \Big)
        + \order*{t/U}^3,
        \nonumber
\end{align}
where $H_t^{\vb{n}\vb{n'}} = \mel{\vb{n}}{H_t}{\vb{n'}}$ is the hopping Hamiltonian evaluated between cavity number states. Using \cref{eq:sw_operator}, the effective low energy theory of \cref{eq:cavity_hubbard} takes the form of the photon dressed Heisenberg model \cref{eq:sw_hamiltonian} with highly non-linear light matter coupling. Note that this expansion is only valid for $U \neq n\Omega, n \in \mathbb{Z}$, but already becomes unreliable close to these resonances due to diverging energy denominators~\cite{grunwaldCavitySpectroscopy2024}. In this work we are far detuned from this resonance condition.

\section{Magnetic Exchange Resummation}
\label{sec:appdx_resummation}

\subsection{Scaling Derivation}
\label{ssec:appdx_resummation_scaling}

We provide an alternative expression for the magnetic exchange interaction \cref{eq:resum_magnetic_J_multinomial} in the thermodynamic limit, derived using an approach inspired by the volume scaling argument in \cite{liEffectiveTheory2022}. After Laplace decoupling this approach reproduces our result from the main text \cref{eq:resum_laplace_intermediate}. To start, notice that we can decompose the summation $\sum_{\vb{k}}$ into distinct terms containing $s$ virtual photons, via $\sum_{\vb{k}} = \sum_{s = 0}^{\infty} \sum_{\{ \sum_\lambda k_\lambda = s \} }$. The exchange interaction can then be written as
\begin{align}
    J_{ij} &= J_0 e^{-\sum_\lambda |g_\lambda|^2} \sum_{s = 0}^{\infty} \sum_{\{ \sum_\lambda k_\lambda = s \} } \nonumber \\
    &\quad \times \left| 
        \mel{\vb{0}}{\frac{(-i)^s}{s!} 
    \left( \sum_\lambda g_\lambda^\dagger a_\lambda \right)^s }{\vb{k}_s} 
    \right|^2
    \frac{U}{U + \vb*{\omega} \cdot \vb{k}_s}
    \label{eq:resum_magnetic_J_scaling_1}
\end{align}
%
where we used that the state $\ket{\vb{k}_s}$ appearing in the expectation value contains exactly $s$ photons, so that only $(\dots)^s$ in the Laurent series of the Peierls phase exponential can yield a finite contribution. The advantage of this approach is that the occupation sums can be transformed into sums over the modes, order-by-order. To understand this, consider the first non-trivial order $s = 1$, viz. processes with a single virtual photon that can occupy any of the modes
\begin{align}
    J_{ij}^{(s = 1)} = J_0 e^{-\sum_\lambda |g_\lambda|^2}
    \sum_\lambda |g_\lambda|^2
    \frac{U}{U + \omega_\lambda}.
\end{align}
The next order is more interesting and can be directly evaluated, by expanding the quadratic brace $(\dots)^2$ to derive
\begin{align}
    J_{ij}^{(s = 2)} &= J_0 e^{-\sum_\lambda |g_\lambda|^2}
    \Big( \frac{1}{2!} \sum_{\lambda} |g_{\lambda}|^4 \frac{U}{U + 2\omega_\lambda} + \ \\
        & \hspace{1cm} + \frac{1}{2!} \sum_{\lambda_1 \neq \lambda_2}
        |g_{\lambda_1}|^2 |g_{\lambda_2}|^2 \frac{U}{U + \Omega_{\lambda_1} + \Omega_{\lambda_2}}
    \Big),
    \nonumber
\end{align}
where the second term described processes, with the two photons in distinct cavity modes, while the first captures two photons occupying the same mode. Because of the mode function normalization, the effective couplings \cref{eq:resum_coupling_constant} scale as $|g|^2 \sim V^{-1}$, such that only the first term gives a finite contribution in the thermodynamic limit (TD). This is valid at all orders $s$, such that we can write the $s$-photon contribution as
\begin{align}
    J_{ij}^{(s)} &= J_0 e^{-\sum_\lambda |g_\lambda|^2}
        \frac{1}{s!}
        \underset{\lambda_1 \neq \lambda_2 \neq \dots \neq \lambda_s}{\sum \dots \sum}
        |g_{\lambda_1}|^2 \dots |g_{\lambda_s}|^2 \\
        & \hspace{2.5cm}\times
        \frac{U}{U + \Omega_{\lambda_1} + \Omega_{\lambda_2} + \dots + \Omega_{\lambda_s}},
        \nonumber
\end{align}
where the restriction $\lambda_1 \neq \lambda_2 \neq \dots \lambda_s$ becomes a point measure in the TD limit and can be dropped.
Finally, the magnetic exchange interaction takes the rather complicated form of a coupled
sum
\begin{align}
    J_{ij} &= J_0 e^{-\sum_\lambda |g_\lambda|^2}
    \sum_{s = 0}^\infty \frac{1}{s!} \\
    &\times
    \left(
    \sum_{\lambda_1} \dots \sum_{\lambda_s}
    |g_{\lambda_1}|^2 \dots |g_{\lambda_s}|^2
    \frac{U}{U + \sum_{k = 1}^s \Omega_{\lambda_s}}
    \right).
    \nonumber
\end{align}
As with the derivation in the main text, the fraction couples the different summations making the problem exponentially hard. This can be alleviated by Laplace transformation \cref{eq:resum_laplace_decoupling} to write the expression as
\begin{align}
    \frac{J_{ij}}{J_0} &= \int_0^\infty \dd{x} e^{-x -\sum_\lambda |g_\lambda|^2}
    \sum_{s = 0}^\infty \frac{1}{s!}
    \left( \sum_\lambda
    |g_{\lambda}|^2 e^{-x U^{-1} \omega_\lambda}
    \right)^s,
    \nonumber
\end{align}
which reduce to our previous result \cref{eq:resum_laplace_intermediate} by recognizing the exponential Laurent series in $s$. This provides a good consistency check for our result.

\subsection{Regularization}
\label{ssec:appdx_resummation_regularization}

Our result \cref{eq:resum_laplace_intermediate} has to be regularized due to the UV divergence of the PDOS $\rho(\omega) \sim \omega^2$ in the UV $\omega \to \infty$. Here, we provide two viewpoints and show their (perturbative) equivalence.

\paragraph*{Ab-Initio Perspective} In the main text, we argued that the coupling of our theory to the free space electromagnetic field leads to a finite electronic mass and emergence of the Coulomb interaction \cite{ruggenthalerUnderstandingPolaritonic2023}. In particular the UV behavior of the PDOS is responsible for the electronic mass. This is an \textit{ab-initio} perspective towards regularizing the theory, where we take into account that the coupling to the free space PDOS is intrinsic to the model parameters we observe without the cavity. We can only ask question relative to free space, hence replacing $\rho \to \rho - \rho_0$. The resulting theory is UV regular as the high-energy behavior is absorbed into the finite free-space electronic mass \cite{eckhardtCavitronicsLowDimensional2024}. This approach is equivalent to only explicitly taking into account modes up to the plasma frequency of the cavity mirrors, while reabsorbing the unconstrained modes into the electronic mass \cite{luCavityEngineering2025}.

\paragraph*{Model Perspective}

Independent of the specific derivation of the low-energy effective theory in \cref{eq:sw_hamiltonian}, one may treat the model as a standalone framework and introduce an appropriate regularization at this level. While this leads to a non-renormalizable theory that is only sensitive to the UV behavior of the PDOS in the limit of vanishing cutoffs, maintaining a finite cutoff—as physically required—recovers the regularization of the \textit{ab-initio} approach.

In this context, we must address the physical interpretation of the parameter $J_0$ appearing in \cref{eq:resum_laplace_intermediate}. While the expression provides a framework to evaluate the magnetic exchange interaction for a given cavity mode structure, it remains incomplete in isolation. This stems from the fact that $J_0$ represents a bare parameter, which is physically unobservable and formally divergent \cite{zinn-justinQuantumFieldTheory2002}, analogous to the bare electronic mass \cite{ruggenthalerUnderstandingPolaritonic2023}.

In free space, we obtain a fully analogous expression to \cref{eq:resum_laplace_intermediate} with free space coupling constants and mode functions $g^0_\lambda$ [\cref{eq:resum_coupling_constant}] alongside the corresponding free space density $\rho_0(\omega)$. The physically measurable quantities are then defined as $J_{ij}^{\mathrm{vac}} = J_0 J_{ij}(g^0_\lambda)$ for vacuum and $J_{ij}^{\mathrm{cav}} = J_0 J_{ij}(g_\lambda)$ for the cavity, where $J_{ij}(g)$ is given by \cref{eq:resum_magnetic_J_multinomial}. By combining these relations, we can eliminate the bare coupling $J_0$ to determine the relative modification of the magnetic exchange relative to vacuum
\begin{align}
    \frac{J^{\mathrm{cav}}}{J^{\mathrm{vac}}} =
    \frac{
        \int_0^\infty \dd{x} e^{Z(x, \eta)}
        e^{M(x)}
    }{
        \int_0^\infty \dd{x} e^{Z(x, \eta)}
    } = \expval{e^{M(x)}}_{e^{Z(x, \eta)}},
    \label{eq:resum_fractional}
\end{align}
where we suppressed the site indices for notational convenience. The previously introduced cavity modification function $M(x)$ [\cref{eq:resum_cavity_mod}] appears in an expectation value with respect to a probability measure $Z(x, \eta)$, given as
\begin{gather}
    \label{eq:resum_prob_measure}
    Z(x, \eta) = -x + P_0 \int_0^\infty 
    \hspace{-0.1cm}
    \dd{\omega} \omega^{-1} \rho_0(\omega) e^{ -\omega x U^{-1}} e^{-\eta \omega},
\end{gather}
with the free density of states $\rho_{ij, 0}(\omega) = 1 / (3 \pi^2 c^3) \omega^2 = \rho_0 \omega^2$. We introduced an exponential regularizer $e^{-\eta \omega}$ that enables explicit calculations. Note that the cavity only enters through the function $M(x)$, while the probability measure $Z(x, \eta)$, which physically captures the coupling to vacuum modes, is universal and can be readily evaluated at finite exponential cutoff $g_\eta(x) = e^{-\eta x}$ to find
\begin{align}
    Z(x, \eta) = -x + \frac{P_0 \rho_0 U^2}{(x + U\eta)^2}.
    \label{eq:resum_prob_measure_evaluated}
\end{align}
It exhibits an essential singularity at $x = 0$ as $\eta \to 0$, while $M(x)$ is regular and bounded. Since the same divergence appears both in numerator and denominator, we expect an explicit cancellation and a finite result. At the same time, all spectral weight resides at $x = 0$ in this limit, such that we only probe $M(0) = 0$, leading to vanishing modification of the magnetic coupling. In other words, this implies that the physics is entirely dictated by the UV behavior of the PDOS, which is unaltered by optical elements. This conclusion seems physically problematic, as even our description of matter in terms of a Hubbard model breaks down at energy scales much larger than $U$, calling into question the inclusion of photonic modes at that energy scale.

At finite cutoff, e.g. set at the plasma frequency $\eta \sim (20\rm{~eV})^{-1}$, this regularization scheme perturbatively coincides with the ab-initio regularization scheme discussed in the last section. Practically, we would like to consider the coupling of light to matter up to a physically relevant energy scale, in general set by the energy where the cavity PDOS approaches free space. 

To investigate this further, we consider the sub-wavelength limit of a surface-plasmon cavity, where bulk modes can be neglected and the surface PDOS is approximately given by a delta-distribution. The resulting cavity modification function takes the form [see Main text and \cref{ssec:appdx_single_mode}]
\begin{align}
    M(x) = \tilde{g}^2 \left(e^{-x \theta} - 1 \right) \approx -x \theta \tilde{g}^2 + \order{\theta}^2
\end{align}
with $\theta = \omega_\infty / U$ and effective light-matter coupling $\tilde{g}^2$. The full cavity modification function is well approximated by its leading order Taylor expansion in $x \theta$. The integrals appearing in the evaluation of the magnetic modification \cref{eq:resum_fractional} then take the form ($\lambda > 0$)
\begin{align}
    \label{eq:appdx_freg_expansion}
    \mathcal{I}(\lambda) &= \int_0^\infty \dd{x} e^{-x \left(1 + \lambda \right)} e^{\alpha / (x + U \eta)^2} \\
    &= \frac{1}{1 + \lambda} + \alpha \bigg[ \frac{1}{\eta U} + (1 + \lambda) \Big( \gamma + \log[ (1 + \lambda) \eta U ] \Big) \bigg] \nonumber \\
    &\quad + \order{\alpha^2, \eta U}
    \nonumber
\end{align}
with $\lambda = \tilde{g}^2 \theta$ for the numerator and $\lambda = 0$ in the denominator of \cref{eq:resum_fractional}. In the second term we have evaluated the expression perturbatively in $\alpha = P_0 \rho_0 U^2 \sim 10^{-7}$ and $\eta U$, the former of which is a small universal constant appearing in the measure function $Z(x, \eta)$ [cf. \cref{eq:resum_prob_measure_evaluated}]. Note that we recover the ab-initio regularized result for $\alpha \to 0$, and can hence identify it with the first term in \cref{eq:appdx_freg_expansion}. Taking the fraction [\cref{eq:resum_fractional}], we arrive at our final result
\begin{align}
    \left( J / J_0 \right)\_{model}
    =  \left( J / J_0 \right)\_{ab-initio} + \order{\frac{\alpha}{\eta U}, \alpha \log(\eta U)},
    \nonumber
\end{align}
that perturbatively coincides with the regularization scheme used in the main text. In particular for realistic cutoffs $\eta \sim (20~\rm{eV})^{-1}$ the correction terms can be neglected against the ab-initio term for surface cavities. Note however that in this regularization scheme, the cutoff dependence is much more explicit and not automatically build into our formalism.

\section{Mode Functions and Photonic Density of States}
\label{sec:appdx_mode_functions}

The mode functions $\vb{f}_\lambda(\vb{r})$ of a cavity structure [see \cref{eq:gauge_vector_decomp}] define the photonic density of states [\cref{eq:photonic_density_of_states}], that in turn controls the magnetic exchange modification \cref{eq:resum_J}. They are determined as solutions of the Helmholtz equation
\begin{align}
    \frac{\varepsilon(\vb{r}, \omega_\lambda)\omega_\lambda^2}{c^2} \vb{f}_\lambda(\vb{r}) -
    \nabla \cross \nabla \cross \vb{f}_\lambda(\vb{r}, t) = 0,
    \label{eq:appdx_helmholtz}
\end{align}
subject to the generalized gauge condition $\nabla \cdot \left( \varepsilon(\vb{r}, \omega) \vb{f}_\lambda(\vb{r}) \right)= 0$ and Hopfield normalization
\begin{align}
    \int_\Omega \dd[3]{\vb{r}}
    \left( \varepsilon(\vb{r}, \omega_\lambda) +
    \frac{\omega_\lambda}{2} \pdv{\varepsilon(\vb{r}, \omega_\lambda)}{\omega_\lambda}  \right)
    \left| \vb{f}_\lambda(\vb{r}) \right|^2 = 1,
    \label{eq:quant_hopfield_normalization}
\end{align}
with the Hopfield function $ \mathcal{V}(\vb{r}, \omega_\lambda) = (\dots)$ that reduces to the conventional mode function normalization for frequency independent $\varepsilon(\omega)$. We consider a planar cavity with longitudinal confinement along the z-direction, while maintaining translational symmetry in the $(x, y)$-plane. \Cref{eq:appdx_helmholtz} is supplemented with metallic boundary conditions in $z$, periodic boundary conditions in plane $\vb{f}_\lambda(\vb{r} + L_\parallel \vb{e}_{x,y}) = \vb{f}_\lambda(\vb{r})$ as well as jump conditions at the interface and metallic boundary
\begin{align}
    \label{eq:appdx_helmoltz_interface1}
    \vb{n} \cdot \llbracket \vb{D} \rrbracket &= 0, \\
    \label{eq:appdx_helmoltz_interface2}
    \vb{n} \cross \llbracket \vb{H} \rrbracket &= 0, \\
    \label{eq:appdx_helmoltz_interface3}
    \vb{n} \cross \llbracket \vb{E} \rrbracket &= 0, \\
    \label{eq:appdx_helmoltz_interface4}
    \vb{n} \cdot \llbracket \vb{B} \rrbracket &= 0,
\end{align}
where $\vb{n}$ is the surface normal and $\llbracket A \rrbracket(x) = \lim_{\epsilon \to 0} \left[ A(x + \epsilon) - A(x - \epsilon) \right]$. Solving this partial differential equation yields the cavity dispersion as well as the mode functions for a given geometry \cite{eckhardtCavitronicsLowDimensional2024}.

\subsection{Fabry P\'erot Cavity}
\label{ssec:appdx_pdos_fp}

Consider an idealized Fabry P\'erot (FP) cavity with co-planer mirrors at $z = 0, d$, illustrated in the inset of \cref{fig:main_magnetic_modification_fp_a}. At these idealized surface, the parallel component of the electric field $\vb{E}_\parallel = 0$ [\cref{eq:appdx_helmoltz_interface3}] and the normal component of the magnetic field $\vb{B}_\perp = 0$ [\cref{eq:appdx_helmoltz_interface4}] vanishes. In a more realistic setup, the metallic mirrors could be modeled as a dispersive Drude metal with frequency dependent dielectric function $\epsilon(\omega)$. In this way the mirrors become transparent above the materials' plasma frequency $\omega_p$, reducing the associated PDOS to free space \cite{svendsenInitioCalculations2024}. In the current context this only leads to minor modifications.

The periodic boundary conditions discretize the in-plane momenta as $\vb{k}_\parallel = \frac{2 \pi}{L} \vb{n}_{||} \in \mathbb{N}^2$ at finite system size, but become continous in the thermodynamic limit. The out-of-plane component $k_z$ remains quantized as $k_z = \omega\_c n \in \mathbb{N}$ (without the 0) with fundamental cavity frequency $\omega\_c = \frac{c \pi}{d}$. The dispersion then reads
\begin{align}
    \omega(n, \vb{k}_\parallel) = \sqrt{(n \omega\_c)^2 + c^2 |\vb{k}_\parallel|^2}.
\end{align}
The cavity spectrum is hence gapped and quadratic for the in-plane momentum around each of the fundamental cavity resonances.

There are two possible polarization for the mode functions, with distinct physical meaning. On the one hand, we find transverse electrical modes (TE-modes)
\begin{align}
    \vb{f}\^{TE}_{\vb{k}_\parallel, n}(\vb{r}_{||}, z) = \sqrt{\frac{2}{L^2 d}} e^{-i \vb{k}_{||} \cdot \vb{r}_{||}} \sin(k_z z) \vb{e}\^{TE},
\end{align}
with $\vb{k}_{||} \cdot \vb{e}\^{TE} = 0$. Secondly, we find transverse magnetic modes (TM-modes) given by
\begin{align}
    \vb{f}\^{TM}_{\vb{k}_\parallel, n}(\vb{r}_{||}, z) = \frac{1}{N\_{TM}} e^{-i \vb{k}_{||} \cdot \vb{r}_{||}}
    \bigg(&
        \sin(k_z z) \vb{e}\^{TM} \\
        & \quad - i \frac{|\vb{k}_{||}|}{k_z} \cos(k_z z) \vb{e}_z
    \bigg),
    \nonumber
\end{align}
with $\vb{e}\^{TM} = \vb{k}_{||} / \norm*{\vb{k}_{||}}$ and the normalization $N\_{TM}^2 = \frac{L^2 d}{2 n^2 \omega\_c^2} \omega(n, \vb{k}_\parallel)^2$. Notice that these modes contain a contribution in $\vb{e}_z$ direction. Special care has to be taken for the $n = 0$ TM-mode, which reads $\vb{f}_{\vb{k}_\parallel, n = 0}\^{TM}(\vb{r}_{||}, z) = \sqrt{\frac{1}{2L^2 d}} e^{- i \vb{k}_\parallel \cdot \vb{r}_\parallel} \vb{e}_z$. It describes a homogeneous mode over the extent of the cavity with polarization in $z$ direction.

\begin{figure}[t]
    \centering
    \includegraphics{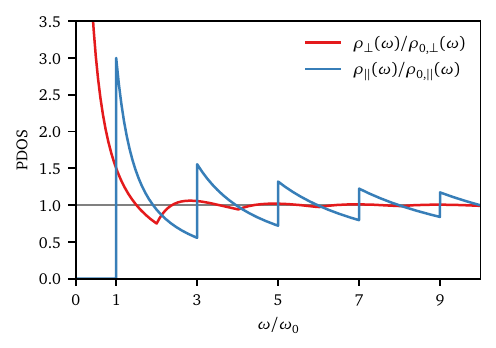}
    \caption{\textbf{Photonic Density of States} of a Fabry P\'erot cavity, measured at the center between the planar mirrors ($z = d / 2$). Shown are the in-plane (blue) and out-of-plane component, relative to their free space counterparts. The former density is relevant for the coupling to embedded 2D materials. Notice that the FP PDOS converges to the free space PDOS in the UV.}
    \label{fig:quant_FP_PDOS}
\end{figure}

We proceed to evaluate the in-plane and out of plane PDOS [\cref{eq:photonic_density_of_states}], the former of which couples to matter. In the thermodynamic limit, one finds
\begin{align}
    \label{eq:quant_fp_pdos_parallel}
    \rho_\parallel(\omega, z) &=
    \rho_0 \omega\_c \omega
    \sum_{n = 1}^{\floor{\omega / \omega\_c}}
    \left[ 1 + \left( \frac{n \omega\_c}{\omega} \right)^2 \right] \sin(\frac{n \pi z}{d})^2 \\
    \rho_\perp(\omega, z) &=
        \rho_0 \omega\_c \omega
        \left(
        \frac{1}{2} +
        \sum_{n = 1}^{\floor{\omega / \omega\_c}}
        \left[ 1 - \left( \frac{n \omega\_c}{\omega} \right)^2 \right]
        \right)
        \cos(\frac{n \pi z}{d})^2
\end{align}
where we defined the constant $\rho_0 = 1 / (3 \pi^2 c^3)$. In the following we will in particular be interested in situations where a 2D material is embedded in the middle of the cavity, viz. $z = d /2$. For this case, the in-plane and out-of-plane PDOS is shown in \cref{fig:quant_FP_PDOS}. The Fabry P\'erot cavity exhibits a gapped spectrum, where the spectral weight is periodically transferred to higher energy regions. The highly structured in-plane PDOS further calls into question commonly used single mode approximations for the FP cavity, a point we return to in \cref{ssec:appdx_single_mode}.

\subsection{Surface Polariton Cavity}
\label{sec:appdx_surface_cavity}

\begin{figure}[h]
    \centering
    \includegraphics{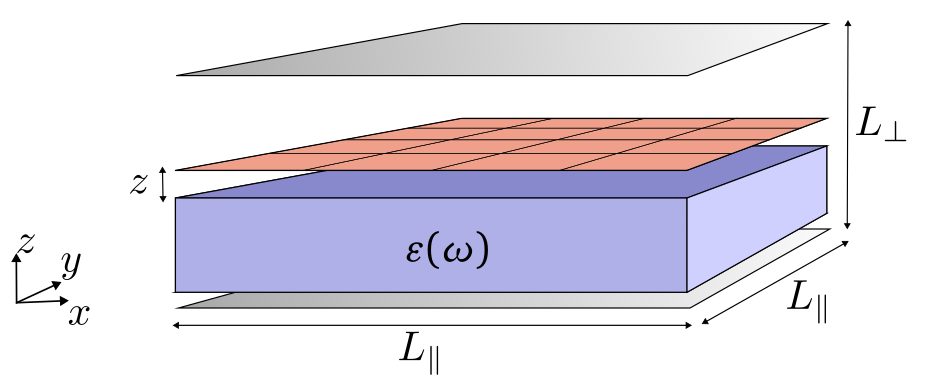}
    \caption{\textbf{Surface Cavity} constituted by an interface between a dispersive medium (blue), modeled via $\varepsilon(\omega)$ in the lower plane $z < 0$ and vacuum in the upper plane $z > 0$. The cavity couples to the cavity embedded material (red) with distance $z$ to the substrate. In-plane we use periodic boundary conditions with system size $L_\parallel$, while we impose perfect metallic boundary conditions out-of-plane at $z = \pm L_\perp/2$ to obtain discrete modes. Considering the thermodynamic limit, we obtain a setup with a single interface at $z = 0$.}
    \label{fig:quant_surface_geometry}
\end{figure}

In the main text we discussed the consistent quantization of polaritonic surface modes using a Coulomb gauge Hopfield quantization. In this section we explicitly calculate the mode functions, dispersion and PDOS for such a setup, using the approach introduced in \cite{eckhardtSurfacemediatedUltrastrong2024,eckhardtCavitronicsLowDimensional2024}.

The setup is illustrated in \cref{fig:quant_surface_geometry} and consists of a single interface between a dispersive medium in the lower plane $z < 0$ and vacuum in the upper plane $z > 0$. The substrate is effectively treated as a frequency dependent Lorenzian dielectric function \cite{jacksonClassicalElectrodynamics2009}
\begin{align}
    \varepsilon(z, \omega) &=  \varepsilon(\omega) \theta(z < 0) + \theta(z > 0), \\
    \text{with} \quad
    \varepsilon(\omega) &= \epsilon_\infty
    \left( 1 + \frac{\omega\_{TO}^2 - \omega\_{LO}^2}{\omega^2 - \omega\_{TO}^2} \right),
    \label{eq:quant_surface_dielectric}
\end{align}
that models the response of transverse $\omega\_{TO}$ and longitudinal $\omega\_{LO}$ phonon excitations of the substrate. We included $\varepsilon_\infty$ to model potential high energy excitations of the substrate. When setting $\omega\_{TO} \to 0$ and identifying the plasma frequency $\omega\_{LO} \equiv \omega\_P$, \cref{eq:quant_surface_dielectric} describes the lossless limit of a Drude metal instead.

To obtain discrete modes, we impose perfect metallic boundary conditions out-of-plane at $z = \pm L_\perp/2$. In this way, we obtain a setup with a single interface at $z = 0$ in the thermodynamic limit. Next to propagating solutions  we find modes that decay exponentially as we move away from the interface. In particularly the doubly evanescent mode -- the surface mode -- is of particular interest, but the solution for all modes follows the same principle.

\begin{figure}[b]
    \centering
    \includegraphics{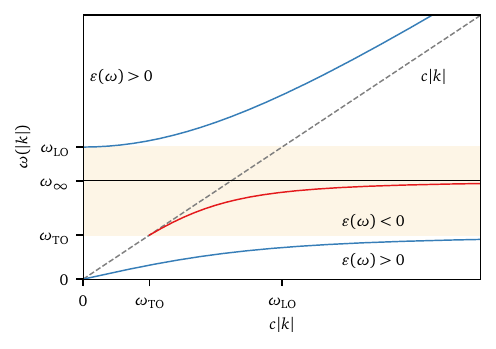}
    \caption{\textbf{Mode Dispersion} at an interface. Shown is the bulk dispersion (blue) as well as the surface dispersion (red). The surface mode only exists in a finite energy window $\omega\_{TO} < \omega < \omega_\infty$ with limit frequency $\omega_\infty$ [\cref{eq:quant_surface_limit_freq}]. A common approximation neglects the weak momentum dependence of the dispersion setting $\omega(|k|) \approx \omega_\infty$ for the surface mode.}
    \label{fig:quant_surface_dispersion}
\end{figure}

\paragraph*{General Structure ---} The in plane momentum $\vb{k}_\parallel = (k_x, k_y, 0)$ is continuous across the substrate interface [ \cref{eq:appdx_helmoltz_interface3}], while the out-of plane component $k_z$ jumps at $z = 0$ [\cref{eq:appdx_helmoltz_interface1}]. Using $\partial_i \vb{f}_\lambda \sim k_i^2 \vb{f}_\lambda$, the relation between $k_{z > 0} \equiv k_>$ and $k_{z < 0} \equiv k_<$ is determined by energy conservation at the interface
\begin{align}
    \omega^2 = \underbrace{c^2 \left( | \vb{k}_\parallel |^2 + k_>^2 \right)}_{z > 0} =
    \underbrace{\frac{c^2}{\varepsilon(\omega)} \left( | \vb{k}_\parallel |^2 + k_<^2 \right)}_{z < 0},
    \label{eq:appdx_surface_dispersion}
\end{align}
for a given energy $\omega$. \Cref{eq:appdx_surface_dispersion} is a quadratic equation for $z < 0$ and hence defines two (bulk) dispersion branches that we illustrated in \cref{fig:quant_surface_dispersion} (blue). The metallic boundary conditions, as well as the jump condition \cref{eq:appdx_helmoltz_interface3,eq:appdx_helmoltz_interface4} determine the general form of the mode functions. The TE polarized modes take the form
\begin{align}
    \vb{f}\^{TE}_{\vb{k}}(\vb{r}_\parallel, z \gtrless 0) &= N\_{TE}^{-1} e^{-i \vb{k}_\parallel \cdot \vb{r}_\parallel} \\
    &\times \sin( k^\gtrless \left( \frac{L_\perp}{2} \mp z \right) ) \sin( k^\lessgtr \frac{L_\perp}{2} ) \vb{e}\_{TE}
    \nonumber
\end{align}
with $\vb{k}_\parallel \cdot \vb{e}\_{TE} = 0$ and normalization constant determined according to \cref{eq:quant_hopfield_normalization}. The upper (lower) sign describes the mode function in the upper (lower) half plane respectively. The remaining interface condition at $z = 0$, viz. \cref{eq:appdx_helmoltz_interface2} leads to a transcendental equation for the out-of plane momenta $k_\gtrless$
\begin{align}
    \label{eq:appdx_interface_TE}
    k_> \cos(k_> \frac{L_\perp}{2})&\sin(k_< \frac{L_\perp}{2}) \\
    + k_< &\cos(k_< \frac{L_\perp}{2})\sin(k_> \frac{L_\perp}{2}) = 0.
    \nonumber
\end{align}
At a given energy $\omega$, \cref{eq:appdx_interface_TE} defines a set of allowed momenta $k_> \in \mathcal{K}\_{TE}(\omega)$, that lead to valid mode functions full-filling boundary and interface conditions as well as energy conservation. The oscillatory nature of \cref{eq:appdx_interface_TE} in general prohibits an easy evaluation of the thermodynamic limit $L_\perp \to \infty$. The set $\mathcal{K}\_{TE}(\omega)$ has hence to be determined via a numerical root finding problem and the out-of-plane system size remains a convergence parameter of our theory.

The interface equations \cref{eq:appdx_interface_TE} gives rise to different types of solutions, depending on whether $k_\gtrless$ are real or complex. The former yield propagating solutions, while the latter describe exponentially decaying modes, so called (r)evanescent modes. We find
\begin{itemize}
    \item \textbf{Propagating Modes} $k_\gtrless \in \mathbb{R}$, which exist at all energy scales $\omega$.
    \item \textbf{Evanescent Modes} $k_< \in \mathbb{R}, ik_> \in \mathbb{R}$, that exist for $\epsilon(\omega) < 1$, and describe modes that propagate only inside the material, while they decay exponentially outside.
    \item \textbf{Revanescent Modes} $k_> \in \mathbb{R}, ik_< \in \mathbb{R}$, also only exist for $\epsilon(\omega) < 1$. They are modes that propagate in free space and whose penetration into the substrate is exponentially suppressed.
    \item \textbf{Surface Modes}: $ik_\gtrless \in \mathbb{R}$, possible for $\epsilon(\omega) < 0$. These modes are exponentially localized at the interface and the dominant contribution in the current context. Their explicit mode functions and corresponding transcendental equation can be solved analytically (see below).
\end{itemize}

Next to the TE polarized modes, we find TM polarized solutions with finite out-of plane component. They read
\begin{gather}
\nonumber
    \vb{f}\^{TM}_{\vb{k}}(\vb{r}_\parallel, z \gtrless 0) = N\_{TM}^{-1} e^{-i \vb{k}_\parallel \cdot \vb{r}_\parallel}
    \left( f_\parallel(z) \vb{e}_{\vb{k}_\parallel} + f_\perp(z) \vb{e}_z \right) \\ 
    f_\parallel(z \gtrless 0) =
    \sin( k^\gtrless \left( \frac{L_\perp}{2} \mp z \right) ) \sin( k^\lessgtr \frac{L_\perp}{2} ) \hspace{-0.4cm} \\ \nonumber
    f_\perp(z \gtrless 0) = \pm i \frac{|\vb{k}_\parallel |}{k_\gtrless}
    \cos( k^\gtrless \left( \frac{L_\perp}{2} \mp z \right) ) \sin( k^\lessgtr \frac{L_\perp}{2} )
\end{gather}
with the in-plane unit vector $\vb{e}_{\vb{k}_\parallel} =  \vb{k}_\parallel / |\vb{k}_\parallel|$ and associated normalization constant $N\_{TM}$ determined according to \cref{eq:quant_hopfield_normalization}. The interface condition \cref{eq:appdx_helmoltz_interface2} takes the form of a transcendental equation
\begin{align}
    \label{eq:appdx_interface_TM}
    \frac{1}{k_>} \cos(k_> \frac{L_\perp}{2})&\sin(k_< \frac{L_\perp}{2}) \\+
    \frac{\epsilon(\omega)}{k_<} &\cos(k_< \frac{L_\perp}{2})\sin(k_> \frac{L_\perp}{2}) = 0,
    \nonumber
\end{align}
that defines the set of allowed momenta $k_> \in \mathcal{K}\_{TM}(\omega)$ for the TM modes. Analogous to the interface equation for the TE modes, \cref{eq:appdx_interface_TM} gives rise to four types of solutions, depending on whether $k_\gtrless$ is real or imaginary. We calculate the bulk modes numerically, using a adaptation of the numerical implementation developed in \cite{eckhardtSurfacemediatedUltrastrong2024}. 

\paragraph*{Surface Modes ---}

The surface modes, which were the focus in the main text can be determined analytically in the thermodynamic limit $L_\perp \to \infty$. The TE interface equation \cref{eq:appdx_interface_TE} does not have solutions for $ik_\gtrless \in \mathbb{R}$ such that there are only TM polarized surface modes.  These are  uniquely labeled by the in-plane momenta $\vb{q} = (q_x, q_y, 0)$. Solving the corresponding interface condition \cref{eq:appdx_interface_TM} gives a simple relation between the out-of-plane momenta in the thermodynamic limit \cite{eckhardtSurfacemediatedUltrastrong2024}
\begin{align}
    |k_>| + \frac{|k_<|}{\varepsilon(\omega)} = 0.
    \label{eq:appdx_interface_surface}
\end{align}
Using the energy conservation across the interface \cref{eq:appdx_surface_dispersion}, we can use \cref{eq:appdx_interface_surface} to determine the surface mode functions and corresponding dispersion, to find
\begin{widetext}
    \begin{gather}
        \vb{f}_{\vb{q}}(\vb{r}_{\parallel}, z \gtrless 0) = N\_S^{-1}
        \exp(\mp |\varepsilon(\omega_q)|^{\mp 1/2} |\vb{q}| z - i \vb{q} \cdot \vb{r}_{\parallel})
        \left( \vb{e}_{\vb{q}} + i |\varepsilon(\omega_q)|^{\pm 1/2} \vb{e}_z \right)
        \label{eq:quant_surface_mode_surface}
        \\
        \omega_q^2 =
        \frac{1}{2\varepsilon_\infty}
         \left(
            (1 + \varepsilon_\infty) c^2 q^2 +
            \varepsilon_\infty \omega\_{LO}^2
            -\sqrt{
            c^4 q^4 +
                \varepsilon_\infty^2 (\omega\_{LO}^2 - c^2 q^2)^2 +
                2\varepsilon_\infty c^2 q^2 (\omega\_{LO}^2 + c^2 q^2 - 2\omega\_{TO}^2)
                }
        \right),
        \label{eq:quant_surface_dispersion}
    \end{gather}
\end{widetext}
with normalization constant obtained form \cref{eq:quant_hopfield_normalization}
\begin{align}
    N\_S^2 &= \frac{\left(1 + |\varepsilon(\omega)| \right)}{2 q \sqrt{|\varepsilon(\omega)|}}
    \left(
        |\varepsilon(\omega)| + \frac{\mathcal{V}_>(\omega)}{|\varepsilon(\omega)|}
    \right), \\
    \text{ with } \quad
    &\mathcal{V}_>(\omega) =
    \varepsilon_\infty
    \left(
    1 - \omega\_{TO}^2
    \frac{\omega\_{TO}^2 - \omega\_{LO}^2}{\left( \omega^2 - \omega\_{TO}^2\right)^2}
    \right).
    \nonumber
\end{align}
The surface dispersion is shown in \cref{fig:quant_surface_dispersion} as a red curve. Note in particular that surface mode only exists in a narrow energy window $\omega \in (\omega\_{TO}, \omega_\infty)$, defined by $\epsilon(\omega) < 0$ and its energy quickly approaches its limit frequency
\begin{align}
    \omega_\infty = \lim_{q \to \infty} \omega_q = \sqrt{\frac{\varepsilon_\infty \omega\_{LO}^2 + \omega\_{TO}^2}{1 + \varepsilon_\infty}}.
    \label{eq:quant_surface_limit_freq}
\end{align}

\paragraph*{Photonic Density of States ---}

With the numerical, or in the case of surface modes analytical, evaluation of the mode function at hand, we proceed to derive the photonic density of states (PDOS). It is defined by
\begin{align}
    \rho_x(z, \omega) = \sum_{\lambda} \sum_{\vb{k}_\parallel} \sum_{k_> \in \mathcal{K}_\lambda(\omega)}
    \left| \vb{f}_{\vb{k}, \lambda}(\vb{r}) \cdot \vb{e}_x \right|^2 \delta( \omega - \omega_\lambda(\vb{k}) ),
    \nonumber
\end{align}
where $\lambda$ labels the type of mode and polarization. Note that $k_>$ is dependent on the in-plane momentum $\vb{k}_\parallel$ via the interface equations \cref{eq:appdx_interface_TE,eq:appdx_interface_TM}. Taking the thermodynamic limit in plane to transform the $\vb{k}_\parallel$ summation into an integral, we resolve the delta-distribution via variable transformation $\nu = \omega(\vb{k}_\parallel)$, leaving us to evaluate the functional determinant $\dd{\nu} = (\partial_\omega |\vb{k}_\parallel|)^{-1} \dd{|\vb{k}_\parallel|}$. For the bulk modes, we can't determine $k_>(\vb{k}_\parallel)$ analytically, so that the ensuing derivative has to be evaluated numerically. To that end, it is convenient to rewrite \cite{eckhardtSurfacemediatedUltrastrong2024}
\begin{align}
    \partial_\omega |\vb{k}_\parallel| = \frac{\omega}{c^2 |\vb{k}_\parallel|}
    \left(1 - \frac{c^2}{\omega}k_> \pdv{k_>}{\omega} \right),
\end{align}
where we have used the energy conservation across the interface [\cref{eq:appdx_surface_dispersion}] to reexpress the functional determinant in terms of a derivative of $k_>$ directly. Since $k_>$ is the solution of our transcendental interface equation that we implemented numerically, its derivate $\partial_\omega k_>$ can be directly evaluated via automatic differentiation. Finally, the PDOS can be determined via \cref{eq:quant_surface_pdos_bulk}
\begin{widetext}
    \begin{align}
    \rho_{ij}(z, \omega) &= \frac{L_\parallel^2}{(2\pi)^2}
    \frac{\omega}{c^2} \sum_\lambda \sum_{k_> \in \mathcal{K}_\lambda(\omega)}
    \left(\int_0^{2\pi} \dd{\varphi} \left| \vb{f}_{\vb{k}, \lambda}(\vb{r}) \cdot \vb{e}_{ij} \right|^2 \right)
    \frac{\omega}{c^2 |\vb{k}_\parallel|}
    \left(1 - \frac{c^2}{\omega}k_> \pdv{k_>}{\omega} \right),
    \label{eq:quant_surface_pdos_bulk} \\
    \rho_{ij}\^{surf}\left(z > 0, \omega\right)
    &= \rho_{0, ij}(\omega) \frac{3\pi}{2}
    \frac{1}{
    \sqrt{|\varepsilon(\omega)|}
    \left(1 + |\varepsilon(\omega)| \right)
    \left(1- |\varepsilon(\omega)|^{-1}\right)^{5/2}
    } 
    e^{- \frac{2z}{\sqrt{|\varepsilon(\omega)| - 1}} \left(\frac{\omega}{c}\right)}
    \Theta \big(\omega \in (\omega\_{TO}, \omega_{\infty})\big),
    \label{eq:quant_surface_pdos}
\end{align}
\end{widetext}

where the angular integration runs over the in plane direction of $\vb{k}_\parallel$ and the out-plane momentum summation over the set of allowed vectors $\mathcal{K}_\lambda(\omega)$ we determined numerically. For the surface modes, \cref{eq:quant_surface_pdos_bulk} can be evaluated analytically to find the lengthy expression in \cref{eq:quant_surface_pdos}.
\begin{figure*}[t]
    \centering
    \includegraphics{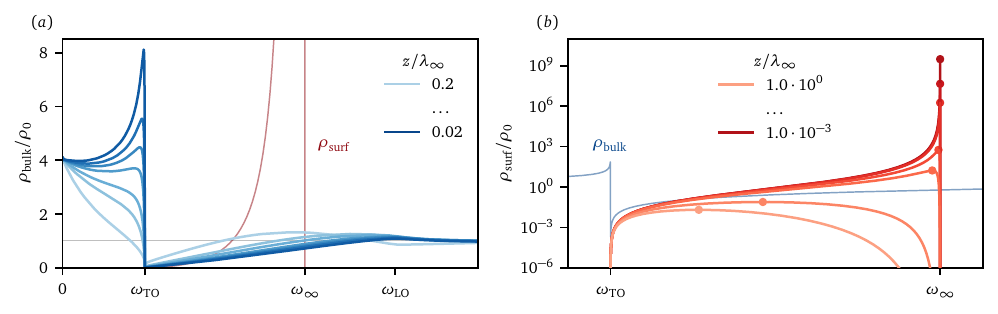}
    \caption{\textbf{Mode resolved photonic density of states of SrTiO$_3$} as a function of frequency for various distances above the substrate in natural units $\lambda_\infty = 2\pi c / \omega_\infty$. Results are shown relative to free space. (a) Bulk mode contribution to the PDOS. As we approach the surface, a peak emerges at the limit frequency ($k \to \infty$) of the bulk polaritonic dispersion [\cref{fig:quant_surface_dispersion}]. Note however the strong separation of scale between the bulk and surface PDOS, which is illustrated in red for the closest substrate distance $z  = 0.02\lambda_\infty$. (b) Surface mode PDOS in a logarithmic scale, illustrating the strong enhancement at the surface mode limit frequency $\omega_\infty$ [see \cref{eq:quant_surface_limit_freq}]. It outweighs the bulk contribution shown in blue for the closest substrate distance $z = 10^{-3} \lambda_\infty$ by many orders of magnitude.}
    \label{fig:appdx_interface_pdos}
    \label[fig_a]{fig:appdx_pdos_bulk}
    \label[fig_b]{fig:appdx_pdos_surface}
\end{figure*}

The PDOS for the surface modes of SrTiO$_3$, with $\omega\_{TO} = 7.92$ THz and $\omega\_{LO} = 32.04$ THz \cite{eckhardtSurfacemediatedUltrastrong2024}, separated into bulk and surface contributions, is shown in \cref{fig:appdx_interface_pdos}. Results are shown  as a function of frequency for various distances above the substrate in natural units $\lambda_\infty = 2\pi c / \omega_\infty$. We solved the transcendental interface equations for $L_\perp = 10$ m, which leads to results converged on the scale of the plots. In \cref{fig:appdx_pdos_bulk} the bulk PDOS close to the interface $z \leq 0.2\lambda_\infty$ is shown. As we approach the surface, a peak emerges at the limit frequency ($k \to \infty$) of the bulk polaritonic dispersion [\cref{fig:quant_surface_dispersion}]. Note however the strong separation of scale between the bulk and surface PDOS shown in \cref{fig:appdx_pdos_surface} and that exhibits a large peak at the limit frequency $\omega_\infty$. Since both enter additively [\cref{eq:quant_surface_pdos_bulk}], $\rho(\omega) - \rho_0(\omega) \approx \rho\^{surf}(\omega)$ provides an excellent approximation for $z \lesssim \lambda_\infty$ in many applications [\cref{sec:appdx_resum}]. Only at distances $z \gtrsim \lambda_\infty$ do the bulk modes become relevant \cite{eckhardtSurfacemediatedUltrastrong2024}, while at small distance to the substrate the strong weight increase of the surface PDOS at the limit frequency $\omega_\infty$, gives rise to dominant light-matter couplings that outweighs bulk contributions by many orders of magnitude.

\subsection{Single Mode Limit}
\label{ssec:appdx_single_mode}

The inclusion of many cavity modes, let alone all of them, is practically impossible in many applications because of the exponential complexity scaling. Single mode cavity approximations are hence commonly used in practice. Formally, the contribution of a single cavity mode vanishes in the thermodynamic limit, but the contribution of a narrow frequency band around a given resonance frequency can be finite \cite{svendsenTheoryQuantum2023}.

In the current discussion, a single mode cavity limit corresponds to a $\delta$-distribution in the difference photonic density of states at a resonance frequency $\omega_\star$, viz.
\begin{align}
    \rho(\omega) - \rho_0(\omega)  \approx
    \rho_0 \omega_\star^3 \bar{K}(\omega_\star) \delta(\omega - \omega_\star)
    \label{eq:resum_delta_approximation}
\end{align}
where we introduced a dimensionless weight function $\bar{K}(\omega_\star)$. There is no formally correct way to determine the weight function uniquely. However, for any given physical observable, the weight can be fixed by matching the observable's value evaluated with both sides of \cref{eq:resum_delta_approximation}. In principle, different observables will yield different weights, implying that the effective single-mode light-matter coupling is observable-dependent. A broad class of observables can be captured by defining the weight function as moments of the PDOS, viz. polynomial expectation values. In this case the weight function, determined from the $n$-th moment reads
\begin{align}
    \bar{K}^{(n)}(\omega_\star) = \rho_0^{-1} \int_0^\infty \dd{\omega} \frac{\omega^n}{\omega_\star^{n + 3}}
    \Big( \rho(\omega) - \rho_0(\omega) \Big).
    \label{eq:resum_weight_function}
\end{align}
Note that e.g. $n = 1$ would be proportional to the electrical field fluctuations. Importantly, if the extracted weight remains approximately constant across different polynomial orders, this indicates that the single-mode approximation is valid for the system under consideration.

\begin{figure}[b]
    \centering
    \includegraphics{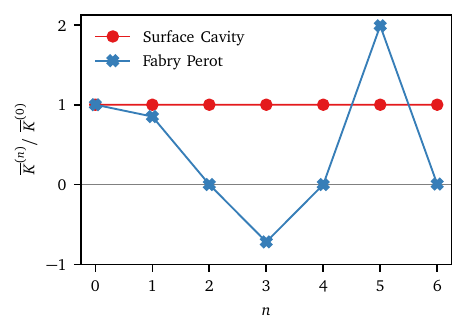}
    
    \caption{\textbf{Dimensionless Weight function} determined from the $n$-th moment of the PDOS [\cref{eq:resum_weight_function}], relative to the $n = 0$ weight. While the weights are approximately constant for the surface polariton cavity (red), we observe strong variations for the Fabry P\'erot (blue). In particular, the sign of the weight function depends on the order, implying that off-resonant coupling to a Fabry P\'erot cavity can not be approximated in single mode picture.}
    \label{fig:quant_single_mode}
\end{figure}

We illustrate this in \cref{fig:quant_single_mode} for the Fabry P\'erot (blue) and surface polariton cavity (red). The FP shows strong variation and in particular a sign flip as a function of moment order implying that no single mode approximation can be made. The moments of the surface cavity on the other hand are approximately constant implying that a $\delta$-approximation can be of high fidelity. We can understand this intuitively due to the strongly peaked nature of the surface PDOS [\cref{fig:appdx_pdos_surface}], while the FP leads to a more delocalized modification of the PDOS [\cref{fig:quant_FP_PDOS}].

The weight function of the surface cavity shows a scaling behavior as a functoin of the system parameters. With resonance at $\omega_\star = \omega_\infty$ we find for $\alpha, \beta \in \mathbb{R}$
\begin{align}
    \bar{K}\_{surf}(\omega_\infty, z) = (\alpha \beta)^{3} \bar{K}\_{surf}(\alpha \omega_\infty, \beta z),
    \label{eq:appdx_single_mode_weight_function_scaling}
\end{align}
a relationship very useful for extracting the scaling behavior of the magnetic exchange modification used in the main text. We used the single mode approximation of the surface cavity in \cite{fanCavityControl2026} to obtain realistic light-matter coupling constants in single mode calculations. Note that in order to obtain the weight of the single mode approximation, knowledge of the full $\rho(\omega)$ is necessary!

\section{Resummation}
\label{sec:appdx_resum}

\subsection{Evaluation for single mode limit}
\label{ssec:appdx_resum_single_mode}

For the surface mode contribution, the single-mode limit [\cref{ssec:appdx_single_mode}] provides high-fidelity results. The delta-distributed PDOS allows an explicit evaluation of the cavity correction function \cref{eq:resum_cavity_mod},
\begin{align}
    M(x) = \bar{g}^2 \big(e^{-x \theta} - 1\big),
\end{align}
with $\theta = \omega_\infty / U$ and effective single-mode coupling $\bar{g}^2 = P_0 \rho_0 \omega_\infty^2 \bar{K}\_{surf}(\omega_\infty, z)$ --- compare with \cref{eq:resum_taylor_g_eff}. Inserting into \cref{eq:resum_J} yields a closed-form expression in terms of incomplete gamma functions,
\begin{align}
    J = \tilde{J}_0 e^{-\bar{g}^2} \frac{\Gamma(\theta^{-1}) - \Gamma(\theta^{-1}, -\bar{g}^2)}{\theta\, \bar{g}^{1 / \theta}},
\end{align}
which admits two useful limiting expansions
\begin{align}
    J = \tilde{J}_0
    \begin{cases}
        \left(1 + \bar{g}^2 \theta + \order{\theta}^2 \right)^{-1}, 
        & \theta \ll 1, 
        \\
        \displaystyle\sum_{n} (-1)^n \bar{g}^{2n} \frac{\Gamma(1 + \theta^{-1})}{\Gamma(1 + \theta^{-1} + n)},
        & \bar{g}^2 \ll 1.
    \end{cases}
\end{align}
The expansion in $\theta$ is accurate for Fabry-P\'erot cavities and provides direct physical intuition, but breaks down for surface polariton setups where $\theta \sim \order{1}$, where the expansion in $\bar{g}^2 \lesssim 0.01$ is appropriate instead. While the two limits yield distinct leading-order corrections as a function of $\theta$, both are governed by the effective coupling $\bar{g}^2$ determined by the frequency-integrated PDOS. From the scaling of $\bar{K}\_{surf}$ [\cref{eq:appdx_single_mode_weight_function_scaling}] we extract $\Delta J \sim z^{-3}$, as used in the main text.

\subsection{Bulk Contribution}
\label{ssec:appdx_bulk_modes}

The vast separation in scale between bulk [\cref{fig:appdx_pdos_bulk}] and surface [\cref{fig:appdx_pdos_surface}] PDOS, suggest that bulk modes are irrelevant for the evaluation of the cavity induces magnetic exchange modifications $J$ [\cref{eq:resum_J}]. But while the surface modes are constrained to a small energy window, bulk modes couple on a much larger energy scale. In this section we explicitly show that bulk contributions can still be neglected against the surface modes, motivating the commonly used approximation $\rho(\omega) - \rho_0(\omega)\approx \rho\_{surface}(\omega)$.

To that end, we investigate this explicitly for a SrTiO$_3$ substrate and illustrate the results in \cref{fig:appdx_interface_pdos}. Specifically, we find that the surface modes (red) cause a decrease in the magnetic exchange $J$, while the bulk modes (blue) lead to an enhancement
\footnote{Note that while the PDOS enter additively in the cavity modification function \cref{eq:resum_cavity_mod}, this is not true for the magnetic modification \cref{eq:resum_J}!}. Critically, we find that the surface mode contribution (red) to the magnetic exchange is orders of magnitude larger compared to the bulk contribution (blue). Consequently, the net change (black dashes) is accurately modeled by considering only the surface modes leading to the previously stated approximation and results from the main text.
\begin{figure}[t]
    \centering
    \includegraphics{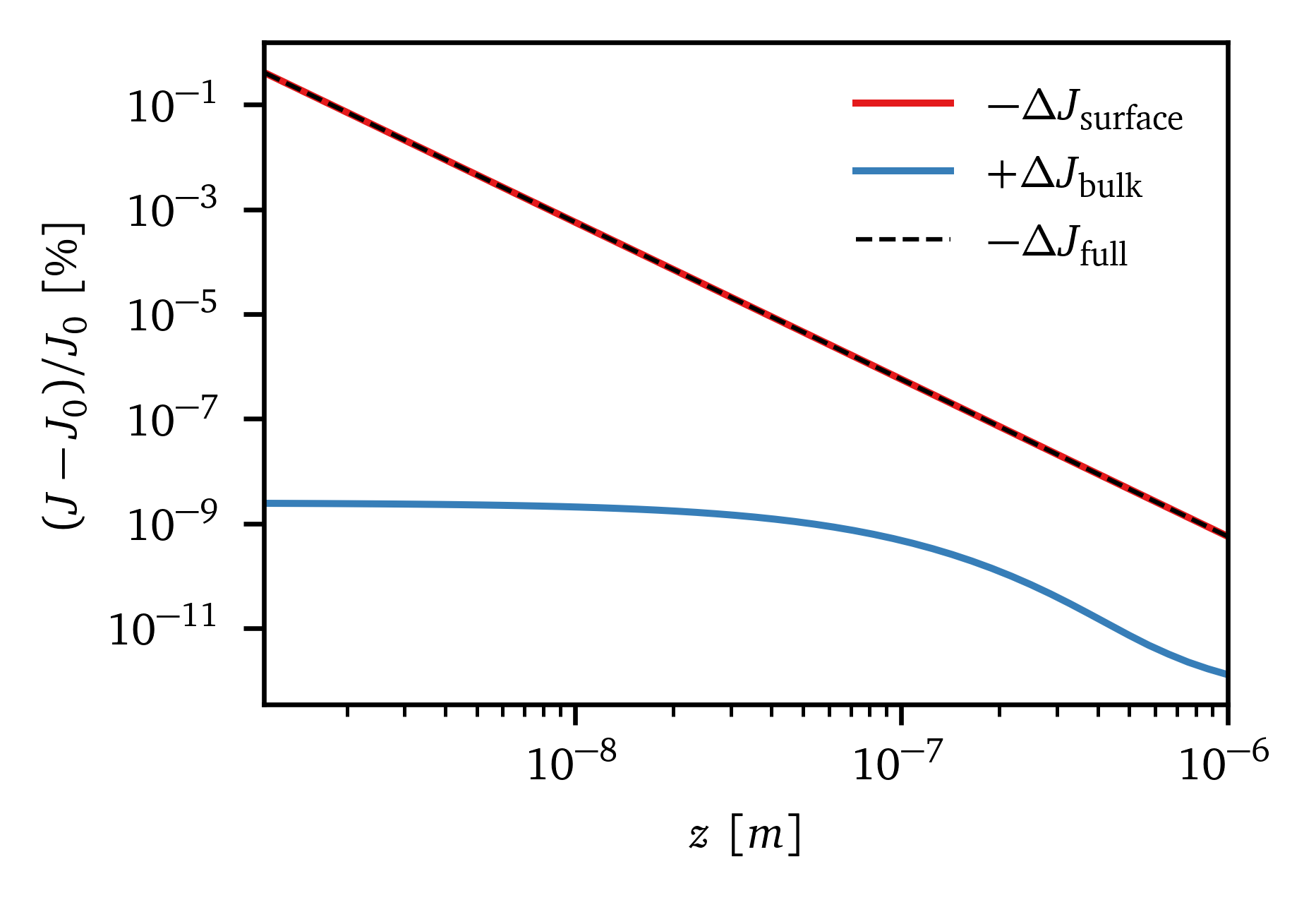}
    \caption{\textbf{Bulk and surface resolved magnetic modifications} due to coupling to surface (red) or bulk (blue) modes of a SrTiO$_3$ substrate previously introduced \cref{fig:appdx_interface_pdos}. The surface modes lead to a decrease of $J$, while the bulk modes enhance the magnetic exchange. Because of the exponentially large spectral density peak of the surface mode, its contribution dominates by many orders of magnitude. The total change (black dashes) can hence be evaluated considering only the surface modes.}
    \label{fig:appdx_interface_pdos}
\end{figure}

\subsection{Scaling of Surface Cavity}
\label{ssec:scaling_surface}

\begin{figure}[t]
    \centering
    \includegraphics{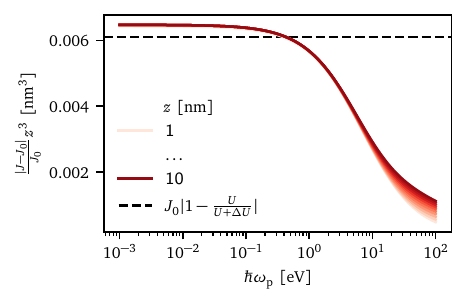}
    \caption{\textbf{Gauge contribution resolved scaling of dressed magnetic exchange} due to surface cavity coupling as a functions of the substrates plasma frequency $\omega\_p$ for $a_{ij} = 6\;$\AA and $U = 5\;$eV for various substrate distances $z$. Shown is the dynamical dressing (reds) and the static screening contribution (black) for fixed position. Notice the scaling collaps with $z^{3}$.}
    \label{fig:appdx_interface_magnetic_scaling}
\end{figure}

The contribution resolved scaling of the surface cavity induced magnetic exchange modification $J - J_0$ is shown in \cref{fig:appdx_interface_magnetic_scaling}. The dynamical dressing contribution shows a saturation behavior with weak linear decrease for small plasma frequencies $\hbar \omega\_p \lesssim U\;$eV, while it is strongly quenched at larger $\omega\_p$. In this regime the total effect is strongly dominated by the static screening. Note that both contributions enter the final result multiplicatively [\cref{eq:resum_J}], such that $ J = J_0 + \Delta J_{\vb{A}} + \Delta J_{\Delta U} + \dots$ only perturbatively, but accurate and useful for our intuitive understanding. 


\section{Truncated Exact Diagonalization for the Cavity-Coupled $t$-$J$ Polaron}
\label{app:tj_truncated_ed}

In this appendix we give the details of the truncated exact-diagonalization calculation used in \cref{sec:tj}. We first consider the nearest-neighbor $t$-$J$ Hamiltonian in the absence of cavity
\begin{align}
    H_\mathrm{tJ} = -t\sum_{\langle ij\rangle\sigma}
    \left(\tilde{c}_{i\sigma}^{\dagger}\tilde{c}_{j\sigma}+\mathrm{h.c.}\right)
    +J\sum_{\langle ij\rangle}
    \left(\vb{S}_i\cdot\vb{S}_j-\frac{1}{4}n_i n_j\right),
    \label{eq:app_tj_electron}
\end{align}
where $\tilde{c}_{i\sigma}=c_{i\sigma}(1-n_{i\bar{\sigma}})$. In the hole representation $d_{i\sigma}=\tilde{c}_{i\bar{\sigma}}^{\dagger}$, this becomes
\begin{align}
    H_\mathrm{tJ} = t\sum_{\langle ij\rangle\sigma}
    \left(d_{j\sigma}^{\dagger}d_{i\sigma}+\mathrm{h.c.}\right)
    +J\sum_{\langle ij\rangle}
    \left(\vb{S}_i\cdot\vb{S}_j-\frac{1}{4}n_i n_j\right).
    \label{eq:app_tj_hole}
\end{align}
We work in the single-hole sector. The truncated Hilbert space is generated by repeated action of the kinetic part on a parent state $\ket{\psi_0}$,
\begin{align}
    \ket{\psi_l}=H_t^l\ket{\psi_0}
    =\left[t\sum_{\langle ij\rangle\sigma}
    \left(d_{j\sigma}^{\dagger}d_{i\sigma}+\mathrm{h.c.}\right)\right]^l\ket{\psi_0}.
    \label{eq:app_tj_truncation}
\end{align}

\begin{figure}[t!]
    \centering
    \includegraphics[width=0.8\linewidth]{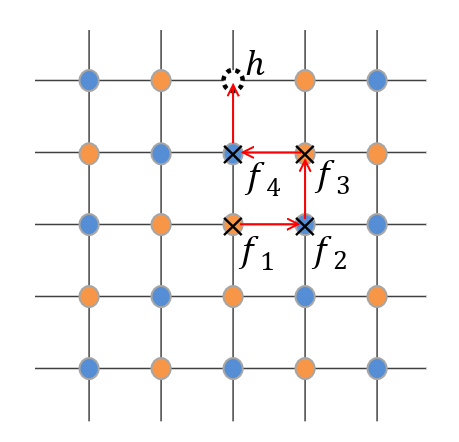}
    \caption{Illustration of a string state with $l=4$. Blue and orange circles indicate the N\'eel spin background. Black crosses indicate spin flips relative to the background, the flips from a string that connects the origin and the current hole position, which is indicate by a dashed circle.}
    \label{fig:string_state}
\end{figure}

\subsection{String basis and translational invariance}

The square lattice sites are labeled by integer coordinates $(m,n)$ with $-N_h\leq m,n\leq N_h$. We retain the diamond-shaped region $\mathcal{R}$ satisfying $|m|+|n|\leq N_h$, containing $L=N_h^2+(N_h+1)^2$ sites. The parent state has the hole at the origin. Applying $H_t$ once moves the hole to $\pm\hat{\vb{x}}$ or $\pm\hat{\vb{y}}$ and flips the spin left behind relative to the N\'eel background. A basis state with fixed origin is therefore represented as $\ket{h,f}$, where $h$ is the integer label of the hole position and $f$ is the array of spin-flip positions. One such string state is illustrated in Fig.~\ref{fig:string_state}. 

When the hole moves from $h$ to a new position $h'$, the spin-flip array is updated as follows. If $h'\in f$, the spin flip at $h'$ is removed, $f'=f\setminus\{h'\}$. Otherwise, the previous hole position is added, $f'=f\cup\{h\}$. Enumerating all hopping paths with $l\leq N_h$ and removing duplicates gives the truncated string basis $\ket{h,f}$.

To preserve translational invariance, we form Bloch states
\begin{align}
    \ket{\vb{k};h,f}=\frac{1}{\sqrt{N}}\sum_{\vb{R}}
    e^{i\vb{k}\cdot\vb{R}}T_{\vb{R}}\ket{h,f},
    \label{eq:app_tj_bloch}
\end{align}
where $T_{\vb{R}}$ translates both the hole and all spin flips by $\vb{R}=m\hat{\vb{x}}+n\hat{\vb{y}}$. For any translationally invariant operator $A$,
\begin{align}
    \bra{\vb{k};h',f'}A\ket{\vb{k};h,f}
    =\sum_{\vb{R}}e^{i\vb{k}\cdot\vb{R}}
    \bra{h',f'}A T_{\vb{R}}\ket{h,f}.
    \label{eq:app_tj_transinv_matrix}
\end{align}
The Bloch string states are not orthonormal. The overlap matrix is obtained from \cref{eq:app_tj_transinv_matrix} by setting $A=\mathbb{I}$, where each fixed-origin overlap $\bra{h',f'}T_{\vb{R}}\ket{h,f}$ is either zero or one. This overlap is retained in the sparse generalized eigenvalue problem at each momentum $\vb{k}$.

\subsection{Projected $t$-$J$ matrix elements}

Let $\vb{r}$ and $\vb{r}'$ denote the real-space positions associated with the hole labels $h$ and $h'$. The projected Hamiltonian matrix element is
\begin{equation}
\begin{aligned}
    &\bra{\vb{k};h',f'}H_\mathrm{tJ}\ket{\vb{k};h,f}
    =\\
    &\quad\quad\sum_{\bm{\tau}}e^{i\vb{k}\cdot(\vb{r}'-\vb{r}+\bm{\tau})}
    \bra{h',f'}H_\mathrm{tJ}T_{\vb{r}'-\vb{r}+\bm{\tau}}\ket{h,f},
    \label{eq:app_tj_projected_hamiltonian}
\end{aligned}
\end{equation}
with $\bm{\tau}=\vb{0},\pm\hat{\vb{x}},\pm\hat{\vb{y}}$. The nonzero matrix elements are evaluated as follows.

First, if $\ket{h',f'}=T_{\vb{r}'-\vb{r}}\ket{h,f}$, then $\bm{\tau}=\vb{0}$ and the diagonal exchange energy is
\begin{equation}
\begin{aligned}
   H^\mathrm{diag}_\mathrm{tJ}(h,f)
    &=2J\sum_i(f_i+h_i)-\frac{J}{2}\sum_{\langle ij\rangle}f_i f_j\\
    &-\frac{J}{2}\sum_{\langle ij\rangle}(f_i+h_i)(f_j+h_j).
    \label{eq:app_tj_diag}
\end{aligned}
\end{equation}
where $f_i$ is the occupation vector with entry one on spin-flip sites and zero otherwise, and $h_i$ is the corresponding occupation vector for the hole position.

Second, consider $\ket{h',f'}\neq T_{\vb{r}'-\vb{r}}\ket{h,f}$. Define the translated flips $f_1=T_{\vb{r}_{h'}-\vb{r}_h+\bm{\tau}}f$. If any element of $f_1$ lies outside $\mathcal{R}$, the matrix element is zero. We also define $f\ominus f'$ as the difference between $f$ and $f'$. For $\bm{\tau}=\vb{0}$, the off-diagonal exchange contribution is
\begin{align}
    H^\mathrm{ex}_{\mathrm{tJ}}(f_1,f')=
    \begin{cases}
    J/2, & f_1\ominus f' \in\mathrm{NN},\\
    J/2, & f'\ominus f_1\in\mathrm{NN},\\
    0, & \text{otherwise},
    \end{cases}
    \label{eq:app_tj_exchange_cases}
\end{align}
where $\mathrm{NN}$ is defined as the set of all nearest-neighbor pairs in $\mathcal{R}$. For $\bm{\tau}=\pm\hat{\vb{x}},\pm\hat{\vb{y}}$, the kinetic contribution is
\begin{align}
    H^\mathrm{kin}_\mathrm{tJ}(f_1,f')=
    \begin{cases}
    t, &  f_1\ominus f'=\{\vb{r}'\},\\
    t, &  f'\ominus f_1=\{\vb{r}'+\bm{\tau}\},\\
    0, & \text{otherwise}.
    \end{cases}
    \label{eq:app_tj_kinetic_cases}
\end{align}
The lowest eigenvalue of the Hamiltonian $H_\mathrm{tJ} = H^\mathrm{diag}_\mathrm{tJ}+H^\mathrm{ex}_{\mathrm{tJ}}+H^\mathrm{kin}_\mathrm{tJ}$ at fixed $\vb{k}$ gives the single-polaron dispersion.

\subsection{Surface-cavity matrix elements}

For the surface-cavity calculation, the hole Hamiltonian is dressed by the Peierls phase,
\begin{equation}
\begin{aligned}
    H_{\mathrm{m,eff}}&=t\sum_{\langle ij\rangle\sigma}
    \left(d_{j\sigma}^{\dagger}d_{i\sigma}e^{i\hat{\theta}_{ij}}+\mathrm{h.c.}\right)\\
    &\quad+J\sum_{\langle ij\rangle}
    \left(\vb{S}_i\cdot\vb{S}_j-\frac{1}{4}n_i n_j\right),
    \label{eq:app_tj_cavity_hamiltonian}
\end{aligned}
\end{equation}
where $\hat{\theta}_{ij}=e(\vb{R}_j-\vb{R}_i)\cdot\hat{\vb{A}}/\hbar$. In generating the fixed-origin basis we use
\begin{align}
    H_t^{(s)}=t\sum_{\langle ij\rangle\sigma}
    \left(d_{j\sigma}^{\dagger}d_{i\sigma}e^{is\hat{\theta}_{ij}}+\mathrm{h.c.}\right),
    \label{eq:app_tj_variational_hopping}
\end{align}
with a variational displacement parameter $s\in[0,1]$. After the hole has moved to position $\vb{r}_h$, the parent state is approximated as $\ket{h,f}\ket{\vb{r}_h}_{\mathrm{ph}}$. The photon displacement is assumed to depend only on $\vb{r}_h$, which is valid for the longitudinal surface mode used here. It is
\begin{equation}
\begin{aligned}
    \ket{\vb{r}_h}_{\mathrm{ph}}
    &=\exp\left[is\frac{e}{\hbar}\int_{\vb{0}}^{\vb{r}_h}d\vb{r}\cdot\hat{\vb{A}}(\vb{r})\right]\ket{0}_{\mathrm{ph}}\\
    &\approx \exp\left[is\frac{e}{\hbar}\vb{r}_h\cdot\hat{\vb{A}}\right]\ket{0}_{\mathrm{ph}},
    \label{eq:app_tj_displaced_state}
\end{aligned}
\end{equation}
where the last expression is the long-wavelength limit. Under translation,
\begin{align}
    T_{\vb{R}}\ket{\vb{r}_h}_{\mathrm{ph}}
    =\exp\left[is\frac{e}{\hbar}\int_{\vb{0}}^{\vb{r}_h}d\vb{r}\cdot\hat{\vb{A}}(\vb{r}+\vb{R})\right]\ket{0}_{\mathrm{ph}},
    \label{eq:app_tj_ph_translation}
\end{align}
which is invariant in the long-wavelength limit. The overlap factorizes as
\begin{align}
    \bra{h',f';\vb{r}_{h'}}T_{\vb{R}}\ket{h,f;\vb{r}_h}
    =\bra{h',f'}T_{\vb{R}}\ket{h,f}\braket{\vb{r}_{h'}}{\vb{r}_h}.
    \label{eq:app_tj_cavity_overlap}
\end{align}

The cavity-dressed Hamiltonian matrix elements follow the same case distinction as above. For the diagonal exchange part,
\begin{align}
    H_\mathrm{diag}
    =H^{\mathrm{diag}}_\mathrm{tJ}\braket{\vb{r}_{h'}}{\vb{r}_h}
    +\bra{\vb{r}_{h'} }H_\mathrm{ph}\ket{\vb{r}_h},
    \label{eq:app_tj_cavity_diag}
\end{align}
where $H_\mathrm{ph}=\sum_\lambda\omega_\lambda a_\lambda^{\dagger}a_\lambda$ with the zero-point energy removed. For an off-diagonal exchange process,
\begin{align}
     H_\mathrm{ex}
    = H^\mathrm{ex}_{\mathrm{tJ}}\braket{\vb{r}_{h'}}{\vb{r}_h}.
    \label{eq:app_tj_cavity_exchange}
\end{align}
For a kinetic process with $\bm{\tau}=\pm\hat{\vb{x}},\pm\hat{\vb{y}}$,
\begin{align}
    H_\mathrm{kin}
    =H^\mathrm{ex}_{\mathrm{tJ}}e^{-i(e/\hbar)\bm{\tau}\cdot\hat{\vb{A}}}\ket{\vb{r}_h}.
    \label{eq:app_tj_cavity_kinetic}
\end{align}

With $\vb{g}_\lambda=e\vb{f}_\lambda/\sqrt{2\hbar\omega_\lambda\epsilon_0}$, the required photon matrix elements are
\begin{subequations}
\label{eq:app_tj_general_photon_matrix}
\begin{align}
    &\braket{\vb{r}_{h'}}{\vb{r}_h}
    =e^{-\sum_\lambda\frac{[(s\vb{r}_h-s'\vb{r}_{h'})\cdot\vb{g}_\lambda]^2}{4\omega_\lambda}},\\
    \begin{split}
    &\bra{\vb{r}_{h'}}H_\mathrm{ph}\ket{\vb{r}_h}
    =\frac{1}{2}\sum_\lambda (s\vb{r}_h\cdot\vb{g}_\lambda)(s'\vb{r}_{h'}\cdot\vb{g}_\lambda)\times \\
    &\quad\quad\quad\quad\quad\quad\quad e^{-\sum_\lambda\frac{[(s\vb{r}_h-s'\vb{r}_{h'})\cdot\vb{g}_\lambda]^2}{4\omega_\lambda}},
    \end{split}\\
    &\bra{\vb{r}_{h'}}e^{-i(e/\hbar)\bm{\tau}\cdot\hat{\vb{A}}}\ket{\vb{r}_h}
    =e^{-\sum_\lambda\frac{[(s\vb{r}_h-s'\vb{r}_{h'}-\bm{\tau})\cdot\vb{g}_\lambda]^2}{4\omega_\lambda}}.
\end{align}
\end{subequations}
For a surface mode, $\lambda=\vb{q}$, $\omega_{\vb{q}}=\omega_s$, and
\begin{align}
    \vb{g}_{\vb{q}}
    =\frac{e}{\sqrt{2\hbar\omega_s\epsilon_0 S}}
    \frac{\omega_p}{2\omega_s}\hat{\vb{q}}\sqrt{q}e^{-ql}.
    \label{eq:app_tj_surface_gq}
\end{align}
Here $l$ is the distance to the surface in the matrix-element evaluation. Define $\alpha_{\mathrm{ph}}=\alpha\omega_p^2/(2\omega_s^2)$. For a plasmonic cavity, $\alpha_{\mathrm{ph}}=\alpha\approx1/137$. 
And expressing $\vb{r}_h$, $\vb{r}_{h'}$, and $\bm{\tau}$ in units of the lattice constant $a$, one obtains
\begin{subequations}
\label{eq:app_tj_surface_matrix_gtilde}
\begin{align}
    &\braket{\vb{r}_{h'}}{\vb{r}_h}
    =e^{-\frac{1}{4}\tilde{g}^2(s\vb{r}_h-s'\vb{r}_{h'})^2},\\
    &\bra{\vb{r}_{h'}}H_\mathrm{ph}\ket{\vb{r}_h}
    =\frac{1}{2}\tilde{g}^2\omega_s(s\vb{r}_h\cdot s'\vb{r}_{h'})
    e^{-\frac{1}{4}\tilde{g}^2(s\vb{r}_h-s'\vb{r}_{h'})^2},\\
    &\bra{\vb{r}_{h'}}e^{-i(e/\hbar)\bm{\tau}\cdot\hat{\vb{A}}}\ket{\vb{r}_h}
    =e^{-\frac{1}{4}\tilde{g}^2(s\vb{r}_h-s'\vb{r}_{h'}-\bm{\tau})^2}.
\end{align}
\end{subequations}
We have introduced the dimensionless light-matter coupling
\begin{align}
    \tilde{g}^2=\frac{\alpha_{\mathrm{ph}}c}{\omega_s}\frac{a^2}{8l^3},
    \label{eq:app_tj_gtilde}
\end{align}
The numerical calculations use the basis sectors with $s=1$, corresponding to the displaced photon state $\ket{\vb{r}_h}_{\mathrm{ph}}$, together with the $s=0$ photon-vacuum sector. The results in \cref{fig:tj} use $N_h=6$, $J=100\,\mathrm{meV}$, and $t=300\,\mathrm{meV}$ unless otherwise specified. The surface-mode frequency is $\omega_s=6.68\,\mathrm{eV}$ for gold and $\omega_s=100\,\mathrm{meV}$ for electron-doped Ge. At $z=1\,\mathrm{nm}$, these parameters give $\tilde{g}^2\simeq0.01$ for gold and $\tilde{g}^2\simeq0.65$ for electron-doped Ge.

\section{Raman spectra and spin structure factor}\label{app:raman}
We consider a Heisenberg model on the square lattice, given by
\begin{align}
 H = J \sum_{\langle ij\rangle} \bS_i \cdot \bS_j + K \sum_{\langle\!\langle ij\rangle\!\rangle} \bS_i \cdot \bS_j,
\end{align}
where $J$ and $K$ are the nearest and second nearest neighbor interactions, respectively. While the model considered in the main text only includes the nearest neighbor coupling, we also include the second nearest neighbor interaction here for generality. To obtain the magnon dispersion, we note that the square lattice antiferromagnet consists of two magnetic sublattices, and expand the respective spin operators as $S_i^z = S - a_i^\dagger a_i$, $S_i^+ = \sqrt{2S} a_i$ and $S_i^- = \sqrt{2S} a_i^\dagger$, and $S_i^z = b_i^\dagger b_i - S$, $S_i^+ = \sqrt{2S} b_i^\dagger$ and $S_i^- = \sqrt{2S} b_i$. 

The nearest neighbor exchange couples spins on opposite sub-lattices, while the next-nearest neighbor interaction involves spins on the same sub-lattice. We can therefore write
\begin{align}
 H &= J \sum_{\langle ij\rangle} \Big[S_i^z S_j^z + \frac{1}{2} (S_i^+ S_j^- + S_i^- S_j^+) \Big] \nonumber \\
 &+ K \sum_{\langle\!\langle ij\rangle\!\rangle} \Big[S_i^z S_j^z + \frac{1}{2} (S_i^+ S_j^- + S_i^- S_j^+) \Big] \nonumber \\
 &= \sum_\bk \Phi_\bk^\dagger \begin{pmatrix} h_0 + h_z & h_x -i h_y \\ h_x + i h_y & h_0 - h_z \end{pmatrix} \Phi_\bk,
\end{align}
where $\Phi_\bk^\dagger = (a_\bk^\dagger \; b_{-\bk})$ is a Nambu spinor, and the non-zero matrix elements are $h_0(\bk) = 4JS - 4KS + 2KS \sum_{\bf b} \cos(\bk \cdot {\bf b})$ and $h_x(\bk) = 2JS \sum_{\bf a} \cos( \bk \cdot {\bf a})$. Here, the sums are over the linearly independent nearest and next-nearest neighbor lattice vectors ${\bf a}$ and ${\bf b}$. This can be diagonalized via a Bogoliubov transformation $H = U^\dagger H U$ (note that $U^{-1} \neq U^\dagger$ to preserve the bosonic commutation relations)~\cite{VinasBostrom2021}
\begin{align}
 \begin{pmatrix} a_\bk^\dagger \\ b_{-\bk} \end{pmatrix} &= 
 \begin{pmatrix} \cosh\frac{\theta_\bk}{2} & e^{i\phi_\bk} \sinh\frac{\theta_\bk}{2} \\ e^{-i\phi_\bk} \sinh\frac{\theta_\bk}{2} & \cosh\frac{\theta_\bk}{2} \end{pmatrix} 
 \begin{pmatrix} \alpha_\bk^\dagger \\ \beta_{-\bk} \end{pmatrix} \\
 \begin{pmatrix} \alpha_\bk^\dagger \\ \beta_{-\bk} \end{pmatrix} &= 
 \begin{pmatrix} \cosh\frac{\theta_\bk}{2} & -e^{i\phi_\bk} \sinh\frac{\theta_\bk}{2} \\ -e^{-i\phi_\bk} \sinh\frac{\theta_\bk}{2} & \cosh\frac{\theta_\bk}{2} \end{pmatrix} 
 \begin{pmatrix} a_\bk^\dagger \\ b_{-\bk} \end{pmatrix}. \nonumber
\end{align}
The mixing angles are given by $\phi_\bk = -\arctanh(h_y/h_x) = 0$ and $\theta_\bk = -\arctan [(h_x^2 + h_y^2)^{1/2}/h_0]$, and the magnon energies are
\begin{align}
 \epsilon_{\alpha/\beta,\bk} = \sqrt{h_0^2 - h_x^2 - h_y^2} \mp h_z = \sqrt{h_0^2 - h_x^2}.
\end{align}
We note that these equations completely define the energies and eigenstates of the magnon system. \\


\subsection{Magnon Raman scattering}
The Raman scattering cross-section for incident radiation of polarization ${\bf e}_s$ and scattered radiation of polarization ${\bf e}_{s'}$ is given by~\cite{VinasBostrom2021,VinasBostrom2023}
\begin{align}
 P_{ss'}(\omega) = \sum_n |\langle\Psi_n| H_{\rm R}(s,s') |\Psi_0\rangle|^2 \delta (\hbar\omega + E_0 - E_n),
\end{align}
where $\omega = \omega_{\rm in} - \omega_{\rm sc}$ is the frequency difference between the incident and scattered light, $|\Psi_n\rangle$ is an excited magnon state with energy $E_n$, and $|\Psi_0\rangle$ the magnon ground state with energy $E_0$. The Raman Hamiltonian can be derived from the directional derivatives of the magnon Hamiltonian with respect to $\bk$~\cite{VinasBostrom2023}, such that
\begin{align}
 H_{\rm R}(s,s') &= - ({\bf e}_s \cdot \nabla_\bk) ({\bf e}_{s'} \cdot \nabla_\bk) H_\bk \nonumber \\
 &= I_0 \begin{pmatrix} f_0 & f_x \\ f_x & f_0 \end{pmatrix}_{ss'}
\end{align}
Here the prefactor quantifies the strength of the light-matter coupling, which is conventionally neglected by considering a normalized signal. The Hamiltonian is written in the spin-flip basis $(a_\bk^\dagger, b_{-\bk})$ using the structure factors
\begin{align}
 f_0(s,s') &= 2KS \sum_{\bf b} ({\bf e}_s \cdot {\bf b}) ({\bf e}_{s'} \cdot {\bf b}) \cos(\bk \cdot {\bf b}) \nonumber \\
 f_x(s,s') &= 2JS \sum_{\bf a} ({\bf e}_s \cdot {\bf a}) ({\bf e}_{s'} \cdot {\bf a}) \cos(\bk \cdot {\bf a}) \nonumber
\end{align}
Rotating this Hamiltonian into the magnon basis, using $U^\dagger H_R U$, we find a Hamiltonian of the form
\begin{align}
 H_{\rm R}(s,s') &= R_0 \sum_\bk \begin{pmatrix} \alpha_\bk^\dagger & \beta_{-\bk} \end{pmatrix}
 \begin{pmatrix} r_0 & r_x \\ r_x & r_0 \end{pmatrix}_{ss'}
 \begin{pmatrix} \alpha_\bk \\ \beta_{-\bk}^\dagger \end{pmatrix}.
\end{align}
Assuming low temperature, and starting from the magnon ground $|\Psi_0\rangle = |0\rangle$, the dominant signal will come from pair creation processes $\sim \alpha_\bk^\dagger \beta_{-\bk}^\dagger$, and we have 
\begin{align}
 P_{ss'}(\omega) = \sum_\bk |r_x(s,s',\bk)|^2 \delta (\omega - [\epsilon_{\alpha\bk} + \epsilon_{\beta\bk}]/\hbar).
\end{align}
Since the magnon bands are flat close to the Brillouin zone edges, resulting in a high density of states, we expect the spectrum to be dominated by these points. Further, as the magnon frequency $\omega_{\rm max} \approx 4J$, this signal will probe the cavity renormalization of $J$.


\subsection{Spin structure factor}
We also evaluate the spin structure factor, which is directly related to the magnon dispersion, and measurable both in inelastic neutron scattering or resonant inelastic X-ray scattering experiments. More precisely, the spin structure factor is defined by
\begin{align}
 S_\bq^{\mu\nu}(\omega) &= \int dt e^{-i\omega t} \sum_{ij} e^{-i\bq \cdot (\br_i - \br_j)} \langle\Psi_0| S_i^\mu(t) S_j^\nu |\Psi_0\rangle,
\end{align}
where $|\Psi_0\rangle$ is the magnon ground state, and $\mu$ and $\nu$ denote components of the spin vector. We here focus on the transverse component $\mu = +$ and $\nu = -$, which corresponds to a spin flip and thereby probes magnon excitations.

Assuming that the spin at lattice site $\br_i$ lives on sub-lattice A, we can break the sum into sites $j \in B$ and $j \in A$ such that
\begin{align}
 &S_\bq^{+-}(\omega) = \int dt e^{-i\omega t} 2S \bigg[ \sum_{ij} e^{-i\bq \cdot (\br_i - \br_j)} \langle\Psi_0| b_i^\dagger(t) a_j^\dagger |\Psi_0\rangle \nonumber \\
 &\hspace*{3.4cm}+ \sum_{ij} e^{-i\bq \cdot (\br_i - \br_j)} \langle\Psi_0| a_i(t) a_j^\dagger |\Psi_0\rangle \bigg] \nonumber \\
 &\hspace*{0.2cm}= \int dt e^{-i\omega t} 2S \big[ \langle\Psi_0| b_{-\bq}^\dagger(t) a_\bq^\dagger |\Psi_0\rangle + \langle\Psi_0| a_{\bq}(t) a_\bq^\dagger |\Psi_0\rangle \big]. \nonumber
\end{align}
We note that evaluated in the magnon vacuum, the corresponding term with $a^\dagger \to b$ will vanish, but that an analogous contribution is expected to appear in $S_\bq^{-+}$. To evaluate the frequency integral, we insert a complete set of states $|\Psi_n\rangle$ between the operators, giving
\begin{align}
 S_\bq^{+-}(\omega) &= 2S \sum_n \big[ \langle\Psi_0| b_{-\bq}^\dagger |\Psi_n\rangle \langle \Psi_n| a_\bq^\dagger |\Psi_0\rangle \\
 &+ \langle\Psi_0| a_{\bq} |\Psi_n\rangle \langle \Psi_n| a_\bq^\dagger |\Psi_0\rangle \big] \delta(\omega - E_n + E_0). \nonumber
\end{align}
Using the notation $\Psi_\bk = U \Phi_\bk$ and $\Psi_\bk^\dagger = \Phi_\bk^\dagger U^\dagger$, we can write this in terms of the magnon operators as
\begin{align}
 S_\bq^{+-}(\omega) &= 2S \sum_{n} \big[ U_{12}^{*} U_{11} + U_{11}^{*} U_{11} \big] \delta(\hbar\omega - \epsilon_{\alpha\bk})
\end{align}
A completely similar expression is obtained by starting with a spin on sub-lattice $B$ (by $a \to b^\dagger$), giving
\begin{align}
 S_\bq^{+-}(\omega) &= 2S \sum_{n} \big[ U_{12}^{*} U_{21} + U_{11}^{*} U_{21} \big] \delta(\hbar\omega - \epsilon_{\alpha\bk})
\end{align}

Putting the pieces together, we find the transverse spin structure factor
\begin{align}
 S_\bq^{+-}(\omega) &= 2S \sum_{n} \delta(\hbar\omega - \epsilon_{\alpha\bq}) \\
 &\times \big[ U_{12}^{*} U_{11} + U_{11}^{*} U_{11} + U_{12}^{*} U_{21} + U_{11}^{*} U_{21} \big].\nonumber
\end{align}
The other transverse component $S^{-+}$ can be obtained by replacing $a \leftrightarrow b$ in all equations above. This leads to a similar expression but probing the $\beta$-branch of the magnons. Physically, we rationalize this as the $\alpha$-magnons carrying a spin angular momentum $\hbar$, while the $\beta$-magnons carry a spin angular momentum $-\hbar$. The end result is
\begin{align}
 S_\bq^{-+}(\omega) &= 2S\sum_{n} \delta(\omega - \epsilon_{\beta\bq}/\hbar) \\
 &\times \big[ U_{12} U_{22}^{*} + U_{22} U_{22}^{*} + U_{12} U_{21}^{*} + U_{22} U_{21}^{*} \big]. \nonumber
\end{align}


\clearpage

\bibliographystyle{apsrev4-1}
\bibliography{bibliography_automatic.bib,bibliography.bib}

\end{document}